\documentclass[acmtog]{acmart}
\acmSubmissionID{TOG-19-0069.R1}

\usepackage{booktabs} 

\setcitestyle{square}

\usepackage[ruled]{algorithm2e} 
\SetAlFnt{\small}
\SetAlCapFnt{\small}
\SetAlCapNameFnt{\small}
\SetAlCapHSkip{0pt}
\IncMargin{-\parindent}

\newcommand{\R}{\mathbb{R}}
\DeclareMathOperator{\Sym}{Sym}
\DeclareMathOperator{\stab}{stab}
\DeclareMathOperator{\tr}{tr}
\DeclareMathOperator{\Span}{span}
\DeclareMathOperator{\rank}{rank}
\DeclareMathOperator*{\argmin}{arg\,min}

\DeclareMathOperator{\dist}{dist}
\DeclareMathOperator{\Ltrans}{L}
\DeclareMathOperator{\Rtrans}{R}
\newcommand{\VOcta}{F}

\newcommand{\VOdeco}{\tilde{F}}
\newcommand{\SO}{\mathrm{SO}}
\newcommand{\GL}{\mathrm{GL}}
\newcommand{\Oct}{\mathrm{O}}
\newcommand{\tran}{^{\mathstrut\scriptscriptstyle{\top}}}

\usepackage{todonotes}
\usepackage{pgfplots}
\usepackage{bbm}
\usepackage{subcaption}
\usepackage{siunitx}
\usepackage{graphbox,tabu}
\usepackage{etoolbox}

\usetikzlibrary{cd}

\usepackage{mystyle}

\setcopyright{rightsretained} 
\acmJournal{TOG}
\acmYear{2020} \acmVolume{39} \acmNumber{2} \acmArticle{16} \acmMonth{3} \acmPrice{}\acmDOI{10.1145/3366786}

\received{August 2019}
\received[revised]{December 2019}
\received[accepted]{December 2019}

\begin{document}
\title{Algebraic Representations for Volumetric Frame Fields}

\author{David Palmer}
\orcid{0000-0002-1931-5673}
\affiliation{%
  \institution{Massachusetts Institute of Technology}
  \streetaddress{77 Massachusetts Avenue}
  \city{Cambridge}
  \state{MA}
  \postcode{02139}
  \country{USA}}
\email{drp@mit.edu}

\author{David Bommes}
\affiliation{%
  \institution{University of Bern}
  \streetaddress{Hochschulstrasse 6}
  \city{Bern}
  \postcode{3012}
  \country{Switzerland}}
\email{david.bommes@inf.unibe.ch}

\author{Justin Solomon}
\orcid{0000-0002-7701-7586}
\affiliation{%
  \institution{Massachusetts Institute of Technology}
  \streetaddress{77 Massachusetts Avenue}
  \city{Cambridge}
  \state{MA}
  \postcode{02139}
  \country{USA}}
\email{jsolomon@mit.edu}

\renewcommand\shortauthors{Palmer et al.}

\begin{abstract}
Field-guided parametrization methods have proven effective for quad meshing of surfaces; these methods compute smooth \emph{cross fields} to guide the meshing process and then integrate the fields to construct a discrete mesh.
A key challenge in extending these
methods to three dimensions, however, is \emph{representation}
of field values.
Whereas cross fields can be represented by tangent vector fields that form
a linear space, the 3D analog---an octahedral frame field---takes values in a nonlinear manifold.
In this work, we describe the space of octahedral frames in the
language of differential and algebraic geometry.
With this understanding, we develop geometry-aware tools for
optimization of octahedral fields, namely
geodesic stepping and exact projection via semidefinite relaxation.
Our algebraic approach not only provides an elegant and mathematically-sound description
of the space of octahedral frames but also suggests a generalization to frames
whose three axes scale independently, better capturing the singular behavior we expect
to see in volumetric frame fields.
These new \emph{odeco frames}, so-called as they are represented by orthogonally
decomposable tensors, also admit a semidefinite program--based 
projection operator.
Our
description of the spaces of octahedral and odeco frames suggests
computing frame fields via manifold-based optimization algorithms;
we show that these algorithms efficiently produce high-quality fields
while maintaining stability and smoothness.
\end{abstract}

\ccsdesc[500]{Computing methodologies~Computer graphics}
\ccsdesc[500]{Computing methodologies~Mesh geometry models}
\ccsdesc[500]{Computing methodologies~Volumetric models}

\keywords{Hexahedral meshing, octahedral frame fields, convex relaxations,
convex algebraic geometry}

\begin{teaserfigure}
\newcommand{\imageheight}{0.18\textwidth}
\centering
\includegraphics[height=\imageheight]{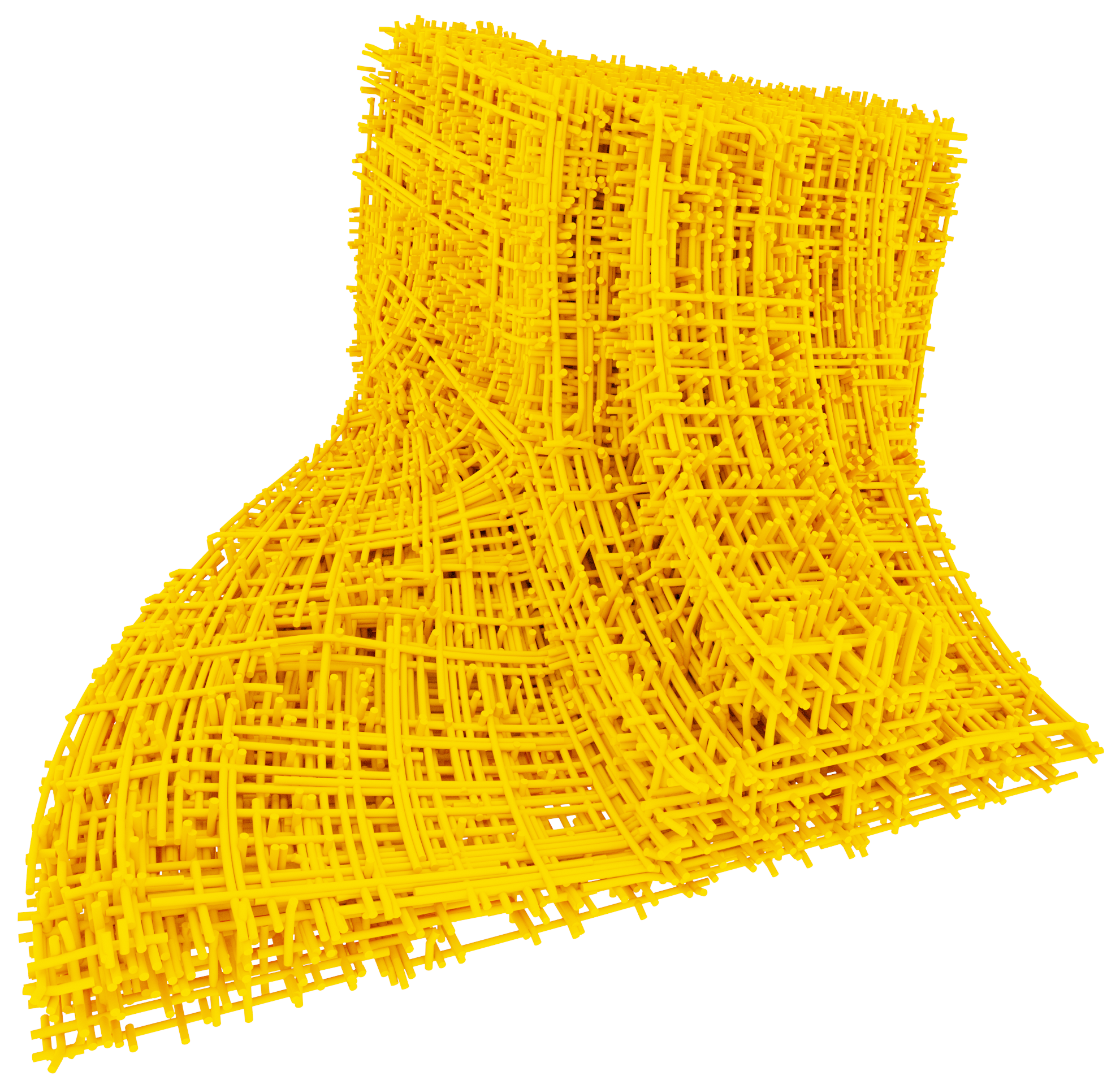}%
\hfill%
\includegraphics[height=\imageheight]{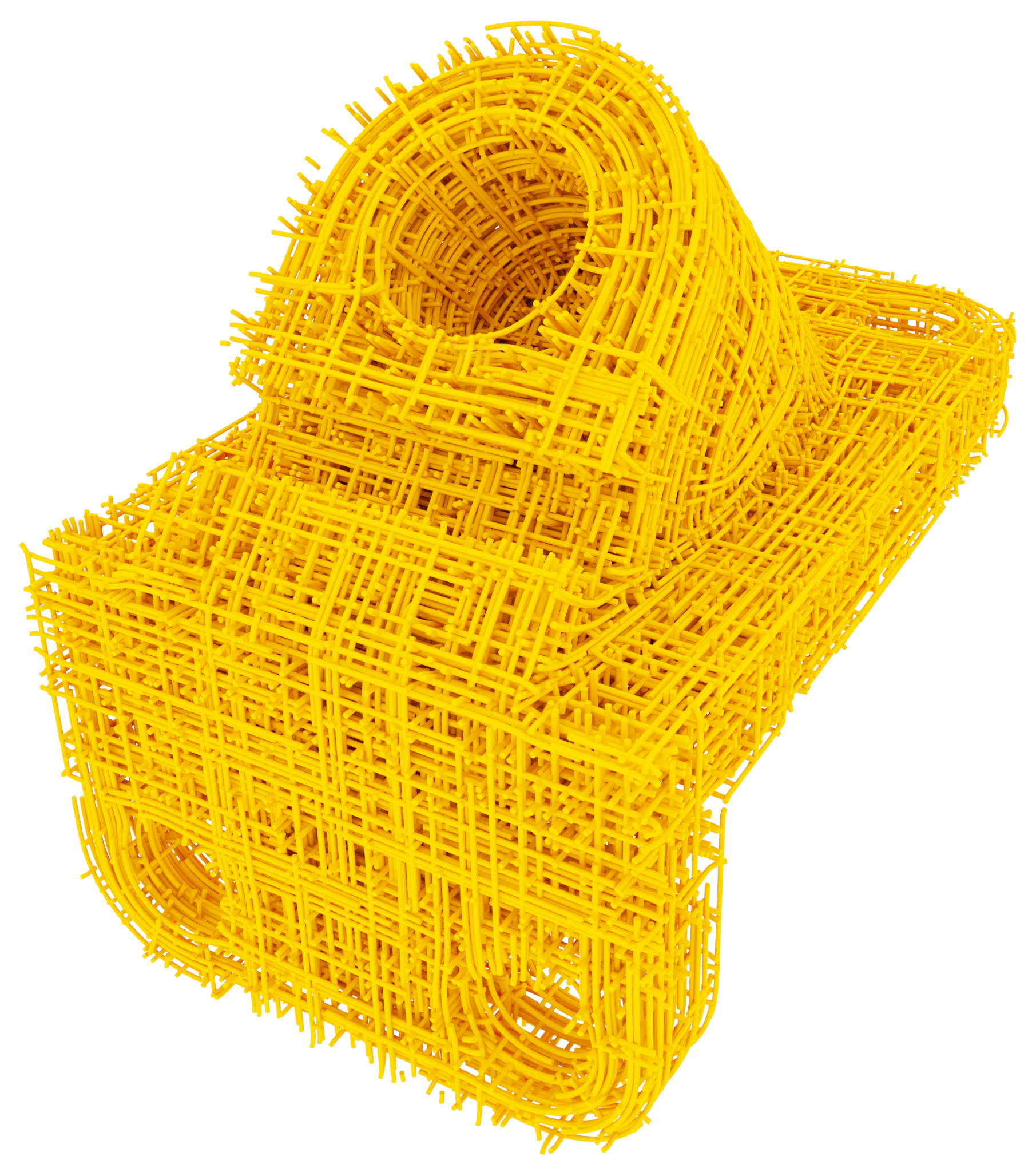}%
\hfill%
\includegraphics[height=\imageheight]{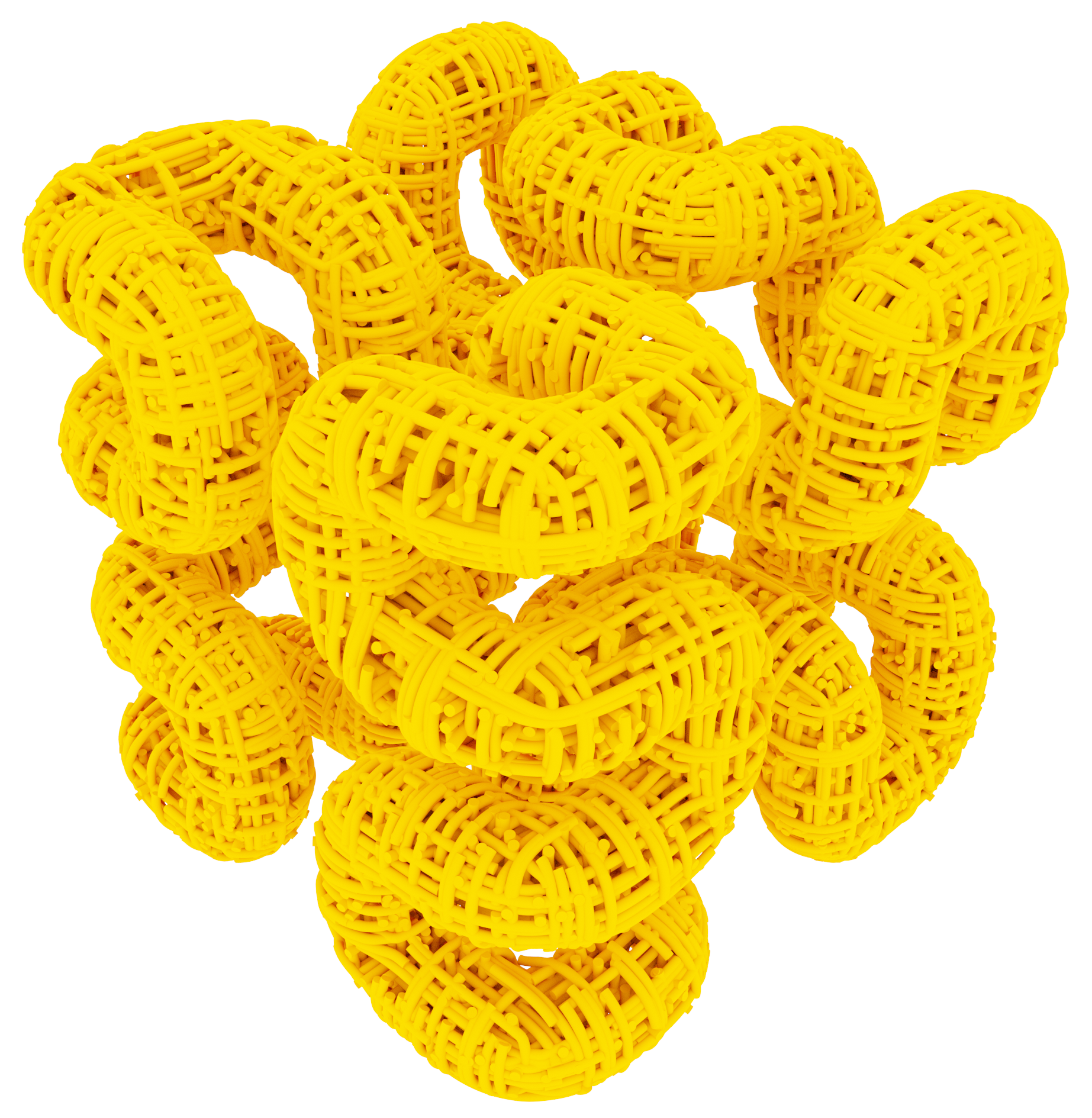}%
\hfill%
\includegraphics[height=\imageheight]{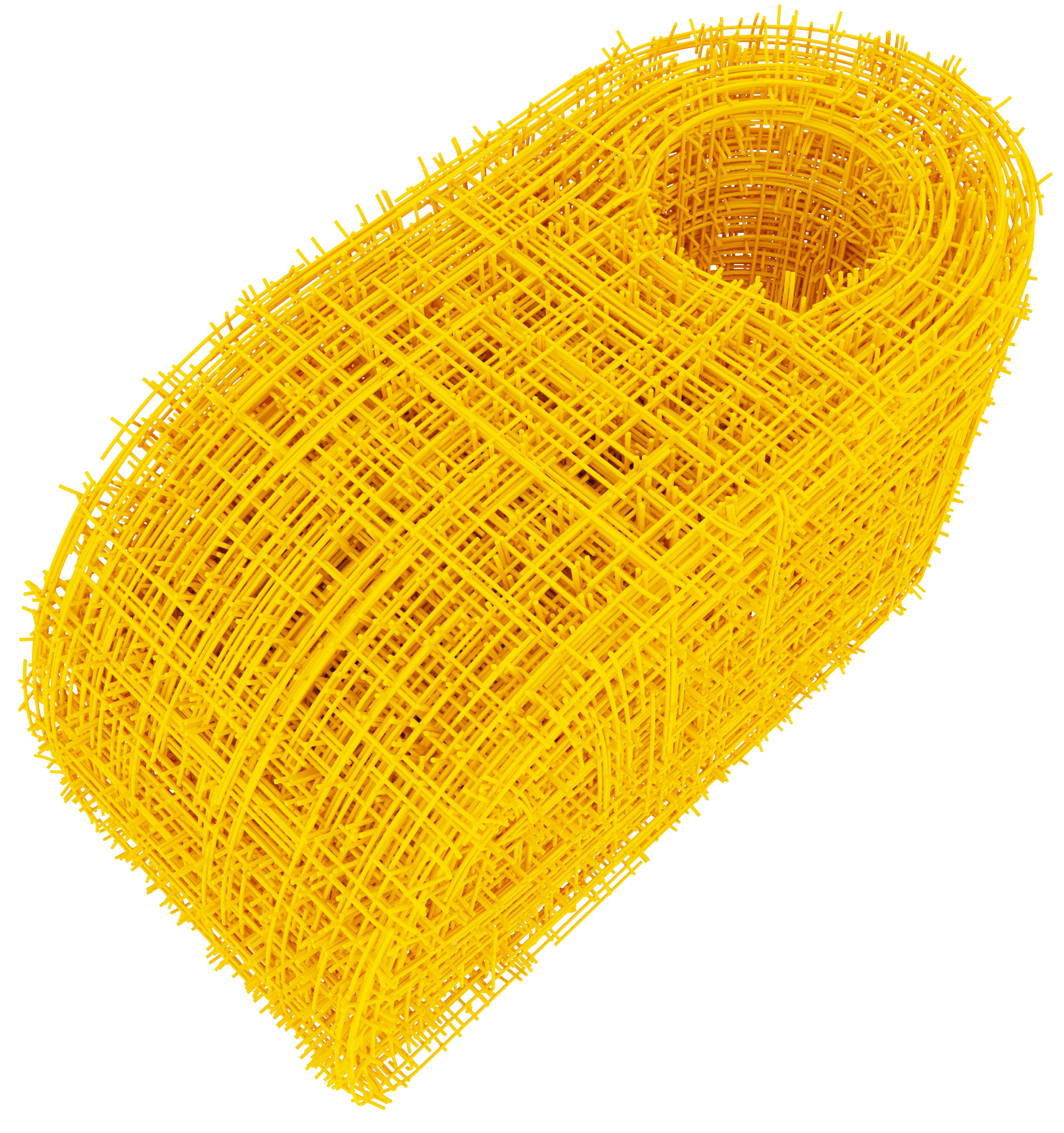}%
\hfill%
\includegraphics[height=\imageheight]{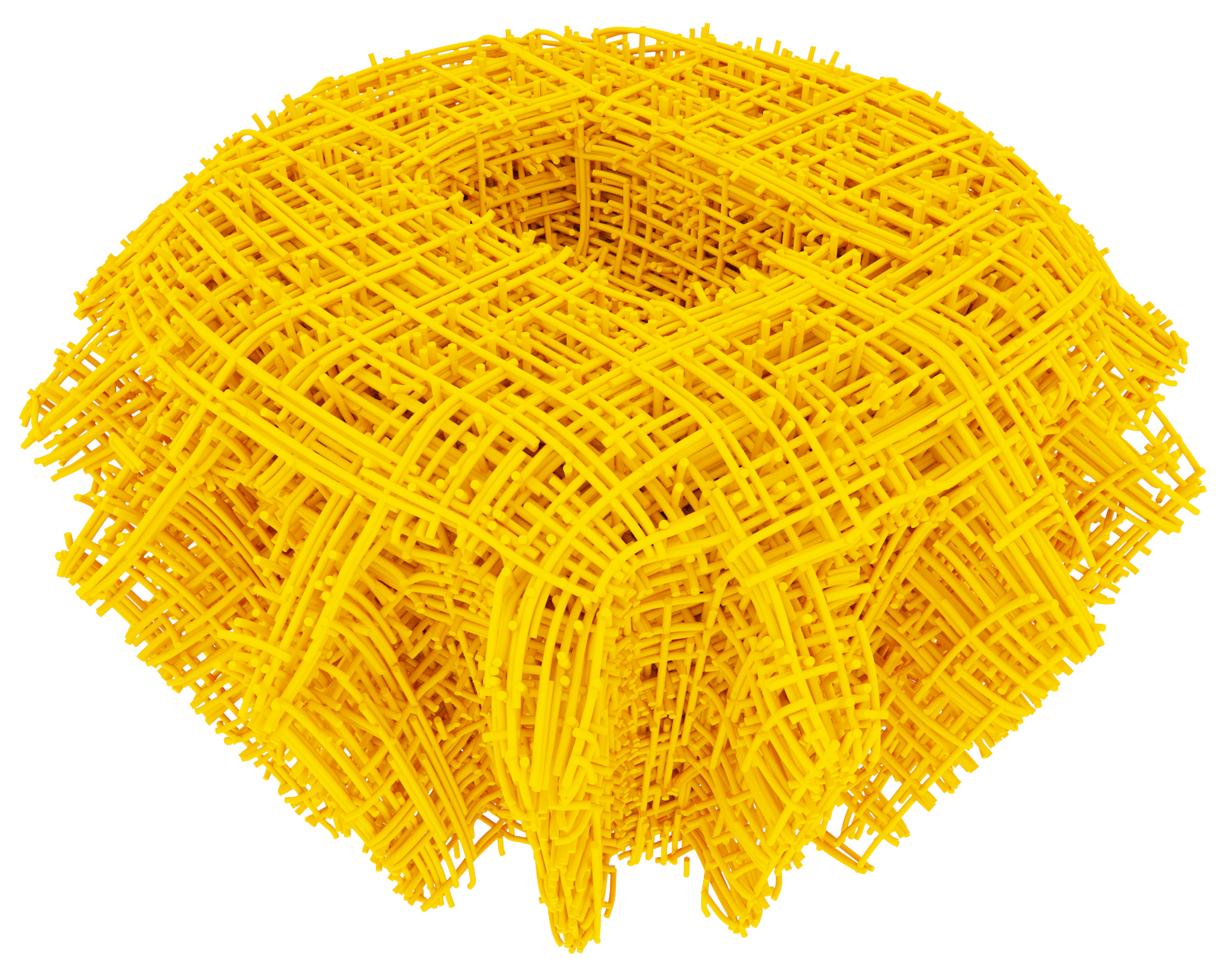}%
\caption{Octahedral fields generated with our methods on various models.}
\label{fig.teaser}
\end{teaserfigure}

\maketitle

\section{Introduction}

\begin{figure}
\centering
\includegraphics[width=0.7\columnwidth]{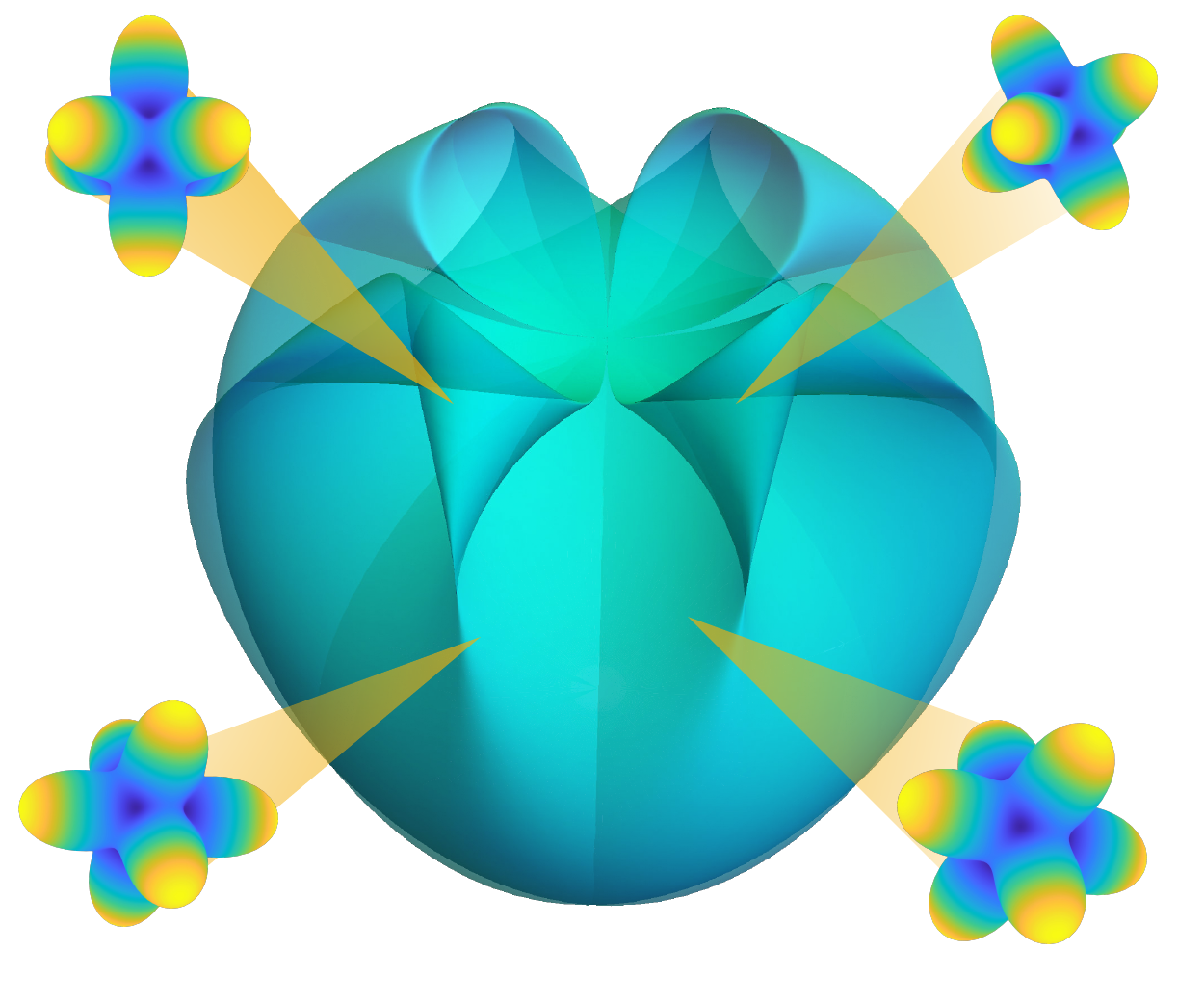}
\vspace{-0.2cm}
\caption{We compute frame fields as maps into the \emph{octahedral} and \emph{odeco varieties}, a projected slice of which is depicted here.}
\label{fig.variety-slice}
\vspace{-0.5cm}
\end{figure}

Inspired by the success of field--based approaches to quadrilateral meshing
on surfaces (\emph{cf.} \cite{vaxman2016}), recent research in applied geometry has focused on developing an
analogous approach to hexahedral meshing.
Motivated by applications in finite-element modeling, hexahedral meshing
is the problem of dividing a given volume into hexahedral elements (deformed cubes)
with minimal distortion and such that mesh boundary faces are aligned to the boundary
of the volume.

Hexahedral meshing couples a geometry problem---minimizing distortion
of mesh elements---to a combinatorial problem---placing mesh elements to achieve a desired
connectivity structure.
As in 2D, field-based approaches first ignore combinatorial constraints
and solve for a \emph{frame field}, which represents the local alignment
and singular structure of a mesh (see Figure \ref{fig.teaser}).
Then, they integrate that field to guide the placement of hex elements~\cite{nieser2011cubecover}.
So as not to impose unnatural constraints, the space
of frame fields must be expressive enough to represent the range of possible singularities that
may appear in hexahedral meshes.
These singularities are described by gluing relations
restricted to the symmetries of a cube, i.e., the octahedral group.

One might hope that 2D field-based methods would extend easily to 3D.
There are at least two obstructions to transferring ideas from the 2D case.
First, the singular structure of
a 3D frame field can be much more complicated than that of a cross field,
comprising an embedded graph rather than a set of isolated points. 
Second, we must understand the geometry of the space of frames to measure and optimize smoothness of a frame field.

Complementing the recent results of \citet{liu} on frame field singularities
and necessary conditions for hex-meshability,
in the present work we study the second challenge, namely characterizing the geometry of
the space of frames.
A 3D frame field is, intuitively, an assignment of three mutually orthogonal directions
to each point in a volume. An orthonormal basis of three
vectors---comprising an orthogonal matrix---is sufficient to specify a frame,
but thanks to octahedral symmetry, multiple orthogonal matrices can specify the same frame.
Formally, the space of octahedral frames can be viewed as
the quotient of the group of 3D rotations
$\SO(3)$ by the right action of
the octahedral group $\Oct$ comprising the rotational symmetries of a cube
(see \S\ref{sec.frame-space}).
The non-commutativity of 3D rotations makes the geometry of this quotient more complicated
than that of its 2D counterpart.
Recent attempts to lift ideas and techniques from 2D have either ignored the
geometry of the frame space entirely or treated it as a black box for
nonlinear optimization.

Consider perhaps the simplest form of optimization on a space, projection---finding
the closest point in the space to a given point in an ambient space.
Previous work on frame fields has treated this projection problem
as a nonlinear, nonconvex optimization problem
over frames parametrized by Euler angles, with no guarantees on
convergence or global optimality.
Our description of the octahedral frame space as an algebraic variety
suggests a different approach to projection based on
semidefinite programming, which yields a certificate of global optimality in polynomial time.
Our semidefinite relaxation of projection is exact in a neighborhood of
the octahedral variety, and we conjecture---with strong empirical evidence---that it is
so universally.

Even when conducting local optimization on the space of octahedral frames, parametrization
by Euler angles may not be the best approach. We show that the map
from $\SO(3)$ to the octahedral variety is a local isometry, enabling us to compute
geodesics on the octahedral variety in closed form. Manifold-based optimization
that moves along geodesics can then be used to accelerate local optimization dramatically.

Beyond precisely characterizing the space of octahedral frames, our algebraic approach admits a generalization to frames whose axes scale independently.  This larger space better captures frame field geometry, e.g.\ allowing for a nonzero direction aligned to singular arcs even if the directions orthogonal to the arcs must vanish.  We call these new objects \emph{odeco frames}, thanks to their construction using orthogonally decomposable tensors, and we derive relevant projection operators.

Our experiments show how the theoretical objects we study enable volumetric frame field design in practice.  In particular, we apply standard manifold-based optimization algorithms to field design, built on our differential and algebraic descriptions of octahedral and odeco frames (see Figure \ref{fig.variety-slice}).  The end result is an efficient suite of techniques for producing smooth fields that obey typical constraints for our target application.

\paragraph{Outline} In \S\ref{subsec.octa} we reintroduce the spherical harmonic representation of octahedral frames and demonstrate how this amounts to an equivariant \emph{isometric} embedding of the quotient $\SO(3)/\Oct$ into $\R^9$. In \S\ref{subsec.odeco} we introduce the \emph{odeco} frames, whose axes can scale independently, and we exhibit the spaces of octahedral and odeco frames as nested varieties cut out by quadratic equations. In \S\ref{sec.ingredients} we describe essential primitives for optimization over octahedral and odeco frames, namely geodesics and projection via semidefinite programming. In \S\ref{sec.fields} we formulate the frame field optimization problem. In \S\ref{sec.alg} we describe two algorithms for optimizing fields. \S\ref{sec.exper} describes experiments using these algorithms. A discussion and conclusion follow in \S\ref{sec.concl}. Further experimental results are included in the supplemental material. In summary, our contributions are
\begin{itemize}[leftmargin=*]
	\item a proof of isometric embedding of $\SO(3)/O$ in $\R^9$;
	\item descriptions of the spaces of octahedral and more general odeco frames as nested algebraic varieties; and
	\item new state-of-the-art optimization techniques for volumetric frame fields valued in both varieties, featuring geodesics and SDP-based projection as primitives.
\end{itemize}

\section{Related Work}
\label{sec.related-work}
\subsection{2D Frame Fields and Quadrilateral Meshing}
Cross fields, and their application to quad meshing, have been studied extensively
in geometry processing (\emph{cf.} \cite{vaxman2016}). A useful insight from
cross field research is that it is advantageous to replace field values defined up to some
symmetry with a \emph{representation vector}---that
is, some function of the field value invariant under the relevant symmetry. In two dimensions,
four vectors forming a right-angled cross can be represented in a unified manner by
their common complex fourth power. This amounts to an embedding of a quotient manifold
into a Euclidean space; we show that the octahedral variety generalizes
this idea to 3D in a natural and isometric manner.

Recently, an effort
has been made to unify and formalize the various algorithmic approaches to cross
fields, borrowing
from the Ginzburg--Landau theory from physics \cite{beaufort2017computing,viertel,osting2017}.
This amounts to replacing
an ill-posed unit norm constraint with a penalty term and taking
the limit as the penalty parameter goes to infinity. \citet{viertel} 
show that local optima of such a procedure have isolated singularities
with indices $\pm 1/4$, as appropriate for quad meshing.
They propose a diffusion-generated algorithm to compute such local optima.
Inspired by the work of Merriman, Bence, and Osher (MBO) \shortcite{MBO1} on
mean curvature flow, this algorithm alternates between finite-time diffusion
and pointwise normalization. \citet{osting2017}
study an analogous method applied to orthogonal matrix--valued fields.
Our projection operators enable
us to develop similar MBO diffusion-generated methods for
optimization of octahedral and odeco fields. In \S\ref{sec.exper} we demonstrate that a modified strategy, where the diffusion parameter is adjusted on-the-fly, frequently helps to avoid local minima.

\subsection{3D Frame Fields and Hexahedral Meshing}

\citet{huang} introduce a representation of frames as functions over the sphere that
exhibit octahedral symmetry, parametrized by coefficients in the spherical
harmonic basis. As an initialization step, they solve a Laplace equation,
resulting in coefficients in the interior that
do not correspond to octahedral frames. These must be projected via nonconvex
optimization over frames parametrized by Euler angles.
They further optimize the results by minimizing a discrete Dirichlet energy
over Euler angles.
\citet{RaySokolov2} refine the three stages of this approach,
with improved boundary constraints in the initialization, an efficient projection technique,
and an L-BFGS optimization algorithm.

\citet{Solomon} reformulate the initialization step of
\citet{RaySokolov2} using the boundary element
method (BEM). This provides a way to harmonically interpolate the
boundary conditions exactly, ignoring the constraints in the interior. Finally,
sampled interior values are
approximately projected onto the constraints as in the previous methods.
As the constraints are ignored at the Dirichlet energy--minimization stage,
there is no sense in which the final frame fields are optimal.

A related but distinct problem is the computation of fields of symmetric
second order tensors (i.e., symmetric matrices) \cite{Palacios2017}.
Every symmetric matrix has an orthonormal basis of eigenvectors, which
corresponds to an octahedral frame. One might thus
think symmetric matrix fields can be used to parametrize octahedral frame fields.
While a symmetric matrix field corresponds to at least one
frame field, the correspondence is not one-to-one---for example, a field of
identity matrices corresponds to all frame fields. Moreover, symmetric matrix
fields can only represent singularities whose indices are multiples of $\pm 1/2$,
while indices $\pm 1/4$ are crucial for hex meshing \cite{liu}. Our fourth order octahedral and odeco fields are rich enough to represent all indices.

The work of \citet{Fang2016} on generalized polycubes imposes an even-more-extreme restriction that singularities may only appear on the boundary.
After cutting handles,
the regular frame field in the interior of the volume may be represented by a
field of rotation (orthogonal) matrices. This is further relaxed to a field of matrices, and
orthogonality is approximately enforced via a penalty term. The resulting field
is used to construct a polycube map of the cut volume, through which a
hex mesh is pulled back. By requiring regularity in the interior,
this method may exclude fields that achieve lower distortion overall.

The main driving force for research on 3D frame fields has been hexahedral mesh generation. For a broader overview we refer the reader to the surveys of \cite{Armstrong:2015:CTM} and \cite{Yu:2015:RAA}, and we limit the subsequent discussion to techniques involving frame fields. The idea of such methods is to construct a volumetric integer-grid map \cite{liu}, through which portions of the Cartesian integer grid are pulled back to a structure-aligned hexahedral mesh in the input domain. \citet{nieser2011cubecover} introduced a parametrization technique that turns a given frame field into an integer-grid map by solving a mixed-integer Poisson problem. 
Extraction of the hexahedral mesh from the map is hampered by degeneracies in the map, motivating the sanitization technique of \citet{Lyon:2016:HRH} to improve robustness.
Several refinements of the above ideas have been proposed, including different guidance for the frame field \cite{li2012} and heuristics to improve the singularity structure by decimation or splitting \cite{li2012,Jiang:2013:AHM} to avoid degeneracies of the map.

While robust hexahedral meshing based on frame fields remains an open problem, recently several hex-dominant meshing algorithms have been proposed \cite{Sokolov:2016:HM,Gao:2017:RHM} that also use frame fields but circumvent the problem of non-meshable singularities. \citet{Gao:2017:RHM} propose a hierarchical optimizer for frame fields that is based on local relaxation.

However, many practical applications demand pure hexahedral meshes---e.g.~for the construction of volumetric spline spaces in isogeometric analysis---and consequently a full understanding of meshable field topology is required. To this end, \citet{liu} enumerate the singular vertex types that may occur in a hex mesh with bounded edge valence; they also develop a topological index formula analogous
to the Poincar\'e-Hopf formula for vector fields on surfaces. These local and global
constraints being established, they propose an algorithm to generate a (meshable)
frame field from a prescribed (meshable) singular structure.
The theoretical portion of our work complements the topological
work of \citeauthor{liu} with a closer study of the geometry of spaces of frames.

In concurrent work, \citet{Corman2019} propose an
alternative approach to computing a frame
field with prescribed singular structure via a discretized connection on a frame bundle.
This approach does not immediately extend to a method for \emph{de novo} frame field design.
However, the connection associated to a field is a natural object for the study of such
properties as integrability. We leave to future work the study of connections
associated to fields valued in the octahedral and odeco varieties and the extraction of such
connections directly from the field coefficients.

\subsection{Alternative Frame Representations}

Complementing the spherical harmonic--based representation in the papers above on octahedral frame fields, \citet{chemin} propose an equivalent representation of octahedral frames as
certain symmetric tensors of order four. They introduce algebraic equations
characterizing the octahedral frames among symmetric fourth-order tensors. These equations
are equivalent under a change of basis to our defining equations for the octahedral variety.
However, by using a basis suggested by the structure of $\SO(3)$, we are able to
not only present the defining equations of the octahedral variety,
but also illuminate how it is an isometric embedding
of the quotient space $\SO(3)/\Oct$, enabling us to compute geodesics.
Additionally, we use our algebraic description of the octahedral variety to introduce a
semidefinite relaxation of projection and to place it in the context of
the more general odeco variety.

In concurrent work, \citet{golovaty2019variational} also represent frames by fourth-order symmetric tensors subject to some algebraic conditions, and they propose gradient flow on a Ginzburg-Landau--type energy to optimize for smooth fields.

An additional alternative representation is proposed by \citet{Gao:2017:RHM}, who use quaternions to encode octahedral frames.  Their representation uses relatively few values, but a matching procedure has to be embedded in their optimization objective to account for the fact that $48$ quaternions correspond to the same frame. The concurrent work \cite{beaufort2019quaternionic} uses quaternions to derive another parametrization of octahedral frames by points of a variety in three complex dimensions.

All previous work on 3D frame fields has only considered octahedral frames, which
do not capture the unidirectional behavior of frame fields near singular curves
(\emph{cf.} Figure \ref{fig.prism}).
In the algebraic geometry community,
\citet{Robeva} and \citet{Boralevi}
have completely characterized a family of algebraic varieties
of \emph{orthogonally decomposable} (odeco) tensors. 
We show how the octahedral variety is embedded in one of the odeco varieties,
and we introduce a technique for optimization over the odeco frames.
For our purposes, odeco frames generalize octahedral frames by allowing independent
scaling of the ``axes'' of a frame, including degeneration to unidirectionality
at singular curves and to zero at singular nodes.

Finally, \citet{shen2016harmonic} consider fields having different local discrete
symmetry groups, such as the tetrahedral and icosahedral groups. They
extend the methods of \cite{huang,RaySokolov2} to compute such fields.

\subsection{Semidefinite Relaxations}

Relaxation of algebraic optimization problems to semidefinite
programs has been studied extensively in the field of real algebraic geometry
and optimization. \citet{blekherman2012semidefinite} provide an introduction to this discipline.
The efficacy of semidefinite
relaxation in computer science was demonstrated dramatically in the seminal paper
of \citet{goemans1995improved} on the maximum cut problem.  Since then,
semidefinite relaxation has continued
to be employed to solve both combinatorial and continuous optimization problems,
such as angular synchronization \cite{singer2011angular}.
In geometry processing and graphics, this machinery has been applied to such
problems as correspondence \cite{kezurer2015tight}, consistent mapping \cite{huang2013consistent}, registration \cite{maron2016point}, camera motion estimation/calibration \cite{agrawal2003camera,ozyesil2015stable}, and deformation \cite{kovalsky2014controlling}.

Most relevant to the present work are relaxations of the Euclidean projection problem
onto a variety defined by quadratic equations,
an example of a quadratically-constrained quadratic program (QCQP).
\citet{cifuentes2017} have recently shown a stability
result implying that the semidefinite relaxation of Euclidean projection
onto a smooth, quadratically-defined variety is exact
in a neighborhood of the variety. \citet{cifuentes2018}
have additionally shown that the region in which the relaxation is exact is a
\emph{semialgebraic set}, and they have provided a formula for the degree of
its algebraic boundary. Unfortunately, computing this boundary is generally
intractible for interesting varieties.
A deeper theoretical understanding of when
semidefinite relaxations of Euclidean projection are globally exact is still lacking.

\section{Spaces of Frames}
\label{sec.frame-space}

As previously studied in \cite{huang, RaySokolov2, Solomon, chemin}, 
the basic unknown in the volumetric frame field problem is a tuple of three mutually orthogonal directions at each point in a volume.  These directions
may be represented by an orthonormal basis of vectors, but the signs and order of the vectors
are irrelevant thanks to octahedral symmetry.
This redundancy makes detecting smoothness of a field of tuples difficult.
Hence, it pays to use a unified representation invariant to the symmetries of the frame.

Below, we apply machinery from differential and algebraic geometry to derive a succinct
description of this basic octahedral frame and show how it is related to
rotations of the function $x_1^4+x_2^4+x_3^4$ on the unit sphere expressed
in the spherical harmonic basis
(as used to represent frames in \cite{huang, RaySokolov2, Solomon}) and to tensorial
representations (as used in \cite{chemin}).  Algebraic language not only provides a succinct
description of previous representations but also suggests a means of generalizing to frames
whose three axes scale independently
(e.g.\ rotations of the function $\sum_i \lambda_i x_i^4$ for
varying $\lambda\in\R^3$), whereas previous work requires them to have the same length
($\lambda_1=\lambda_2=\lambda_3$). This broader set better aligns with the realities of
the frame field problem, since singular edges have nontrivial directionality that
cannot be captured by existing representations.
Relevant background material may be found in Appendix \ref{app.background}.

\subsection{Octahedral Variety}
\label{subsec.octa}
Intuitively, an octahedral frame field is a smooth assignment of
(unoriented) orthonormal coordinate axes to each point $x$ in a region
$\Omega \subset \R^3$. Such frames are called octahedral because
they exhibit symmetries described by the octahedral group $\Oct$.\footnote{Technically,
their stabilizers are conjugate to $\Oct$.}
The space of such octahedral frames
can be identified with the quotient space $\mathcal{F}~=~\SO(3)/\Oct$ --- that is, the quotient
of the group of oriented rotations by the right action of the octahedral
group. Since $\Oct$ is not a normal subgroup of $\SO(3)$ (indeed $\SO(3)$
is simple) $\mathcal{F}$ is not a group.
However, as $\Oct$ is a finite group acting freely on $\SO(3)$, $\mathcal{F}$
is a manifold, and the surjective map $\SO(3) \to \mathcal{F}$ is a covering map.
The universal cover of $\mathcal{F}$ is that of $\SO(3)$,
namely $\mathrm{SU}(2)$, and so its fundamental group
$\pi_1(\mathcal{F})$ is the lift of $\Oct$ to $\mathrm{SU}(2)$, the binary octahedral
group $\mathrm{BO}$. In particular, $\mathrm{BO}$ classifies singular curves of octahedral fields \cite{mermin_topological_1979}.

$\SO(3)$ admits a bi-invariant Riemannian metric \eqref{eq.bi_invariant}, which
means that the action of $\Oct$ by right-multiplication is isometric.
Thus the Riemannian metric on $\SO(3)$ descends to $\mathcal{F}$,
making it a Riemannian manifold. In particular, $\mathcal{F}$
has geodesics that lift to
geodesics of $\SO(3)$. We will employ these geodesics to compute optimization sub-steps
on $\mathcal{F}$ (\emph{cf.} \S\ref{subsec.geodesics}).

We have introduced $\mathcal{F}$ as an abstract smooth manifold, but an embedding
of $\mathcal{F}$ in a Euclidean space is necessary to make it
amenable to computation. Previous works have introduced this equivariant
embedding via the
action of $\SO(3)$ on polynomials and spherical harmonic coefficients.
For completeness, we reintroduce it here, but in a slightly different way
that highlights the role of representation theory and the orbit-stabilizer theorem.
Later, we will show that the embedding of $\mathcal{F}$ in $\R^9$ is an \emph{equivariant
isometric embedding} (\emph{cf.} \cite{Moore1976}).

Consider the irreducible representation $\rho: \SO(3) \to \SO(9)$ corresponding
to the fourth band of spherical harmonics.
The $9 \times 9$ orthogonal matrices $\rho(r)$ for $r \in \SO(3)$ are sometimes referred to as Wigner $D$-matrices. Form a linear operator $H\in\R^{9\times9}$ as 
\[ H = \frac{1}{|\Oct|}\sum_{o \in \Oct} \rho(o). \]
This $H$ is a projection operator onto the subspace of $\R^9$ invariant under all octahedral rotations:
\begin{lemma} \label{lem.oct-avg}
	For any $o' \in \Oct$, $\rho(o')H = H$.
\end{lemma}
\begin{proof}
	\[ \rho(o')H = \frac{1}{|\Oct|}\sum_{o \in \Oct} \rho(o')\rho(o)
	= \frac{1}{|\Oct|}\sum_{o \in \Oct} \rho(o' o)
	= \frac{1}{|\Oct|}\sum_{o \in \Oct} \rho(o) = H. \qedhere \]
\end{proof}

\begin{corollary}
	For $q \in \R^9$, $\rho(\Oct)\cdot q = q$ if and only if $q \in \mathrm{Im}(H)$.
\end{corollary}
\begin{proof}
	The forward implication follows by writing $q = Hq$. The
	reverse implication follows directly from Lemma \ref{lem.oct-avg}.
\end{proof}

Because $\Oct$ is a maximal subgroup of $\SO(3)$,
we have the following corollary.
Here $\stab{q}$ denotes the stabilizer of $q$---that is,
the subgroup of $\SO(3)$ leaving $q$ invariant (see Appendix \ref{app.background}).
\begin{corollary}
	Let $q \in \mathrm{Im}(H)$ and nonzero. Then $\stab(q) = \Oct$.
\end{corollary}

It happens that $H \in \R^{9\times 9}$ has rank one, i.e., its image is one-dimensional,
motivating the following definitions.
\begin{definition}
	The \textbf{canonical octahedral frame} is the normalized vector
	\[ q_0 = \left(0, 0, 0, 0, \sqrt{\frac{7}{12}}, 0, 0, 0, \sqrt{\frac{5}{12}}\right)\tran \in \R^9. \]
	such that $H = q_0 q_0\tran$.
\end{definition}
Our canonical frame is the same as the normalized
projection of the polynomial $\sum_i x_i^4$ into the fourth band
of spherical harmonics, as in, e.g., \cite{Solomon}.

\begin{definition}
	The \textbf{octahedral variety} is the orbit of $q_0$ under the action of $\SO(3)$,
	\[ \VOcta = \rho(\SO(3)) q_0 = \{ \rho(r) q_0 \mid r \in \SO(3) \}. \]
\end{definition}

To summarize, $\VOcta$ is the orbit of $q_0$, whose stabilizer
is $\Oct$. A smooth version of the orbit-stabilizer theorem
\cite[Theorem 21.18]{lee2012} shows that there is an equivariant diffeomorphism
$\phi: \mathcal{F} \to \VOcta$:
\begin{center}
\begin{tikzcd}
	\SO(3) \arrow[d, two heads] \arrow[r, "\rho"] & \SO(9) \arrow[d, two heads, "\cdot q_0"] \\
	\mathcal{F} \arrow[r, dashed, "\phi"] & \VOcta
\end{tikzcd}
\end{center}

Next, we will characterize the Riemannian geometry of $\VOcta$ by showing that $\phi$
is an isometry up to a uniform scale factor.
To our knowledge, this observation has not appeared in previous work.

\begin{proposition}
	\label{thm.isometric}
	Let $\alpha = \sqrt{3/20}$ and
	$\VOcta_\alpha \coloneqq \alpha \VOcta = \{\alpha q : q \in \VOcta\}$.
	Let $\pi_\alpha:~\R^{9\times 9}~\to~\R^9$ denote matrix multiplication by
	the scaled canonical octahedral frame $q_\alpha \coloneqq \alpha q_0$. That is, $\pi_\alpha(A) = A q_\alpha$.
	Then the map
	\[\pi_\alpha \circ \rho: \SO(3) \to \VOcta_\alpha \]
	is a local isometry, making
	the induced diffeomorphism $\phi_\alpha : \mathcal{F} \to \VOcta_\alpha$ an isometry.
\end{proposition}
\begin{proof}
	Taking the differential of $\rho$ at the identity yields
	the associated Lie algebra representation
	\[ (D\rho)_I : \mathfrak{so}(3) = T_I \SO(3) \to \mathfrak{so}(9) \subset \R^{9\times 9}.\]
	$(D\rho)_I$ is characterized by
	$L_i\coloneqq(D\rho)_I(l_i)$, the images of the Lie algebra generators $l_i$ (see Appendix \ref{app.background} for definitions and supplemental material for explicit expressions).
	$\pi_\alpha$ is a linear map, so its differential is also multiplication by $q_\alpha$.
	
	To see that $\pi_\alpha \circ \rho$ is a local isometry, first recall that
	the metric on $\SO(3)$ is bi-invariant. In particular, for each $g \in \SO(3)$,
	an orthonormal basis for $T_g\SO(3)$ is given by the right-translated Lie algebra generators
	\[ \{(D\Rtrans_g)_I l_i\}_{i=1}^3. \]
	So it suffices to prove that their images under $\pi_\alpha \circ \rho$ form an orthonormal basis in $T_{\rho(g)q_\alpha}\VOcta_\alpha$.
	But
	\begin{equation}
	\begin{aligned}
	D(\pi_\alpha \circ \rho)_g (D\Rtrans_g)_I l_i
	&= (D\pi_\alpha) (D\rho)_g (D\Rtrans_g)_I l_i \\
	&= (D\pi_\alpha) (D\Rtrans_{\rho(g)})_I (D\rho)_I l_i \\
	&= (D\pi_\alpha) L_i \rho(g) \\
	&= L_i \rho(g) q_\alpha,
	\end{aligned} \end{equation}
	where the second equality follows by differentiating the
	representation property $\rho \circ \Rtrans_g = \Rtrans_{\rho(g)} \circ \rho$
	at the identity.
	So the equation we need to prove is
	\begin{equation} \label{eq.iso_condition}
	\left\langle L_i q, L_j q\right\rangle = \delta_{ij} \quad \forall q \in \VOcta_\alpha.
	\end{equation}
 	This isometry condition can be checked explicitly at $q_\alpha$:
 	\begin{equation}
	\label{eq.iso_at_qalpha}
	   \left\langle L_i q_\alpha, L_j q_\alpha \right\rangle = \delta_{ij}
	   = \left\langle l_i, l_j \right\rangle.
	\end{equation}
	
	For a general $q = \rho(g)q_\alpha \in \VOcta_{\alpha}$, we compute
	\begin{equation}
	\begin{aligned}
	\left\langle L_i \rho(g) q_\alpha, L_j \rho(g) q_\alpha\right\rangle
	&= \left\langle \rho(g)\tran L_i \rho(g) q_\alpha, \rho(g)\tran L_j \rho(g) q_\alpha\right\rangle \\
	&= \left\langle D\rho_I(g\tran l_i g) q_\alpha, D\rho_I(g\tran l_j g) q_\alpha\right\rangle.
	\end{aligned}
	\label{eq.iso_conj}
	\end{equation}
	The first equality follows because $\rho(g) \in \SO(9)$, and the second follows by Lemma \ref{lem.adj_comm}.
	Let $a_{ik} = a_{ik}(g)$ be the coefficients of the adjoint representation of $\SO(3)$ (see Appendix \ref{app.background}), i.e., $g\tran l_i g \eqqcolon \sum_k a_{ik} l_k$; these coefficients form an orthogonal matrix.
	Now using linearity of $D\rho$ along with \eqref{eq.iso_at_qalpha}
	and \eqref{eq.iso_conj}, we obtain
	\begin{equation}\begin{aligned}
	\left\langle L_i \rho(g) q_\alpha, L_j \rho(g) q_\alpha\right\rangle
	&= \left\langle \sum_r a_{ir} L_r q_\alpha, \sum_s a_{js} L_s q_\alpha \right\rangle \\
	&= \sum_{r, s} a_{ir} a_{js} \langle L_r q_\alpha, L_s q_\alpha \rangle \\
	&= \sum_k a_{ik} a_{jk} = \delta_{ij}
	\end{aligned}\end{equation}
	as required.
\end{proof}

In summary, we have described the octahedral variety as an embedded submanifold in
$\R^9$ isometric to $\mathcal{F} = \SO(3)/\Oct$. This isometry means that we can do
manifold optimization over frames by working in the embedding (\emph{cf.} \S\ref{subsec.rtr}).
To show that $\VOcta$ is really an algebraic variety, we will need to exhibit equations
that cut it out. We will delay this until \S\ref{subsec.octa_odeco}, when we
can give a unified description of the octahedral and odeco varieties.

\subsection{Odeco Variety}
\label{subsec.odeco}

\begin{figure}
	\centering
	\begin{subfigure}[t]{0.45\columnwidth}
	\includegraphics[width=\textwidth]{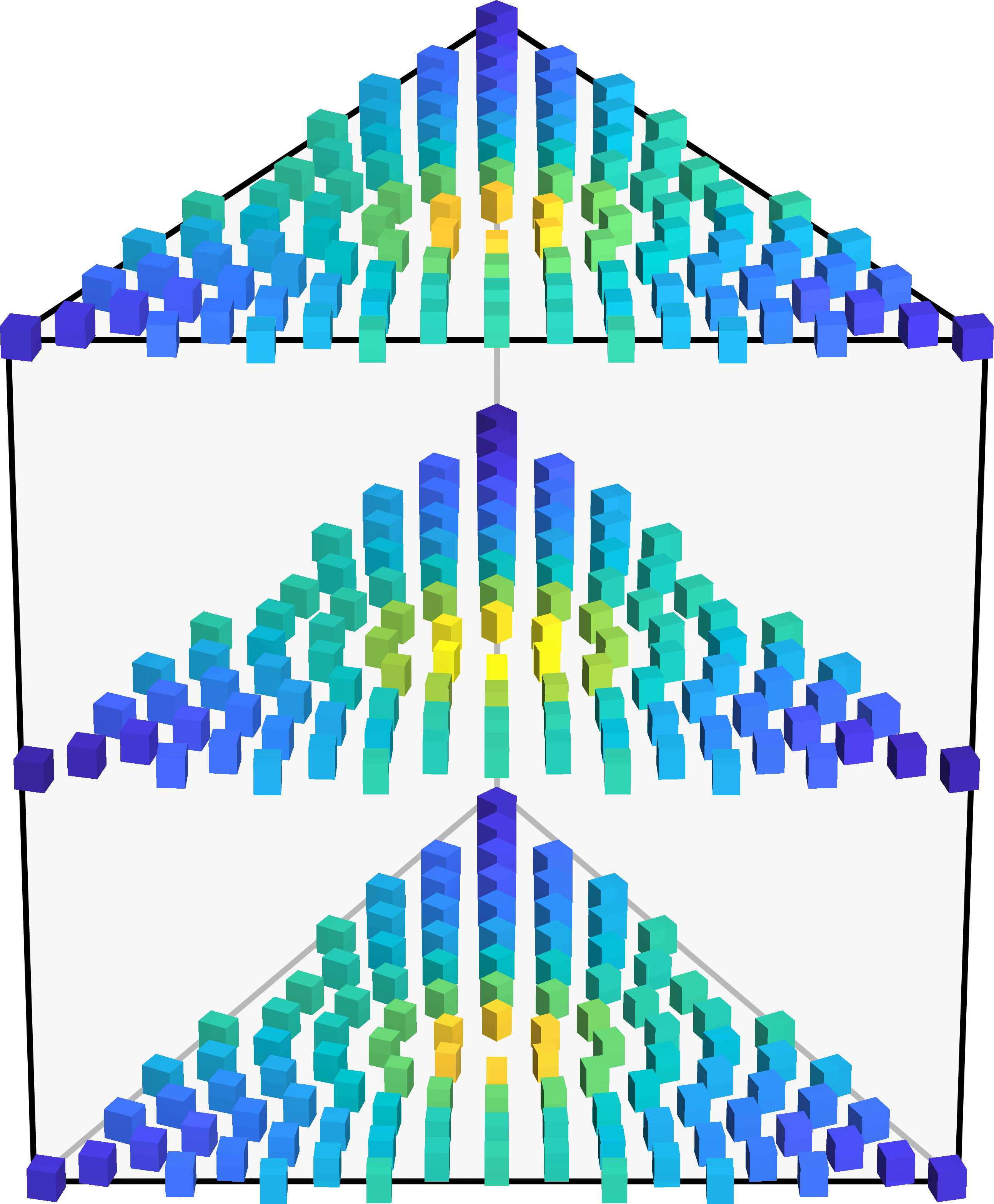}%
	\end{subfigure}
	\qquad%
	\begin{subfigure}[t]{0.45\columnwidth}
	\includegraphics[width=\textwidth]{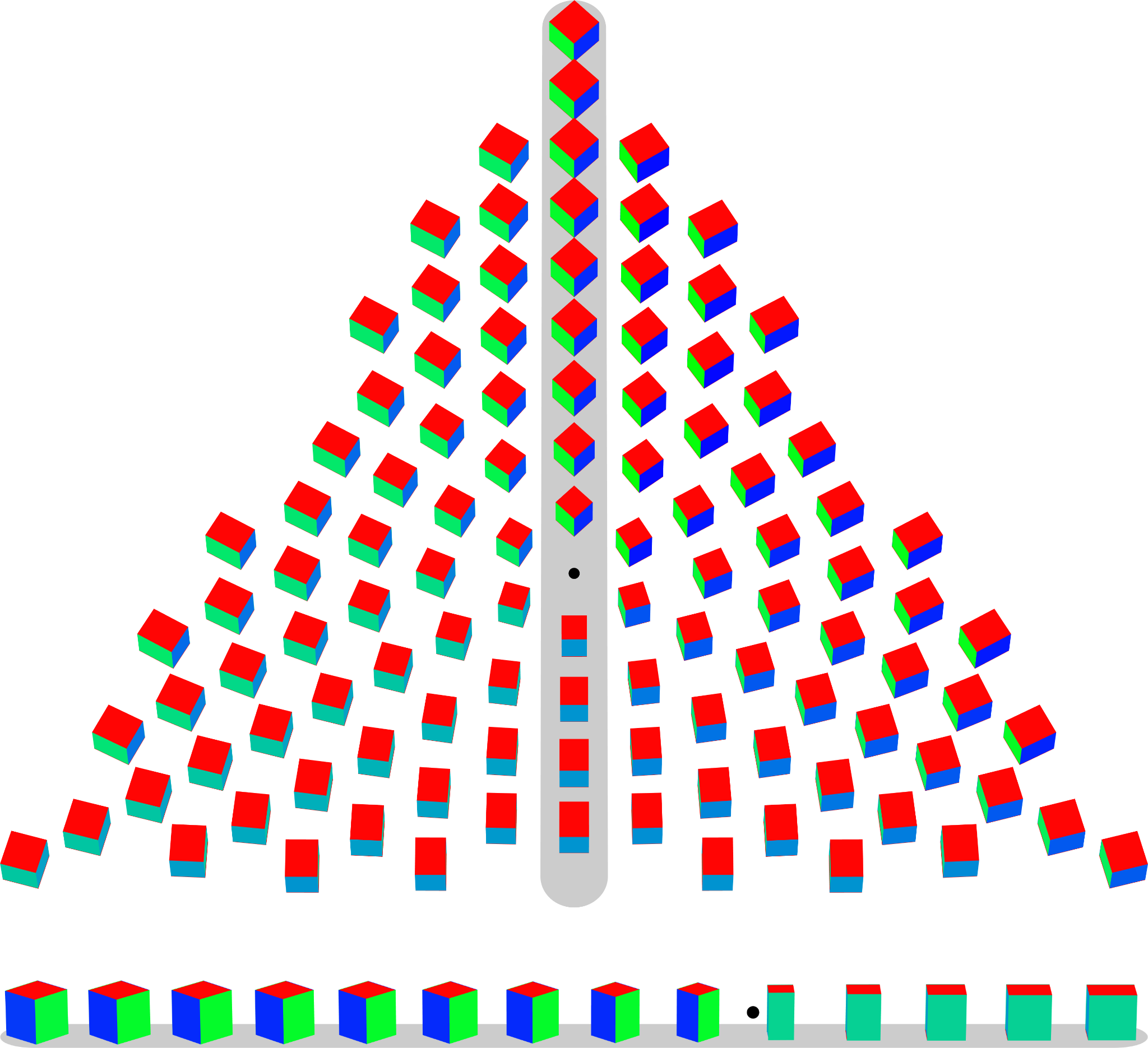}%
	\end{subfigure}
	\caption{An odeco field
	on a triangular prism. The norm of the second band coefficients
	indicates degree of anisotropy (left). Odeco frames close to
	the approximate position of the singular curve, indicated by a dot (right), scale toward
	zero in the directions normal to the curve, but remain nondegenerate in the
	direction along the curve.}
	\label{fig.prism}
\end{figure}

The previous section provides more insight
into the octahedral frames used in all previous work.
Octahedral frames are suitable for representing singularity-free frame fields.
However, frame fields commonly encountered in applications
have singularities comprising an embedded graph.
Consider the triangular prism shown in Figure \ref{fig.prism}.
Near the singular curve, a smooth octahedral field would rotate infinitely
quickly, and it would not have a well-defined value along the curve.
This issue is analogous to the case of \emph{unit} cross fields---note
that for cross fields, the hairy ball theorem \emph{requires}
singularities on simply-connected surfaces.

One solution to this is to replace the hard constraint that the field values
lie in the octahedral variety with a penalty term,
as motivates the MBO method for octahedral fields detailed below (\S\ref{subsec.mbo}).
Another solution is homogenization---i.e., allowing field values to scale,
replacing singularities by zeros, as is considered for cross fields in \cite{knoppel2013}.
But consider the triangular prism again: a scaled octahedral field
would vanish completely at the singular curve since all three orthogonal axes must scale
uniformly. This makes the octahedral representation unable to capture the alignment
of the field to the singular curve.
To cope with this problem and to show the value of our algebraic approach,
we now describe a superset of the octahedral frames.
This set allows the axes to scale independently; for instance,
as shown in Figure \ref{fig.prism},
the frame axes orthogonal to the singular curve scale toward zero
while the axis tangent to the singular curve remains nonzero.

The symmetric orthogonally decomposable (\emph{odeco}) tensor varieties, introduced
in \cite{Robeva}, parametrize symmetric tensors $T \in \Sym^d \R^n$
that can be written
\[ T = \sum_{i=1}^n \lambda_i (v_i)^{\otimes d} \]
for some set of $n$ orthonormal vectors $v_i \in \R^n$,
where $v^{\otimes d}$ denotes the $d$-wise tensor power of the vector $v$.
That is, an odeco tensor
encodes a set of orthogonal vectors up to permutation.
If $d$ is even, $T$ is also invariant under sign changes to the $v_i$.
Moreover, an odeco tensor assigns \emph{weights} $\lambda_i$
to the vectors $v_i$. This property allows us to encode frames whose
axes scale independently.

There is a one-to-one correspondence between symmetric tensors of order $d$
over $\R^n$ and homogeneous polynomials of degree $d$ in $n$ variables
(\emph{cf} \cite[\S1.2]{Robeva}).
This correspondence is given in one direction by taking a polynomial
$p \in \R[x_1, \dots, x_n]$ to
its tensor of $d$th derivatives, and in the other direction by symbolic evaluation
on the vector of formal variables $x = (x_1, \dots, x_n)$:
\[ T \in \Sym^d \R^n \mapsto
  p_T(x) = \frac{1}{d!} T(x, \dots, x) \in \R[x]. \]
This is a generalization of the correspondence between symmetric bilinear forms
(i.e. symmetric $2$-tensors) and quadratic forms (i.e., homogeneous quadratic polynomials).
Note that $T$ is odeco if and only if $p_T$ can be written
as a sum of $d$th powers of linear forms
\[ p_T(x) = \sum_i \lambda_i (v_i\tran x)^d \]
where the $v_i$ are orthonormal as above.
In this case, we also refer to $p_T$ as an odeco polynomial.

\begin{figure}[h]
\centering
\newcommand{\imageheight}{0.2\columnwidth}
\begin{tabular}{cccc}
\includegraphics[align=c,height=\imageheight]{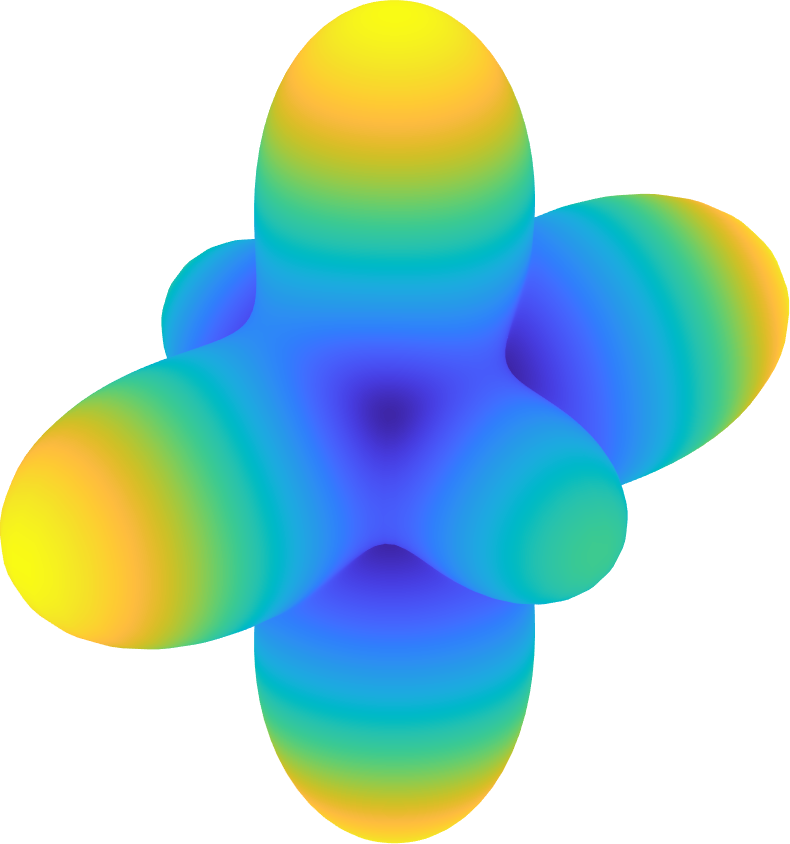} &%
\includegraphics[align=c,height=\imageheight]{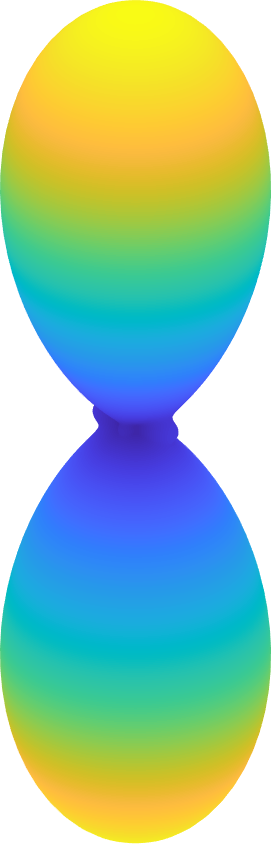} &%
\includegraphics[align=c,height=\imageheight]{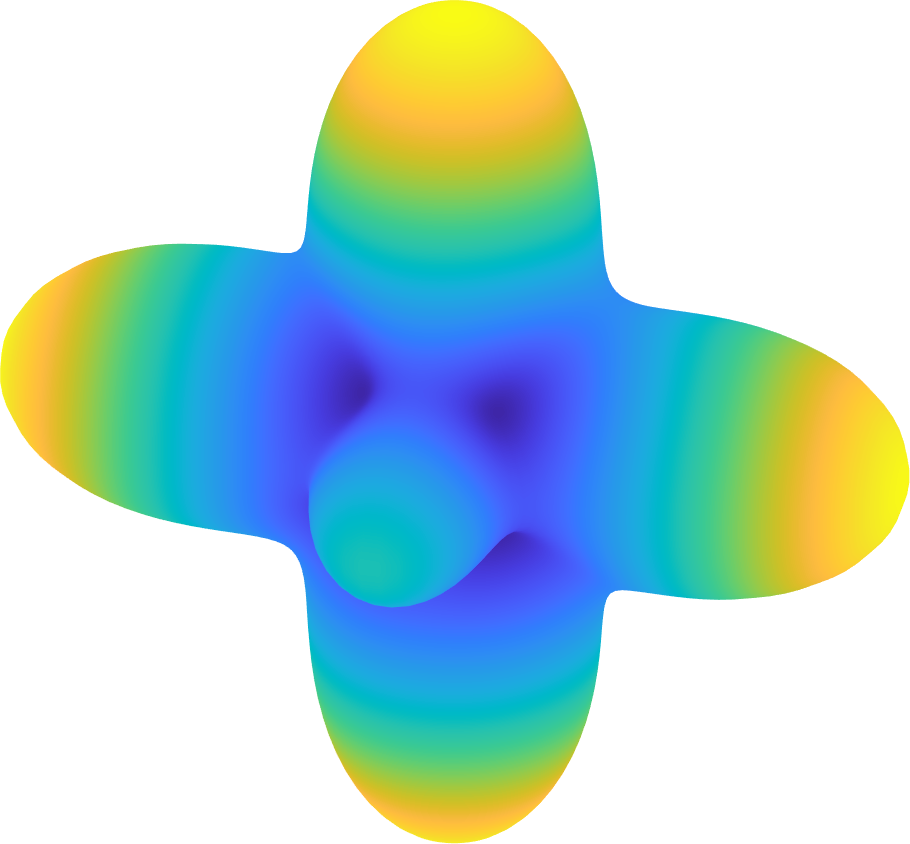} &%
\includegraphics[align=c,height=\imageheight]{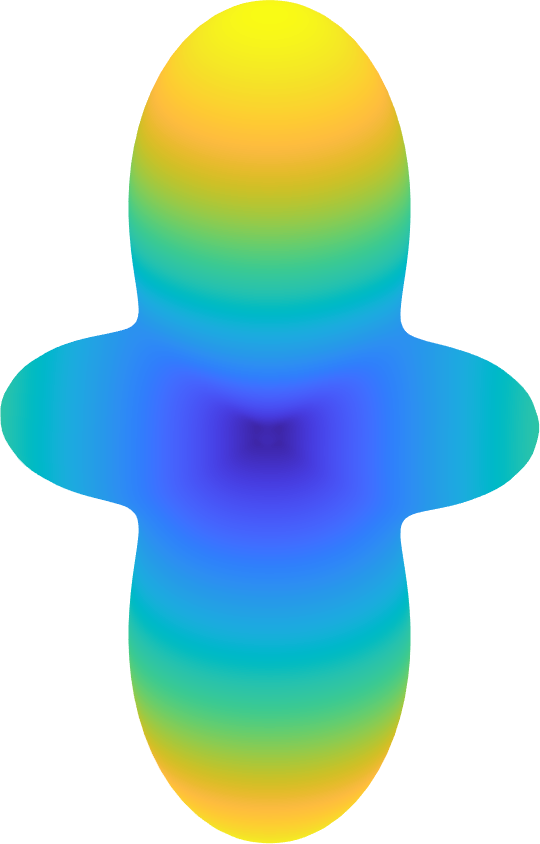}
\end{tabular}
\caption{%
Examples of $\boldsymbol{z}$-aligned odeco polynomials, plotted over the sphere.}
\label{fig.odeco_examples}
\end{figure}

The defining equations of the odeco varieties are quadrics---homogeneous quadratic
equations---in the tensor coefficients \cite[Theorem 4]{Boralevi}
or equivalently in the coefficients of the associated polynomial $p_T$. That is,
a homogeneous polynomial
\[ p(x) = \sum_{d_1 + \dots + d_n = d} u_{d_1, \dots, d_n}
					   x_1^{d_1} \dots x_n^{d_n} \]
is odeco if and only if the coefficients $u$ satisfy
\begin{equation} \label{eq.odeco_quads}
u\tran A_i u = 0 \end{equation}
for a finite set of symmetric matrices $A_i$. In the case
relevant to us, where $n = 3$ and $d = 4$,
there are exactly $27$ such defining equations.
While dimension counting might suggest otherwise, these equations are not redundant---as can be seen by computing a Gr\"obner basis of the ideal they generate.
The matrices $A_i$ are listed explicitly in the supplemental document.
We will henceforth refer to this particular odeco variety simply as
the \textbf{odeco variety} $\VOdeco$. Figure \ref{fig.odeco_examples}
plots several fourth-order odeco polynomials over the unit sphere.

\subsubsection{Octahedral Variety as a Subvariety of Odeco Variety}
\label{subsec.octa_odeco}
Recall that our octahedral frames were represented by coefficients in an irreducible
representation of $\SO(3)$, while the odeco variety was defined using monomial or tensor
coefficients. To see the relationship between the two varieties,
it is beneficial to recast the odeco variety in the irreducible representation basis.
This corresponds to looking at the coefficients of odeco polynomials in
the basis of spherical harmonics.

The quartic polynomials comprising the odeco
variety decompose as linear combinations of the spherical
harmonics of bands $0$, $2$, and $4$. Consider
an odeco polynomial $\sum_{i=1}^3 \lambda_i (v_i\tran x_i)^4$, and let
\[ q = (q_0, q_2, q_4) \in V_0 \times V_2 \times V_4 \]
be its coefficients in this basis of even spherical harmonics. These $15$ coefficients
give us a different representation of odeco frames in $\R^{15}$,
where each band has a clear meaning.
In particular, $q_0 = C_0 \sum_i \lambda_i$, where $C_0$ is a constant, and similarly
\[ \|q_2\|^2 = C_2 \left(\sum_i \lambda_i^2 - \sum_{i < j} \lambda_i \lambda_j \right). \]
The parenthesized expression is the squared distance between $(\lambda_1, \lambda_2, \lambda_3)$
and the line $\lambda_1 = \lambda_2 = \lambda_3$.
Thus $q_2 = 0$ if and only if $q_4$ is a scalar multiple of an octahedral frame.
The set
\[ \{q \in \VOdeco \mid q_2 = 0\} \]
consists of scalar multiples of the octahedral variety indexed by $q_0$.
Fix $q_2 = 0$ and let $q_0$ be a constant such that $\|q_4\| = 1$. The octahedral
variety is the intersection of this affine subspace with the odeco variety.
Reducing the equations \eqref{eq.odeco_quads} of the odeco variety with respect to this subspace
yields $15$ inhomogeneous quadratic equations
\begin{equation} \label{eq.octa_quads}
\begin{pmatrix}
	1 \\ q_4
\end{pmatrix}\tran P_i \begin{pmatrix}
	1 \\ q_4
\end{pmatrix} = 0, \quad i = 1, \dots, 15 \end{equation}
cutting out the octahedral variety $\VOcta$. 
As was the case for the odeco variety,
these equations are not redundant. The symmetric
matrices $P_i$ are listed explicitly in the supplemental document.

\section{Geodesics and Projection}
\label{sec.ingredients}
We have introduced two spaces, the octahedral and odeco varieties, that
can serve as target sets for frame fields. To compute
such fields, we will have to solve optimization problems over products
of many copies of these varieties. Na\"ively, one might
plug the quadratic constraints in \eqref{eq.odeco_quads}
and \eqref{eq.octa_quads} directly into a generic quadratic optimization solver.
However, the $A_i$ or $P_i$ are not positive semidefinite,
nor can their span be rewritten as the span of positive semidefinite matrices.
So the constraints are nonconvex and challenging to enforce.

As an alternative, we use optimization algorithms that are tailored
to the manifold-valued variables in our problem. These algorithms
employ sub-steps such as geodesic traversal and projection. We derive
these operations for a single frame below.

\subsection{Octahedral Geodesics}
\label{subsec.geodesics}
By Proposition \ref{thm.isometric}, the scaled
octahedral variety $\VOcta_\alpha$ is
locally isometric to $\SO(3)$ via the map $r \mapsto \rho(r) \cdot q_\alpha$.
It follows that geodesics of $\SO(3)$ push forward to geodesics of $\VOcta_\alpha$.
The relation \eqref{eq.exp_rep} in Appendix \ref{app.background}
then allows us to compute geodesics on $\VOcta_\alpha$ in closed form without evaluating
the representation map $\rho$ explicitly.
Let $\mathfrak{v} \in T_q \VOcta_\alpha$. $\mathfrak{v}$ can be
written in a basis induced by the $\SO(3)$ action, $\mathfrak{v} = \sum_{i=1}^3 v_i L_i q$,
Here the coefficient vector $v = (v_1, v_2, v_3) \in \R^3$ is
the ``axis-angle'' representation of a rotation, and
the $\SO(3)$ exponential maps it to the corresponding rotation matrix.
This can be computed by conjugation
with a rotation $r$ taking the unit vector $v / \|v\|$
to $(0, 0, 1)$. Composing with the representation map, we have
\begin{equation} \label{eq.octa_exp}
\exp_q(\mathfrak{v})
= \rho(r\tran \exp(\|v\| l_3) r) q
= \rho(r)\tran \exp(\|v\| L_3) \rho(r) q,
\end{equation}
where (unsubscripted) $\exp$ denotes the ordinary matrix exponential.

To compute $\rho(r)$,
define spherical coordinates $\theta, \varphi$ such that
\[ v = \|v\| (\cos \theta \sin \varphi, \sin \theta \sin \varphi, \cos \varphi). \]
Then one possible choice for $r$ is
\[ r = \exp(-\phi l_2)\exp(-\theta l_3)
= r_{23}\tran \exp(-\phi l_3) r_{23} \exp(-\theta l_3), \]
where $r_{23} = \exp((\pi/2)l_1)$.
So
\begin{equation} \label{eq.rot_axis}
\rho(r) = R_{23}\tran \exp(-\phi L_3) R_{23} \exp(-\theta L_3),
\end{equation}
where $R_{23} = \exp((\pi/2)L_1)$.
Combining \eqref{eq.rot_axis} with \eqref{eq.octa_exp},
we can compute geodesics by products of two simple ingredients:
$R_{23}$ and $\exp(t L_3)$ for $t \in \R$.
Closed-form expressions for both appear in the supplemental document, \S2.
Note that a formula similar to \eqref{eq.rot_axis} is common in the graphics literature,
e.g., for rotating BRDFs expressed in spherical
harmonic coefficients (\emph{cf.} \cite{kautz2002fast}, Appendix).

\subsection{Projection via Semidefinite Relaxation}
\label{subsec.proj}
$\VOcta$ and $\VOdeco$ are both varieties defined by quadratic equations.
$\VOcta \subset \R^9$ is
cut out by $15$ inhomogeneous quadratic equations \eqref{eq.octa_quads}, while
$\VOdeco \subset \R^{15}$ is cut out by $27$
homogeneous quadratic equations \eqref{eq.odeco_quads}.
Consider the problem of finding the closest point
in $\VOcta$ to a given point $y \in \R^9$:
\begin{equation} \label{prob.proj} \tag{P$_{\VOcta}$}
	\Pi_{\VOcta}(y) = \argmin_{q \in \VOcta} \|q - y\|^2.
\end{equation}
This problem is called Euclidean projection onto a quadratic variety.
It is an example of a QCQP, and the general recipe for semidefinite
relaxation detailed in \S\ref{subsec.relaxations} automatically applies.
The SDP will have the form
\begin{equation}
\label{sdp.proj}
\begin{alignedat}{4}
	&\argmin_{Q \in \R^{10\times10}} &\langle Y, Q \rangle \\
	&\text{subject to} &Q_{11} &= 1 \\
		&&\langle P_i, Q \rangle &= 0, &\; i = 1, \dots, 15 \\
		&&Q &\succeq 0,
\end{alignedat} \tag{SDP$_\VOcta$}
\end{equation} 
where $P_i$ are the symmetric matrices from \eqref{eq.octa_quads}, and
\[ Y = \begin{pmatrix} \|y\|^2 & -y\tran \\ -y & I_{9\times 9} \end{pmatrix}. \]

Let $Q^*$ be an optimal solution to \eqref{sdp.proj}.
Since \eqref{sdp.proj} is a relaxation of \eqref{prob.proj}, $\langle Y, Q^*\rangle$
is a lower bound on the objective value of \eqref{prob.proj}.
If it happens that $\rank(Q^*) = 1$, then $Q^*$
can be written in the form
\[ Q^* = \begin{pmatrix} 1 \\ q^* \end{pmatrix} \begin{pmatrix} 1 \\ q^* \end{pmatrix}\tran, \]
from which it follows that
\[ \begin{pmatrix} 1 \\ q^* \end{pmatrix}\tran P_i \begin{pmatrix} 1 \\ q^* \end{pmatrix} = 0,
\quad i = 1, \dots, 15, \]
i.e., $q^* \in \VOcta$.
In this case, $q^*$ is the globally optimal solution to \eqref{prob.proj}.
This situation is called exact recovery.
The foregoing discussion also applies, \emph{mutatis mutandis}, to $\VOdeco$.

In our MBO algorithm presented below (\emph{cf.} \S\ref{subsec.mbo}), we alternate
projection with stepping off the variety.
We refer to the following theorem, which suggests
that when taking small enough steps from smooth points of a variety,
projection will be generically exact.
\begin{theorem}[\citet{cifuentes2017}, Theorem 1.2]
\label{thm.proj_stability}
	Consider the Euclidean projection problem
	\[ \argmin_{q \in \mathcal{V}} \|q - y\|, \quad y \in \R^n, \]
	where $\mathcal{V} \subset \R^n$ is a real variety defined
	by quadratic equations $f_1(q) = \dots = f_m(q) = 0$. Let $\bar{y} \in \mathcal{V}$
	be a point at which the rank of the Jacobian $\nabla f(\bar{y})$ is equal
	to the codimension of $\mathcal{V}$.
	Then the semidefinite relaxation of projection is exact for $y \in \R^n$ sufficiently
	close to $\bar{y}$.
\end{theorem}

\begin{figure}
\centering
\begin{tikzpicture}
\begin{semilogxaxis}[
    ymin=-0.05, ymax=1,
    enlarge x limits = false,
    minor y tick num = 4,
    xlabel={\footnotesize $\lambda_2/\lambda_1$},
    ylabel={\footnotesize Probability Density},
    xlabel near ticks,
    ylabel near ticks,
    every tick label/.append style = {font=\tiny},
    legend cell align = left,
    legend style = {font=\tiny, row sep=0.1pt}]
\addplot[const plot,mark=none,red,fill,fill opacity=0.1] table[x index = 0, y index = 1, col sep = comma] {figures/OdecoExactnessTest2Ratio.csv};
\addlegendentry{Odeco}
\addplot[const plot,mark=none,blue,fill,fill opacity=0.1] table[x index = 0, y index = 1, col sep = comma] {figures/OctaExactnessTest3Ratio.csv};
\addlegendentry{Octahedral}
\addplot[const plot,mark=none,green,fill,fill opacity=0.1] table[x index = 0, y index = 1, col sep = comma] {figures/OdecoPosExactnessTest2Ratio.csv};
\addlegendentry{Odeco (positive initial polynomial)}
\addplot[mark=|, blue] coordinates {(2.41e-8, 0)};
\addplot[mark=|, green] coordinates {(1.54e-9, 0)};
\end{semilogxaxis}
\end{tikzpicture}
\caption{Histogram of eigenvalue ratio $\lambda_2(Q^*)/\lambda_1(Q^*)$
for solutions to the SDP
relaxations of Euclidean projection onto $\VOcta$ and $\VOdeco$.
Projections of $10^6$ random points were tested for each. The maximum ratio
for octahedral projection was $\num{2.41e-8}$. The maximum ratio
for odeco projection of positive quartic polynomials was $\num{1.54e-9}$. See \S\ref{subsec.proj}.}
\label{fig.exactness}
\end{figure}
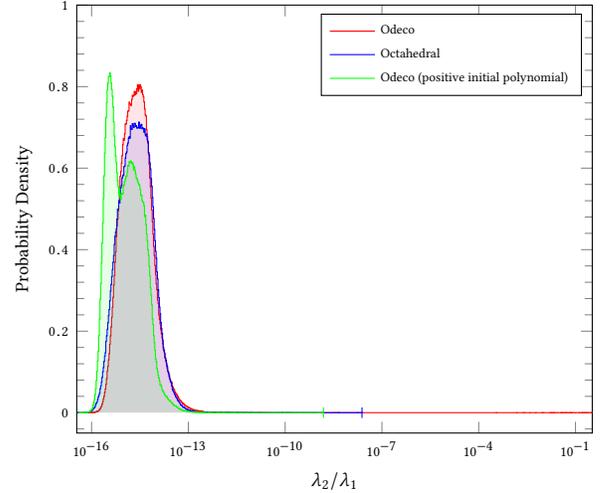

In addition to this theoretical motivation, we have ample empirical evidence
that our relaxations are exact under much more general conditions.
The blue histogram in Figure \ref{fig.exactness}
shows the results of attempting to project $10^6$ random points
onto $\VOcta$ via semidefinite relaxation. We use the ratio of
the second largest eigenvalue $\lambda_2(Q^*)$ to the largest eigenvalue $\lambda_1(Q^*)$
as a proxy for exactness, as it
indicates whether $Q^*$ was rank-one up to machine precision.
For all octahedral projections, $\lambda_2(Q^*)/\lambda_1(Q^*)$ was $\approx 10^{-8}$ or less.
Based on these results, we make the following conjecture.
\begin{conjecture} \label{conj.octa_exact}
	For a generic point $q_0 \in \R^9$,
	the solution to \eqref{sdp.proj} has rank $1$,
	and therefore yields the exact projection $\Pi_\VOcta(q_0)$.
\end{conjecture}

We also tried projecting $10^6$ random quartic polynomials onto
the odeco variety, as shown in the red histogram in Figure \ref{fig.exactness}.
The vast majority of odeco projections were also exact up to numerical precision:
out of $10^6$ solutions, only $60$ had an eigenvalue ratio greater than $10^{-8}$.
Theorem \ref{thm.proj_stability} gives us some intuition as to why this might happen.
The stability result only holds near smooth points of the variety, but whereas
the octahedral variety is a smooth manifold,
the odeco variety has a singularity at the origin, separating polynomials of different
signs.

\begin{conjecture} \label{conj.odeco_pos}
	For a generic point $q_0 \in \R^{15}$ representing a \emph{positive}
	polynomial, the SDP relaxation yields the exact projection $\Pi_{\VOdeco}(q_0)$.
\end{conjecture}

Indeed, the green histogram in Figure \ref{fig.exactness}
shows the results of odeco projection on $10^6$ random
positive quartic polynomials, generated as sums of squares of random quadratic polynomials. For
all such positive initial points, the SDP solution had $\lambda_2(Q^*)/\lambda_1(Q^*) < 10^{-8}$,
supporting Conjecture \ref{conj.odeco_pos}.

\begin{figure}
\newcommand{\imagewidth}{0.2\columnwidth}
\begin{tabular}{cccc}
\includegraphics[align=c,width=\imagewidth]{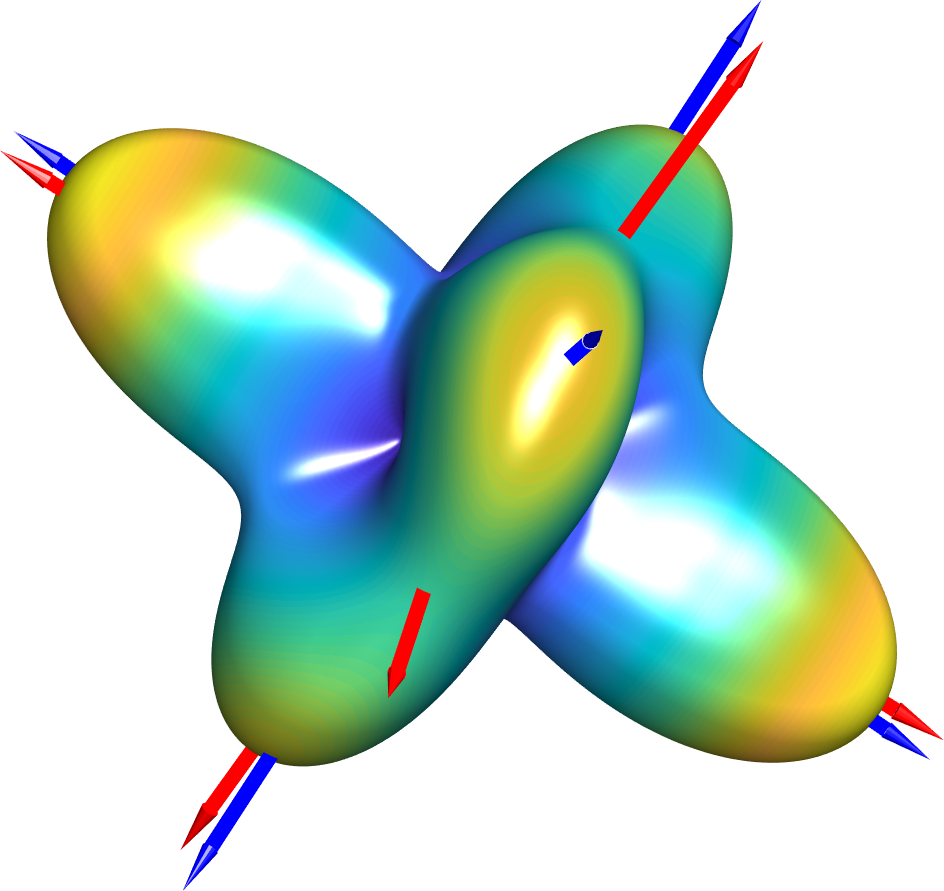} &%
\includegraphics[align=c,width=\imagewidth]{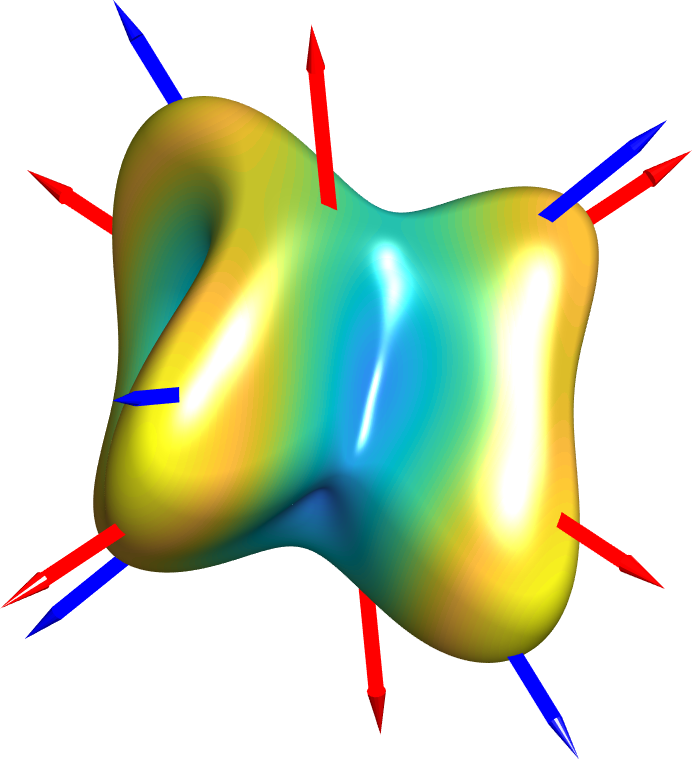} &%
\includegraphics[align=c,width=\imagewidth]{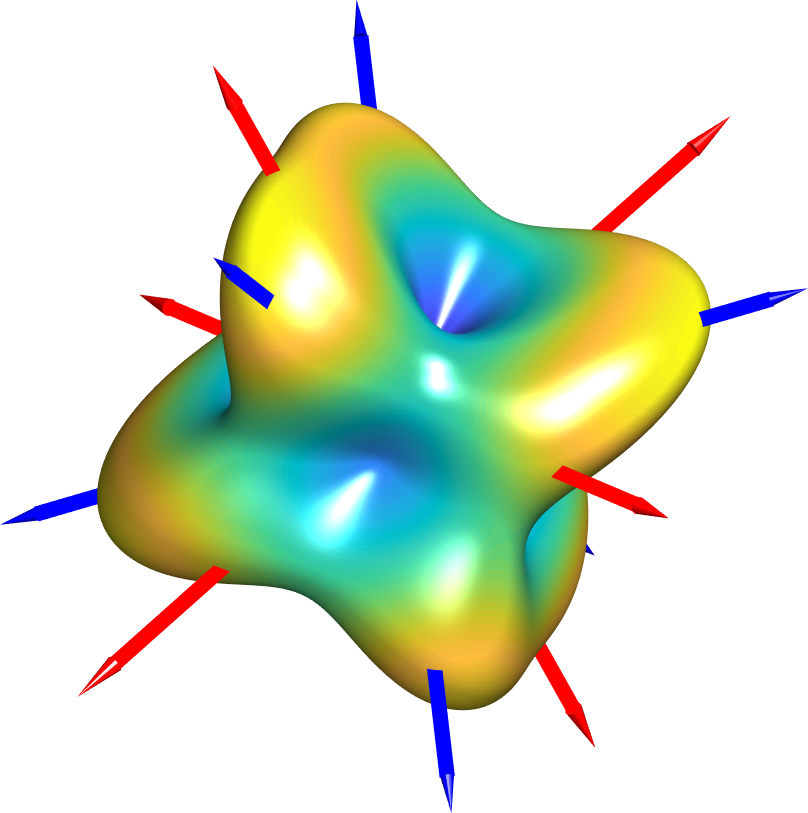} &%
\includegraphics[align=c,width=\imagewidth]{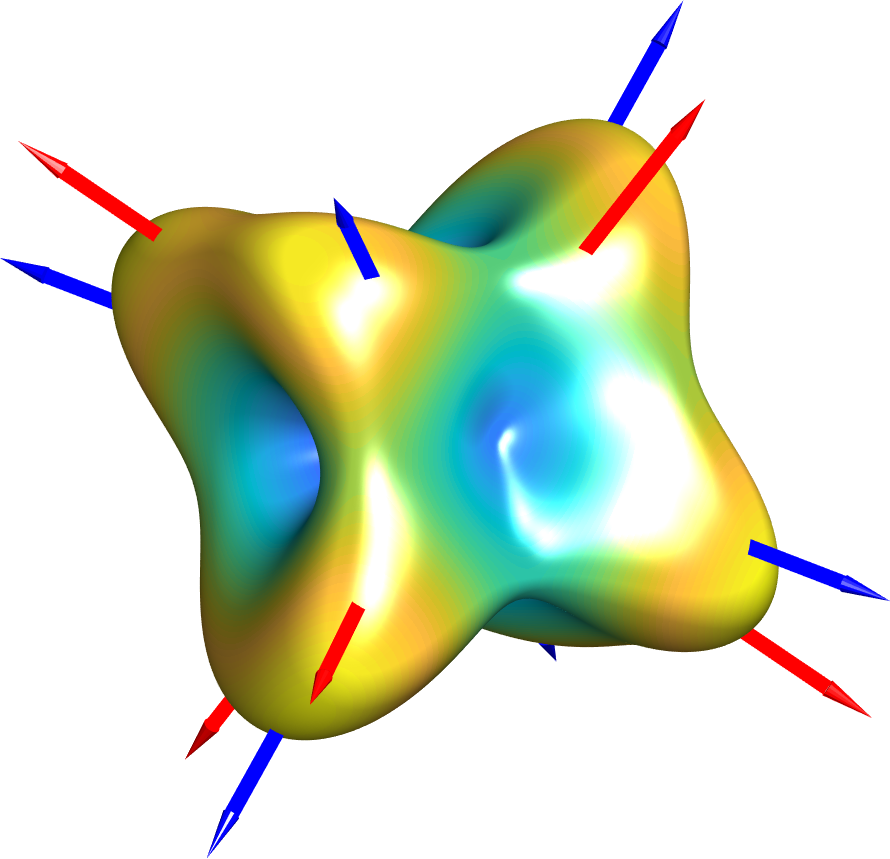} \\
{\color{red} $d_\textrm{\citeauthor{RaySokolov2}} \hfill = 0.963$} & {\color{red} 1.07} & {\color{red} 1.08} & {\color{red} 1.07} \\
{\color{blue} $d_\textrm{Ours} \hfill = 0.496$} & {\color{blue} 0.850} & {\color{blue} 0.857} & {\color{blue} 0.871}
\end{tabular}
\caption{%
For some query polynomials, our globally optimal SDP-based projection onto the octahedral variety (blue) yields dramatically different results from \citeauthor{RaySokolov2}'s approximate projection (red). Our projections are closer to the query points $y$ (distances shown in blue) compared to \citeauthor{RaySokolov2}'s (distances in red). The query polynomials are plotted on the sphere, with color and distance from the center proportional to magnitude.}
\label{fig.proj_comparison}
\end{figure}

For octahedral frames, we also compare our SDP-based projection to the
previous state of the art method proposed by
\citet{RaySokolov2}. Because \citeauthor{RaySokolov2}'s
method is based on nonconvex optimization, we would expect it to get
stuck in local minima. Indeed, out of $\num{100000}$ trials,
we found $\num{600}$ cases for which the result of \citeauthor{RaySokolov2}'s method
was at least $10^{-3}$ further from the initial point than our projection---a
nearly $1\%$ error rate. Moreover, the difference between the computed projections
can be substantial, as illustrated in Figure \ref{fig.proj_comparison}.

\section{From Frames to Frame Fields}
\label{sec.fields}

The overall aspiration of this work is to compute smooth frame fields
in a region $\Omega \subseteq \R^3$ aligned to the normal $\boldsymbol{n}$
on its boundary $\partial \Omega$. We assume $\Omega$ to be compact with
$\partial \Omega$ a union of smooth, embedded surfaces.
We think of a frame field as a map $\phi: \Omega \to \mathcal{V}$
into some space of frames $\mathcal{V}$ satisfying alignment boundary conditions.

We've examined the geometry of two candidates
for the fiber $\mathcal{V}$---the octahedral variety $\VOcta$ and
the odeco variety $\VOdeco$.\footnote{
The term ``fiber'' is intended to be suggestive.
It may be fruitful to consider a bundle $\pi: P \to \Omega$ with fiber $\mathcal{V}$,
whose restriction $\partial P \to \partial \Omega$
is nontrivial and encodes the boundary alignment condition.
The map $\phi$ would be replaced with a section of $P$.} 
It remains to describe the boundary conditions and to define an energy
we want $\phi$ to minimize.

\subsection{Boundary Conditions}
\label{subsec.bdry}

In either $\VOcta$ and $\VOdeco$, the frames aligned to a given direction
$\boldsymbol{n} \in \R^3$ form the intersection of an affine subspace with the
respective variety. In each case, this intersection is a lower-dimensional variety
defined by a different set of quadratic equations. We can impose alignment boundary
conditions by working over this lower-dimensional variety. For boundaries with sharp
creases, we can optionally exclude creased vertices, where the normal direction
is ill-defined, from the alignment constraint. Other constraints
can also be imposed at creases---for example, alignment to the crease tangent.

\subsubsection{Octahedral boundary conditions}
Let $\rho: \SO(3) \to \SO(9)$ be the irreducible representation as in \S\ref{sec.frame-space}.
First consider the octahedral frames aligned to the vertical $\boldsymbol{z} = (0, 0, 1)\tran$.
The space of vertical-aligned frames
$F_{\boldsymbol{z}} \coloneq \rho(A_{\boldsymbol{z}}) q_0$,
where $A_{\boldsymbol{z}}$ is the coaxial subgroup consisting of all rotations about $\boldsymbol{z}$, i.e.,
\[ A_{\boldsymbol{z}} = \{\exp(t l_3) \mid t \in \R\}, \]
and $q_0$ is the canonical octahedral frame.
As derived in \cite{huang, RaySokolov2}, such frames have the form
\begin{equation}
\begin{aligned}
\exp(t L_3) q_0 &= \left(\sqrt{\frac{5}{12}}\sin(4t), 0, 0, 0, \sqrt{\frac{7}{12}}, 0, 0, 0, \sqrt{\frac{5}{12}}\cos(4t)\right)\tran \\	
&= q_{\boldsymbol{z}} + B s_{\boldsymbol{z}},
\end{aligned}
\end{equation}
where $q_{\boldsymbol{z}} = (0, 0, 0, 0, \sqrt{7/12}, 0, 0, 0, 0)\tran$,
$s_{\boldsymbol{z}} = (\cos(4t), \sin(4t))\tran$, and $B_{\boldsymbol{z}}$
is the obvious $9 \times 2$ matrix.
Given $q \in \R^9$, the closest vertical-aligned frame is
\[ \Pi_{\VOcta_{\boldsymbol{z}}}(q) = q_{\boldsymbol{z}} + B_{\boldsymbol{z}}\frac{B_{\boldsymbol{z}}\tran q}{\|B_{\boldsymbol{z}}\tran q\|}. \]

For a general unit normal $\boldsymbol{n}$, the aligned frames can be written
\begin{equation} \label{eq.octa_bdry}
\rho(r_{\boldsymbol{n}})(q_{\boldsymbol{z}} + B_{\boldsymbol{z}} s_{\boldsymbol{z}}), \end{equation}
where $r_{\boldsymbol{n}} \in \SO(3)$ is any rotation taking
$\boldsymbol{z}$ to $\boldsymbol{n}$. The projection of $q \in \R^9$
onto the $\boldsymbol{n}$-aligned frames is then
\begin{equation} \label{eq.proj_aligned} \Pi_{\VOcta_{\boldsymbol{n}}}
= \rho(r_{\boldsymbol{n}})\Pi_{\VOcta_{\boldsymbol{z}}}(\rho(r_{\boldsymbol{n}})\tran q) \end{equation}
During optimization, this projection is used for boundary-aligned frames.

\subsubsection{Odeco boundary conditions} 
Consider the standard odeco frame $\sum_i \lambda_i x_i^4$ rotated by
some $r \in A_{\boldsymbol{z}}$, and fix $\lambda_3 = 1$.
As described in \S\ref{subsec.octa_odeco},
it has coefficients in the even-numbered irreducible representations
(bands of spherical harmonics)
\[ q = (q_0, q_2, q_4) \in V_0 \times V_2 \times V_4 = \R \times \R^5 \times \R^9. \]
Just as in the octahedral case, $q$
can be decomposed into a fixed part and a rotating part parametrized by
a lower-dimensional vector
\begin{equation} \label{eq.odeco_bdry}
q = q_{\boldsymbol{z}} + B_{\boldsymbol{z}} s_{\boldsymbol{z}}, \end{equation}
where now $s_{\boldsymbol{z}} \in \R^5$ and $B_{\boldsymbol{z}} \in \R^{15 \times 5}$.
The equations
\eqref{eq.odeco_quads} reduce to three quadratic equations in the coefficients of
$s_{\boldsymbol{z}}$, which
can be used to construct a semidefinite program to project onto the vertical-aligned
odeco frames, as in \S\ref{subsec.proj}. The details are
given in the supplementary material.
For frames aligned to a direction $\boldsymbol{n}$,
\[ q = \rho(r_{\boldsymbol{n}}) (q_{\boldsymbol{z}} + B_{\boldsymbol{z}} s_{\boldsymbol{z}}), \]
where $\rho$ is now the direct sum representation on $V_0 \times V_2 \times V_4$.
Projection onto the $\boldsymbol{n}$-aligned odeco frames can be constructed
from projection onto the $\boldsymbol{z}$-aligned odeco frames as in \eqref{eq.proj_aligned}.

\subsection{Objective Function}

As we are searching for smoothest fields, a natural choice for the energy
is Dirichlet energy $\frac{1}{2}\int_\Omega \|\nabla\phi\|^2 \; dx$, where the norm
is taken with respect to a metric on $\mathcal{V}$. There are
multiple problems this formulation will need to confront. As we have
seen, $\VOcta$ is a smooth manifold. But being a quotient of the $3$-sphere,
it has positive curvature, making mere existence of harmonic maps
$\Omega \to \VOcta$ a hard problem. Even more fundamentally, it is
not clear how to represent singularities of $\phi$---the map cannot be consistently defined
along singular curves. $\VOdeco$ attempts to
solve this problem by representing singular frames with only one direction,
while allowing the other two to vanish. Along a singular
curve, we would expect an odeco field to take rank-one values of the form
$\lambda v^{\otimes 4}$, where $v$ is tangent to the singular curve.

\subsection{Discretization}
Let $\mathcal{T} = (V, T)$ be a tetrahedral mesh of $\Omega$
with vertices $V$ and tetrahedra $T$. The set of boundary vertices will be denoted
by $\partial V$. A discrete octahedral
frame field on $\mathcal{T}$ is specified by a map $q : V \to \mathcal{V} = \VOcta$ or
$\VOdeco$, giving a frame $q_i$ for each vertex $i$. As the octahedral variety
$\VOcta \subset \R^9$ and the odeco variety
$\VOdeco \subset \R^{15}$, we can think of $q$ as a $d \times n$ matrix, where $n = |V|$
and $d = 9$ or $15$, respectively.

We then define a discretized Dirichlet energy using the Euclidean metric in
the spherical harmonic basis to compare elements
of $\mathcal{V}$. This is equivalent to measuring $L^2$ distance between the
corresponding polynomials over the sphere, as employed in \cite{huang, RaySokolov2, Solomon}.
Note that this would not be the case if we compared odeco frames in the monomial basis. The
discrete energy is
\[ E(q) = \frac{1}{2} \tr\left(q L q\tran\right), \]
where $L$ is the linear finite element Laplacian on $\mathcal{T}$, with the
usual cotangent weights.

\section{Algorithms}
\label{sec.alg}

We now have two \emph{variety-constrained optimization} problems of the form
\begin{equation} \label{eq.overall_problem}
\begin{alignedat}{4}
	&\min \; &&\frac{1}{2}\tr\left(q L q\tran\right) \\
	&\text{s.t. } && q_i \in \mathcal{V}, &&\forall i \in V \\
	&&& q_i \in \mathcal{V}_{\boldsymbol{n}_i}, \quad &&\forall i \in \partial V,
\end{alignedat}
\end{equation}
where the variety $\mathcal{V}$ is either $\VOcta$ or $\VOdeco$, $q_i$
are the columns of $q$, $\partial V$
denotes the set of boundary vertices, and $\mathcal{V}_{\boldsymbol{n}_i}$
is the variety of frames aligned to the normal at boundary vertex $i$.

\subsection{Manifold Optimization Methods}
\label{subsec.rtr}
\begin{wrapfigure}{l}{0.25\columnwidth}
\centering
\includegraphics[width=0.25\columnwidth]{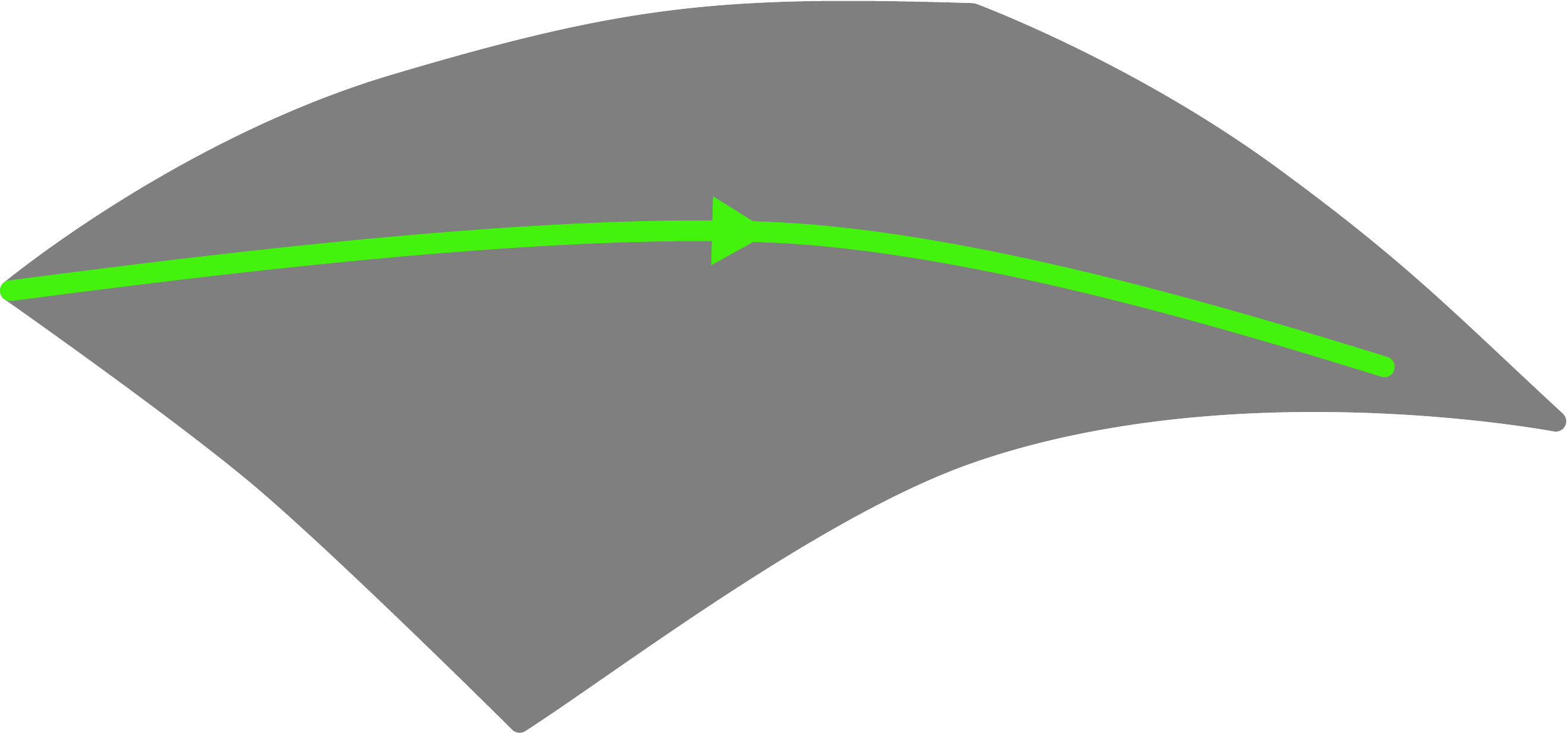}	
\end{wrapfigure}
In the case $\mathcal{V} = \VOcta$, \eqref{eq.overall_problem} becomes a manifold-constrained
optimization problem.
The octahedral variety is a Riemannian manifold,
and we can compute geodesics on it
as described in \S\ref{subsec.geodesics}.
So we can apply the standard Riemannian trust-region (RTR) algorithm \cite{Absil2007}
as implemented in the Manopt toolbox for MATLAB \cite{manopt}.
We consider the field to be a point in the product manifold
\[ q \in \prod_{i \in V \setminus \partial V} \VOcta \times
\prod_{i \in \partial V}\VOcta_{\boldsymbol{n}_i}. \]
In addition to the geodesics computed in \S\ref{subsec.geodesics}, we also need
a way to project vectors (e.g. gradients) in the ambient space $\R^9$ into the
tangent space at a point $q$, $\operatorname{proj}_q: \R^9 \to T_q\VOcta$.
The vectors $\{L_i q\}_{i=1}^3$ form an orthogonal basis for this tangent
space (\emph{cf.} Proposition \ref{thm.isometric}), so projection
is just given by taking the inner product with each of these vectors.

The odeco variety $\VOdeco$ is not a manifold, but it is smooth almost everywhere.
We know of no way to compute exact geodesics on $\VOdeco$, but we can implement
a version of RTR by replacing the exact exponential map with a \emph{retraction}, i.e., a first-order
approximation (\emph{cf.} \cite[Definition 2.1]{Absil2007}).
In doing so, we give up the superlinear local convergence guarantees
of RTR (\emph{cf.} \cite[Theorem 4.12]{Absil2007}) but retain a global convergence
guarantee. In practice, we find that odeco RTR converges at a reasonable rate.

At a smooth point of $\VOdeco$, it is easy to project vectors onto the tangent space using the fact that $\VOdeco$ is defined
by \emph{quadratic} equations. Let $q \in \VOdeco$, and assume the coefficients of $q$
are expressed in spherical harmonic coefficients.
Then differentiating the equations
$q\tran \tilde{A}_i q = 0$ shows that the normal space at $q$ is
\[ N_q \VOdeco = \Span \{\tilde{A}_i q\}, \]
where $\tilde{A}_i$ are expressed in the spherical harmonic basis.
Then $T_q \VOdeco = (N_q \VOdeco)^\perp$,
where the orthogonal complement is taken with respect to the standard metric under which
the spherical harmonic functions are orthonormal.

We compute retractions as follows. The tangent
space to the odeco variety decomposes into a \emph{rotational part} and a \emph{scaling part}:
\[ T^r_q \VOdeco = \Span \{\tilde{L}_i q\}_{i=1}^3, \qquad T^s_q \VOdeco = (T^r_q \VOdeco)^\perp, \]
where the orthogonal complement is taken within $T_q \VOdeco$. We then set
\[ \operatorname{retr}_q(v) = e^{v_r \cdot \tilde{L}} (q + v_s), \]
where $v_s$ is the component of $v$ in $T^s_	q \VOdeco$, $v_r$ is the component of $v$ in
$T^r_q \VOdeco$, and
\[ v_r \cdot \tilde{L} \coloneqq \sum_j (v_r)_j \tilde{L}_j, \]
where $v_r = \sum_j (v_r)_j \tilde{L}_j q$. It is simple to verify that
$\operatorname{retr}_q(v) \in \VOdeco$ and
that
\[ \left. \frac{\partial}{\partial t} \operatorname{retr}_q(tv) \right\rvert_{t = 0} = v. \]

\begin{figure}
\centering
\begin{tikzpicture}
\begin{semilogyaxis}[
    enlarge x limits = false,
    minor y tick num = 4,
    xlabel={\footnotesize Time (s)},
    ylabel={\footnotesize Gradient Norm},
    xlabel near ticks,
    ylabel near ticks,
    every tick label/.append style = {font=\tiny},
    legend style={font=\tiny}]
\addplot[mark=none, color=red!60!yellow] table[x index = 0, y index = 1, col sep = comma] {figures/convergence/bone_80k_raycond_grad_ray.csv};
\addlegendentry{Ray et al. (Bone)}
\addplot[mark=none, color=blue!60!green] table[x index = 0, y index = 1, col sep = comma] {figures/convergence/bone_80k_raycond_grad_rtr.csv};
\addlegendentry{RTR (Bone)}
\addplot[mark=none, color=red] table[x index = 0, y index = 1, col sep = comma] {figures/convergence/cup_85k_raycond_grad_ray.csv};
\addlegendentry{Ray et al. (Cup)}
\addplot[mark=none, color=blue] table[x index = 0, y index = 1, col sep = comma] {figures/convergence/cup_85k_raycond_grad_rtr.csv};
\addlegendentry{RTR (Cup)}
\addplot[mark=none, color=red!60!black] table[x index = 0, y index = 1, col sep = comma] {figures/convergence/rockerarm_91k_raycond_grad_ray.csv};
\addlegendentry{Ray et al. (Rockerarm)}
\addplot[mark=none, color=blue!60!black] table[x index = 0, y index = 1, col sep = comma] {figures/convergence/rockerarm_91k_raycond_grad_rtr.csv};
\addlegendentry{RTR (Rockerarm)}
\end{semilogyaxis}
\end{tikzpicture}
\caption{Convergence behavior of octahedral RTR and the method of \citet{RaySokolov2}
on various models, starting from \citeauthor{RaySokolov2}'s initialization
and using the combinatorial Laplacian.}
\label{fig.convergence-raycond}
\end{figure}
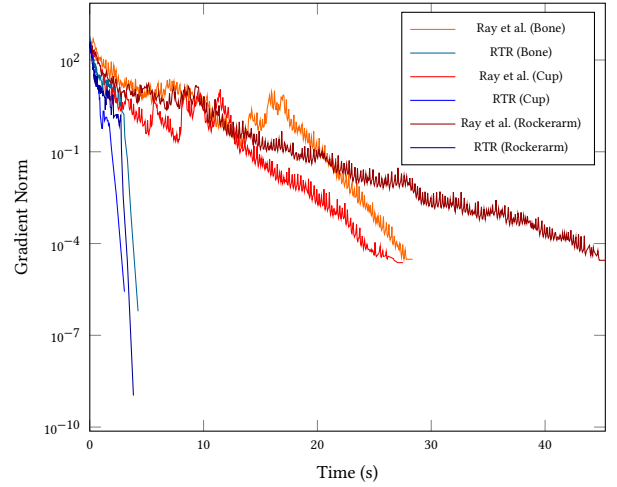

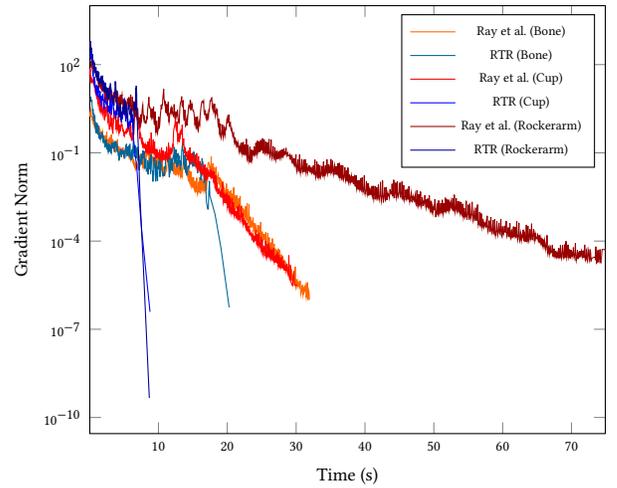
\begin{figure}
\centering
\begin{tikzpicture}
\begin{semilogyaxis}[
    enlarge x limits = false,
    minor y tick num = 4,
    xlabel={\footnotesize Time (s)},
    ylabel={\footnotesize Gradient Norm},
    xlabel near ticks,
    ylabel near ticks,
    every tick label/.append style = {font=\tiny},
    legend style={font=\tiny}]
\addplot[mark=none, color=red!60!yellow] table[x index = 0, y index = 1, col sep = comma] {figures/convergence/bone_80k_ourcond_grad_ray.csv};
\addlegendentry{Ray et al. (Bone)}
\addplot[mark=none, color=blue!60!green] table[x index = 0, y index = 1, col sep = comma] {figures/convergence/bone_80k_ourcond_grad_rtr.csv};
\addlegendentry{RTR (Bone)}
\addplot[mark=none, color=red] table[x index = 0, y index = 1, col sep = comma] {figures/convergence/cup_85k_ourcond_grad_ray.csv};
\addlegendentry{Ray et al. (Cup)}
\addplot[mark=none, color=blue] table[x index = 0, y index = 1, col sep = comma] {figures/convergence/cup_85k_ourcond_grad_rtr.csv};
\addlegendentry{RTR (Cup)}
\addplot[mark=none, color=red!60!black] table[x index = 0, y index = 1, col sep = comma] {figures/convergence/rockerarm_91k_ourcond_grad_ray.csv};
\addlegendentry{Ray et al. (Rockerarm)}
\addplot[mark=none, color=blue!60!black] table[x index = 0, y index = 1, col sep = comma] {figures/convergence/rockerarm_91k_ourcond_grad_rtr.csv};
\addlegendentry{RTR (Rockerarm)}
\end{semilogyaxis}
\end{tikzpicture}
\caption{Convergence behavior of octahedral RTR and the method of \citet{RaySokolov2}
on various models, starting from random initialization
and using the geometric Laplacian.}
\label{fig.convergence-ourcond}
\end{figure}

We compare the convergence behavior of octahedral RTR to that of the method of \citet{RaySokolov2} under
two sets of conditions---\citeauthor{RaySokolov2}'s initialization and combinatorial
Laplacian (Figure \ref{fig.convergence-raycond}); and our random initialization and linear finite element Laplacian (Figure \ref{fig.convergence-ourcond}).
The quadratic local convergence of RTR stands in stark contrast
to the slower linear convergence behavior of
\citeauthor{RaySokolov2}'s method. RTR converges faster but finds solutions
of similar Dirichlet energy to previous work (see Table 1 in the supplemental document).

\subsection{Generalized MBO Methods}
\label{subsec.mbo}
As we have observed, it is possible to do optimization on the octahedral
and odeco varieties by moving along curves that stay on the varieties exactly. However,
this hard constraint sometimes causes pure manifold optimization to get stuck in local minima.
To avoid such local minima, an approach that allows ``tunneling'' is required.
\begin{wrapfigure}{r}{0.25\columnwidth}
\centering
\includegraphics[width=0.25\columnwidth]{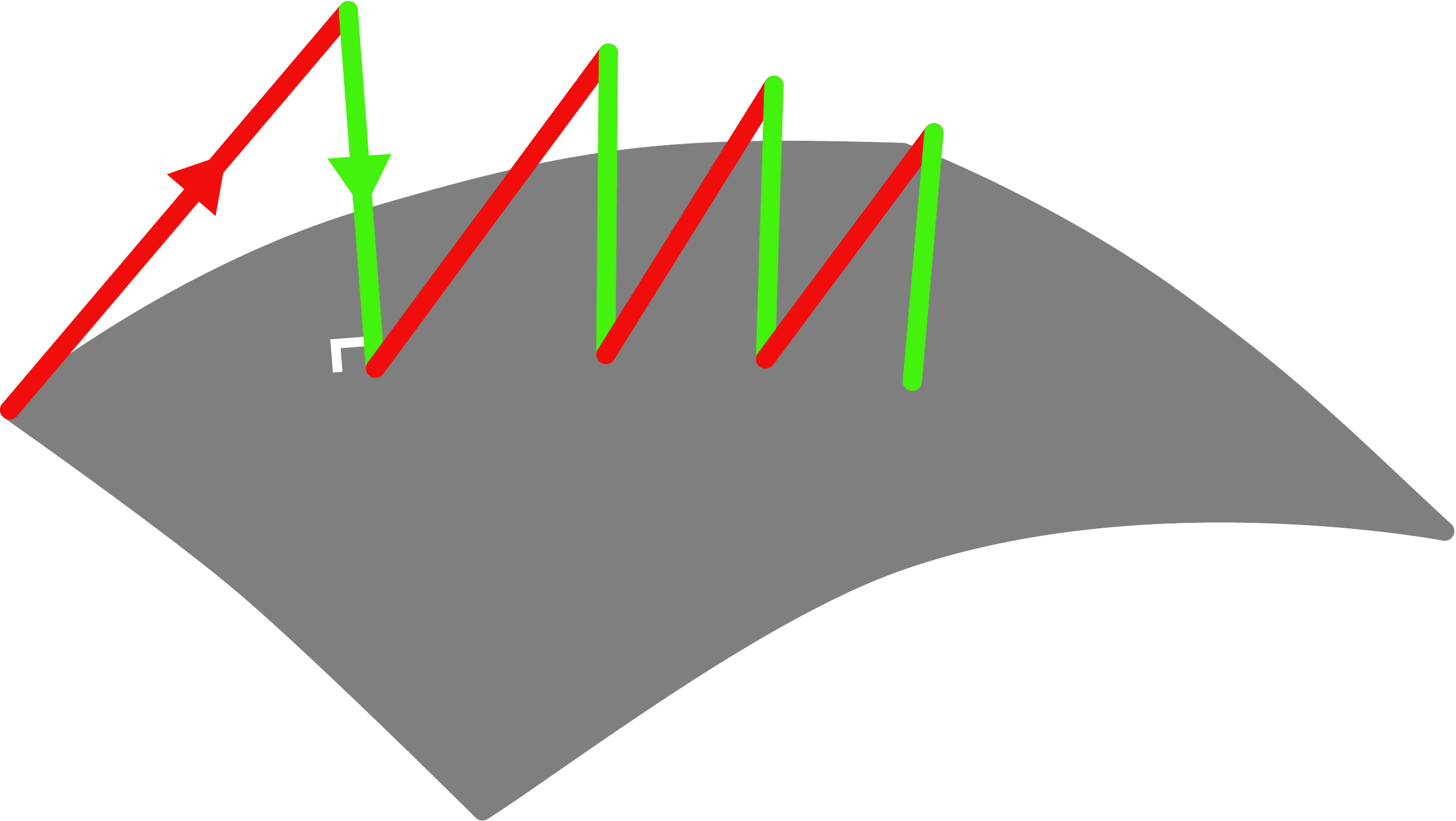}	
\end{wrapfigure}

The method of Merriman, Bence, and Osher \shortcite{MBO1} is a
diffusion-generated method for computing (possibly singular)
maps into a target space embedded
in a Euclidean space.
Following \cite{viertel,osting2017}, we first replace the hard constraint that the
field values lie in the variety with a penalty term, yielding an
energy of Ginzburg-Landau type,
\begin{equation}
	E_\epsilon(q) = \frac{1}{2}\int_\Omega \|\nabla q\|^2 \; dx
	+ \frac{1}{2\epsilon^2}\int_\Omega \dist(q, \mathcal{V})^2 \; dx, \label{eq.g-l}
\end{equation}
where $\mathcal{V}$ is our variety. Consider this energy in the limit $\epsilon \to 0$,
as in \cite[\S4]{viertel}.
The MBO method consists of alternating descent on the two terms of $E_\epsilon$.
Gradient flow on the first term---Dirichlet energy---yields
componentwise heat diffusion,
\begin{equation}
\begin{aligned}
	\partial_t q(x, t) &= \Delta q(x, t) \\
	q(x, 0) &= q_{k - 1}.
\end{aligned}
\end{equation}
with the boundary constraints given in \S\ref{subsec.bdry}.
In practice, we do one step of implicit Euler integration per overall algorithm
step.
Gradient flow on the second term of \eqref{eq.g-l} in the limit $\epsilon \to 0$
amounts to projection onto the variety.
The overall algorithm is shown in Algorithm \ref{alg.mbo}.

\begin{algorithm}
\DontPrintSemicolon
\SetKwInOut{Input}{input}
\SetKwInOut{Result}{result}
\Input{initial $d \times n$ field $q_0$, diffusion parameter $\tau_0$,
	   optimization schedule $\beta(k)$}
\Result{$q_k$}
\BlankLine
$k \gets 1$\;
\Repeat{$\Delta E_k / E_k < \delta$ or $\Delta_k < \delta$}{
	$\tau_k \gets \beta(k) \tau_0$\;
	\textbf{Diffusion step: }Solve the linear system $(M - \tau_k L) \overline{q}_k\tran = M q_{k-1}\tran$
	with columns $(\overline{q}_k)_i$ constrained to be in the affine span
	\eqref{eq.octa_bdry}
	or \eqref{eq.odeco_bdry} for each $i \in \partial V$.\;
	\textbf{Projection step: } Project $\overline{q}_k$ into the variety:
	$(q_k)_i \gets \begin{cases}
	\Pi_{\mathcal{V}}((\overline{q}_k)_i) &i \notin \partial V \\
	\Pi_{\mathcal{V}_{\boldsymbol{n}_i}}((\overline{q}_k)_i) &i \in \partial V.
	\end{cases}$\;
	$\Delta_k \gets \frac{\tr((q_k - q_{k-1}) M (q_k - q_{k-1})\tran)}
						 {\tr(q_k M q_k\tran)}$\;
	$E_k \gets \tr(q_k L q_k\tran)$\;
	$\Delta E_k \gets E_k - E_{k-1}$\;
	$k \gets k + 1$
}
\caption{MBO over a variety $\mathcal{V} \subset \R^d$\label{alg.mbo}}
\end{algorithm}

\begin{figure}
\centering
\newcommand{\imagewidth}{0.3\columnwidth}
\begin{tabular}{ccc}
\includegraphics[align=c,width=\imagewidth]{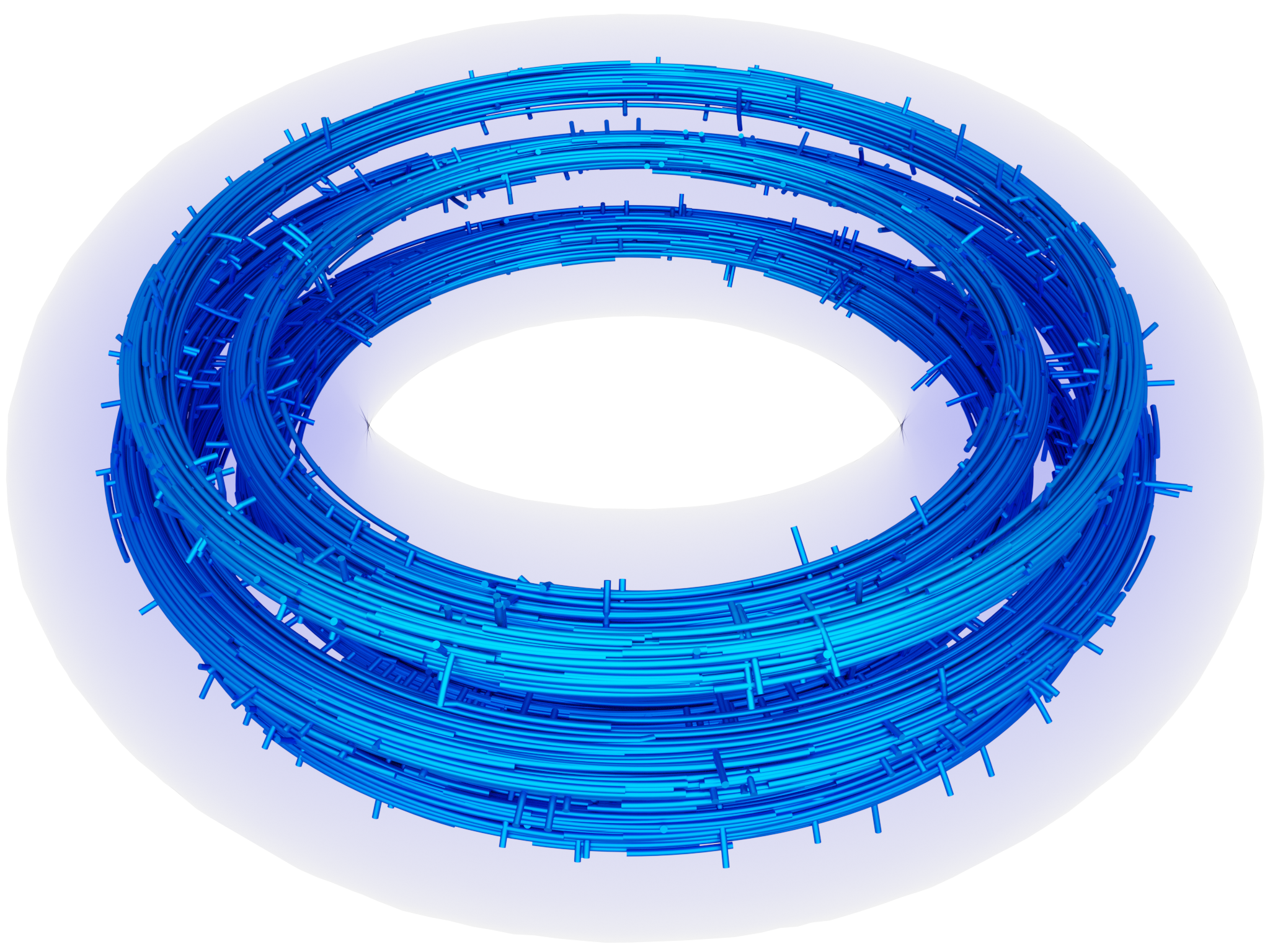} &%
\includegraphics[align=c,width=\imagewidth]{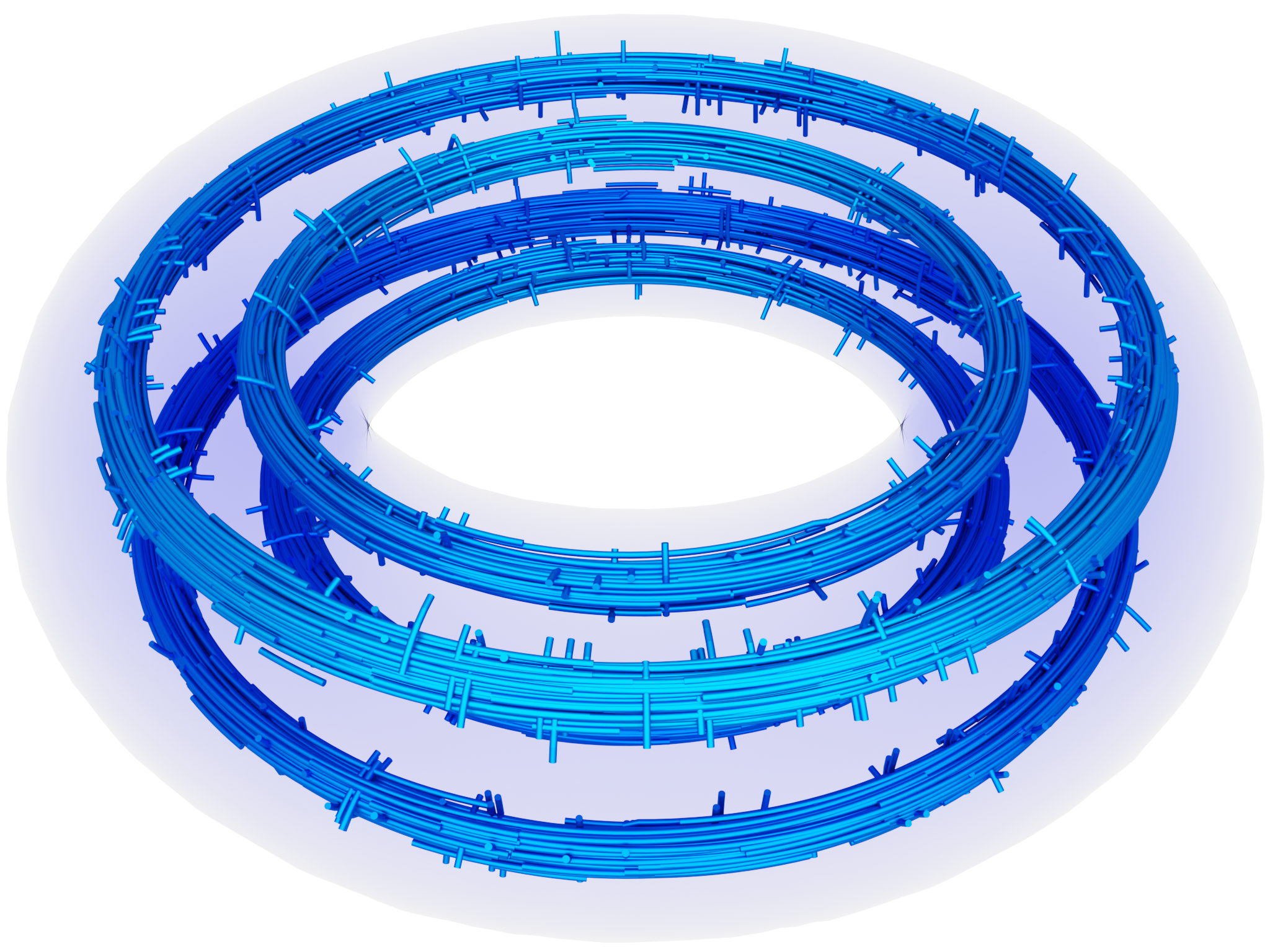} &%
\includegraphics[align=c,width=\imagewidth]{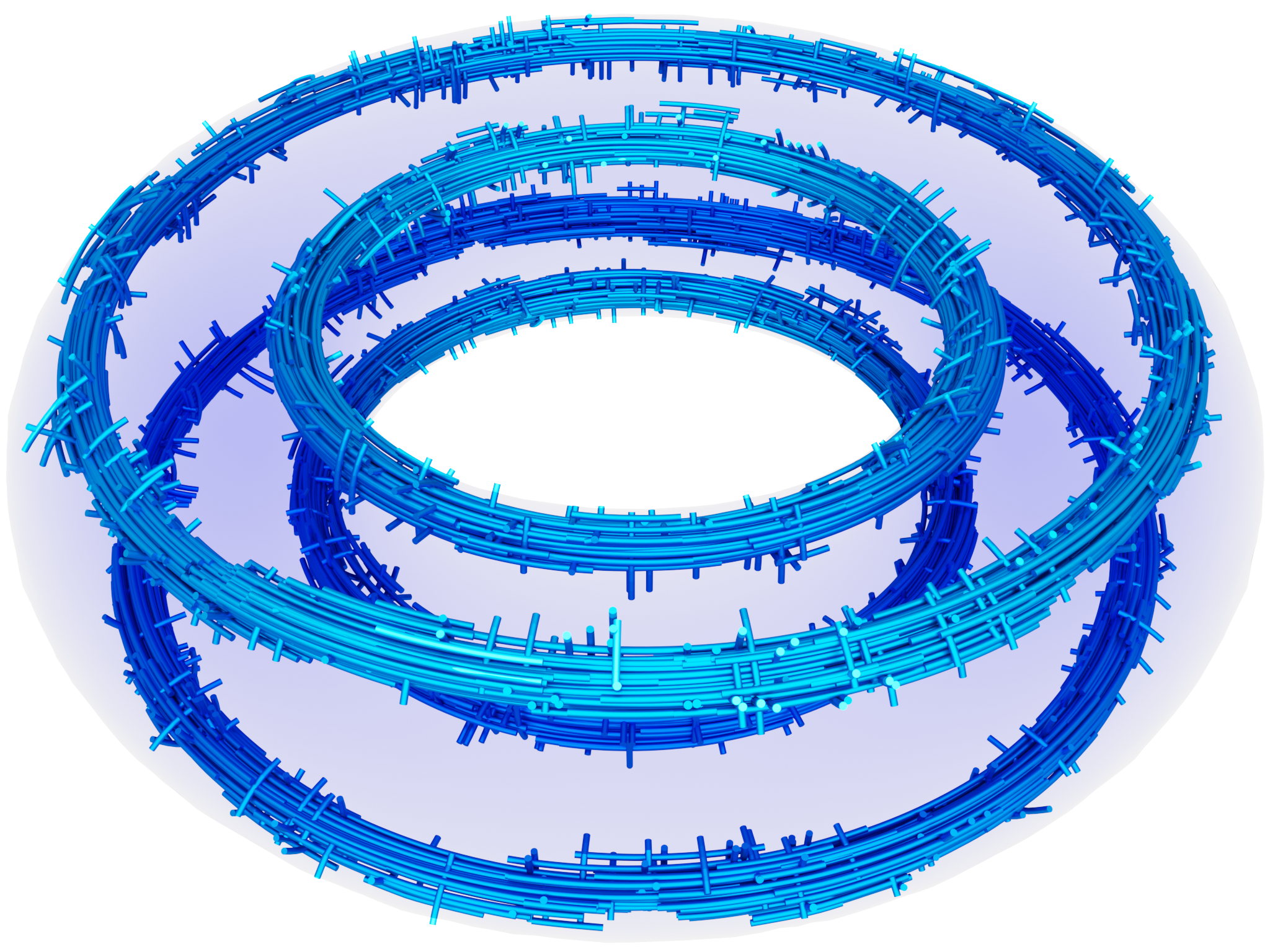} \\
\includegraphics[align=c,width=\imagewidth]{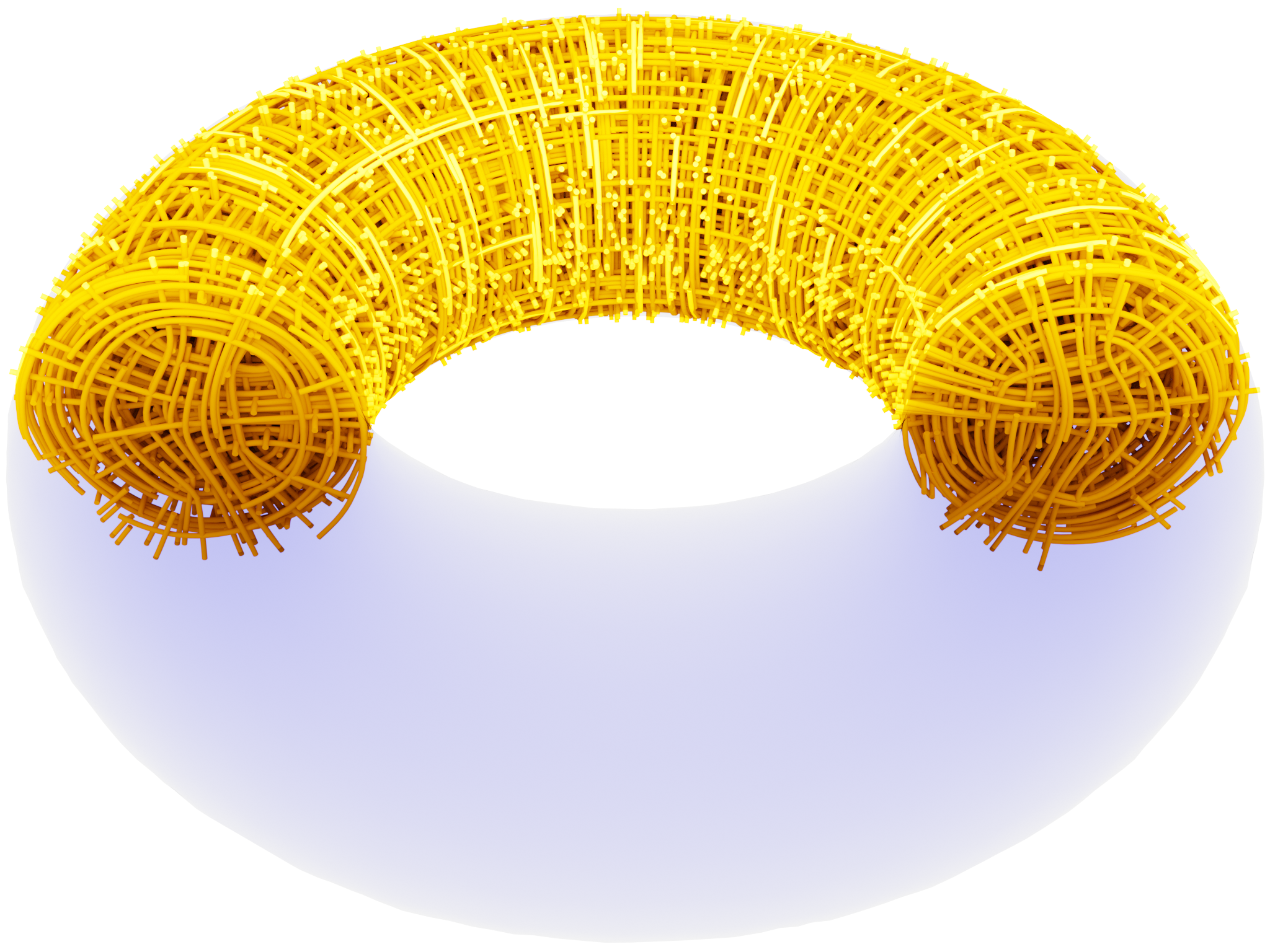} &%
\includegraphics[align=c,width=\imagewidth]{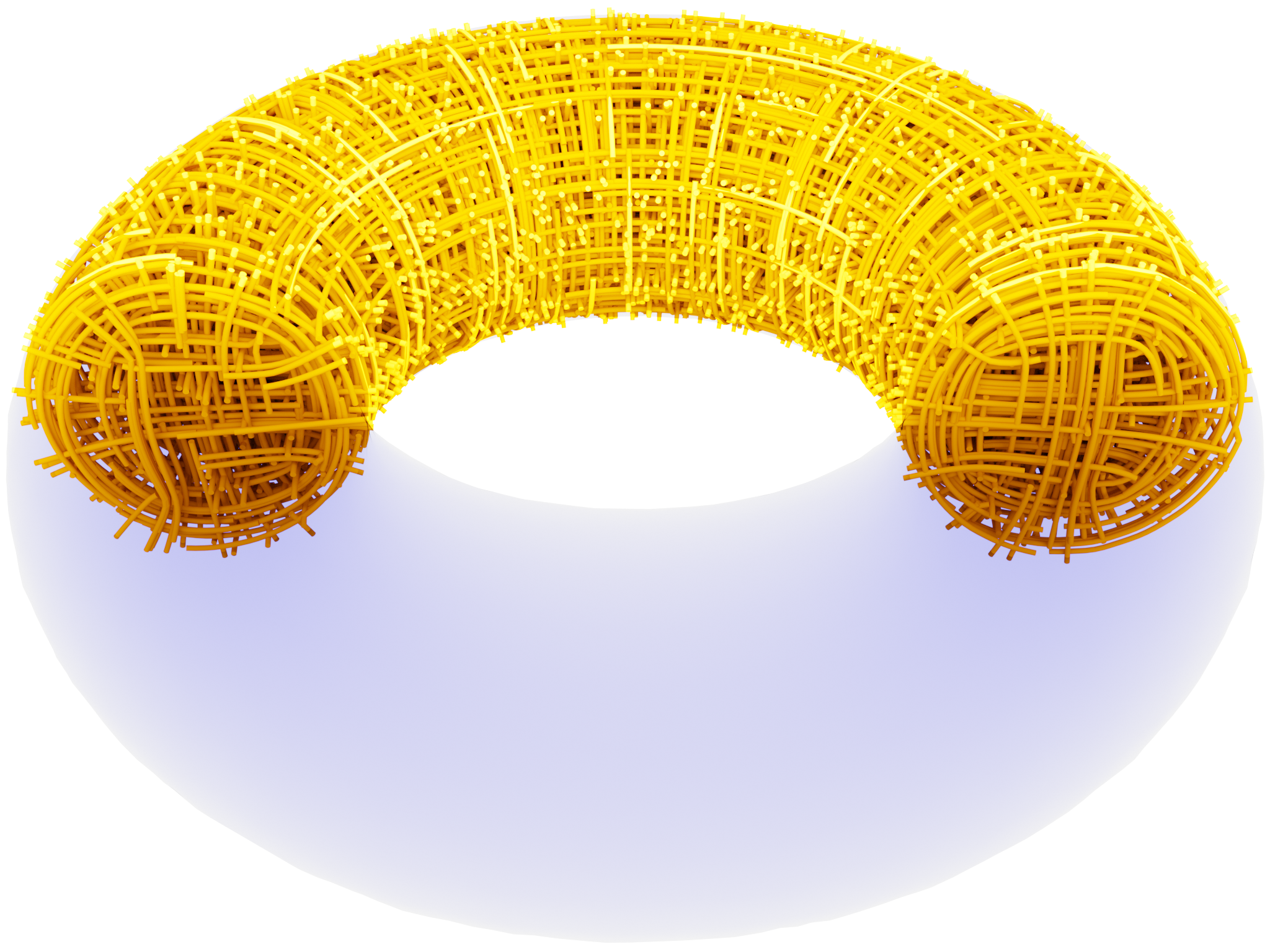} &%
\includegraphics[align=c,width=\imagewidth]{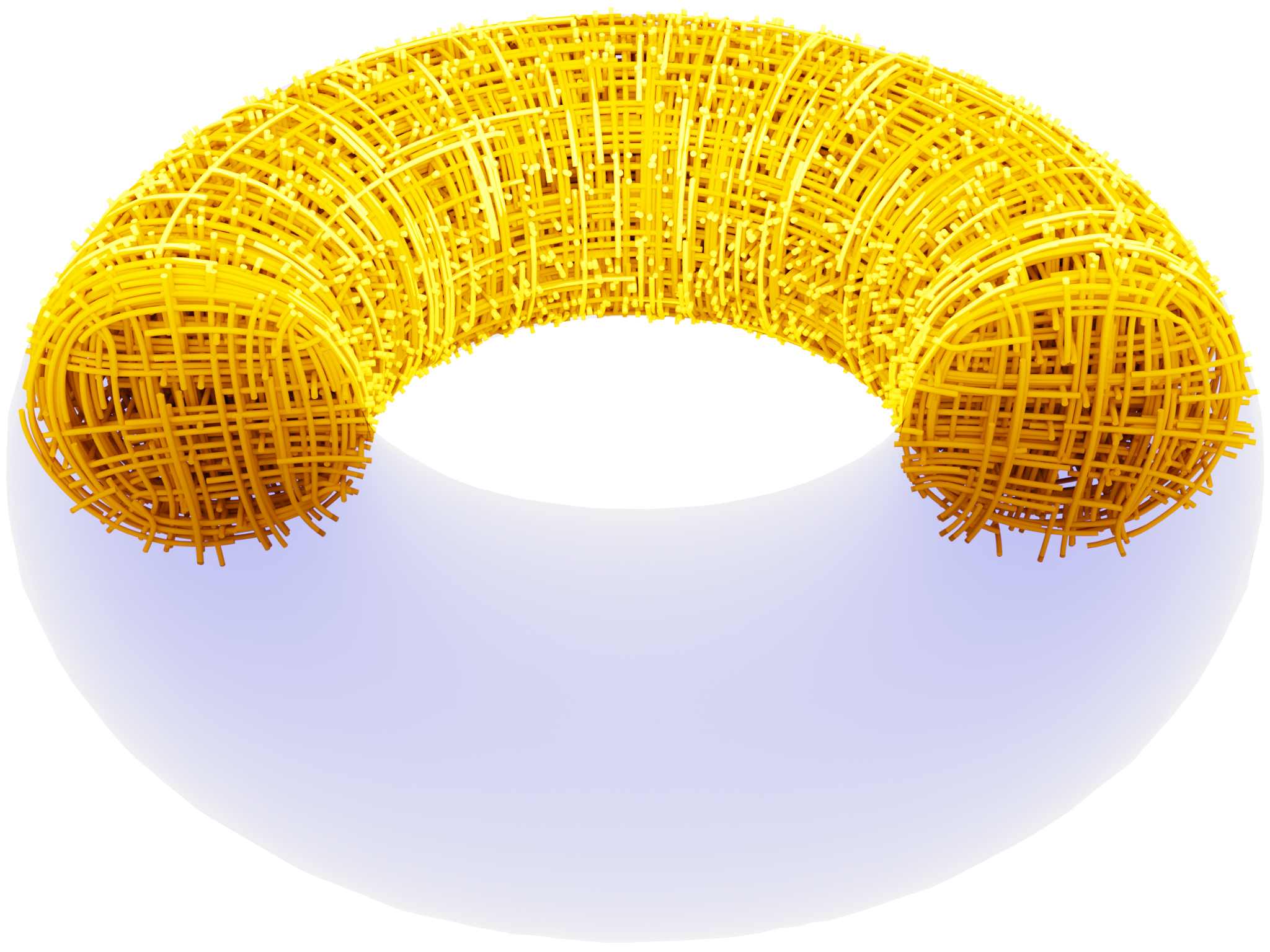} \\
\footnotesize{$\tau \approx 0.5$} &%
\footnotesize{$\tau \approx 0.05$} &%
\footnotesize{$\tau \approx 0.05$ + RTR} \\
\footnotesize{$E = 106.76$} &%
\footnotesize{$E = 106.76$} &%
\footnotesize{$E = 85.99$}%
\end{tabular}%
\caption{Results of octahedral MBO on a torus. With a relatively
large diffusion time $\tau$, MBO produces a field with tightly packed singular
regions (left).
At a smaller value of $\tau$, singular curves relax toward
the boundary, reducing the Dirichlet energy (center). 
RTR reduces the energy even further (right),
pushing the singular curves to the boundary.}
\label{fig.mbo_tau}
\end{figure}

We follow the suggestion of \citep{viertel,vanGennip2014}
to set $\tau_0$ relative to the inverse of the smallest nonzero eigenvalue of the Laplacian.
In ordinary MBO, $\beta(k) = 1$, i.e.,
$\tau$ does not change over the course of the algorithm.
In Figure \ref{fig.mbo_tau}, we test the effects of different values of
$\tau$. As we decrease $\tau$, the field is able to relax more and reduce the Dirichlet energy.
On the other hand, larger values of $\tau$ allow more tunneling so that
the field can escape local minima.
This suggests a modified
MBO scheme (mMBO) in which we start with a large $\tau$ and progressively reduce
it over the course of optimization---analogously to what happens to
the temperature in simulated annealing.
Our mMBO uses an optimization schedule $\beta(k) = 50 k^{-3}$.
This optimization schedule starts with a large diffusion time for
robustness to random initialization, then
sweeps $\tau$ through various orders of magnitude
quickly. We have found that this heuristic produces a good
balance between quick convergence and field quality.

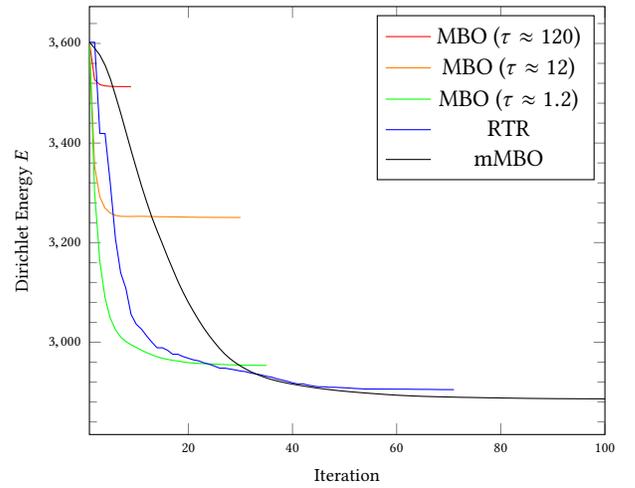
\begin{figure}
\centering
\begin{tikzpicture}
\begin{axis}[
    enlarge x limits = false,
    minor y tick num = 4,
    xlabel={\footnotesize Iteration},
    ylabel={\footnotesize Dirichlet Energy $E$},
    xlabel near ticks,
    ylabel near ticks,
    every tick label/.append style = {font=\tiny},
    cycle list = {red, orange, green, blue, black}]
\addplot+[mark=none] table[x expr = \coordindex + 1, y index = 0, col sep = comma] {figures/MBO_E_only.csv};
\addlegendentry{MBO ($\tau \approx 120$)}
\addplot+[mark=none] table[x expr = \coordindex + 1, y index = 0, col sep = comma] {figures/MBO2_E_only.csv};
\addlegendentry{MBO ($\tau \approx 12$)}
\addplot+[mark=none] table[x expr = \coordindex + 1, y index = 0, col sep = comma] {figures/MBO3_E_only.csv};
\addlegendentry{MBO ($\tau \approx 1.2$)}
\addplot+[mark=none] table[x expr = \coordindex + 1, y index = 1, col sep = comma] {figures/RTR.csv};
\addlegendentry{RTR}
\addplot+[mark=none] table[x expr = \coordindex + 1, y index = 1, col sep = comma] {figures/MMBO.csv};
\addlegendentry{mMBO}
\end{axis}
\end{tikzpicture}
\caption{Energy convergence on the rockerarm\_91k model. MBO's convergence to lower
energies is limited by the diffusion time $\tau$. Decreasing $\tau$ according
to the optimization schedule used in mMBO achieves the best results overall.}
\label{fig.convergence}
\end{figure}

Figure \ref{fig.convergence} depicts the convergence behavior of RTR, MBO with various $\tau$ values, and mMBO
starting from a projection of an approximately harmonic field. While the energy value achieved by
ordinary MBO is limited for each $\tau$, mMBO achieves a lower value by sweeping
through various values of $\tau$.
Note that the iteration counts shown do not reflect wall-clock time; in particular,
RTR runs at least an order of magnitude
faster than the other algorithms.

\section{Experiments}
\label{sec.exper}

We initialize our algorithms with
vertexwise random octahedral fields, generated
by---separately for each vertex---starting with the canonical frame,
rotating it by a random angle between $0$ and $2\pi$
about the positive $z$-axis, and then rotating the positive $z$-axis to
a random direction.
We do this even for optimization of odeco fields to avoid encountering odeco frames that
have negative weights $\lambda_i$ (\emph{cf.} Conjecture
\ref{conj.odeco_pos}).
When starting from octahedral initialization,
the odeco frames we compute always have nonnegative weights.

\begin{figure}[ht]
\newcommand{\imagewidth}{0.48\columnwidth}
\centering
\includegraphics[width=\imagewidth]{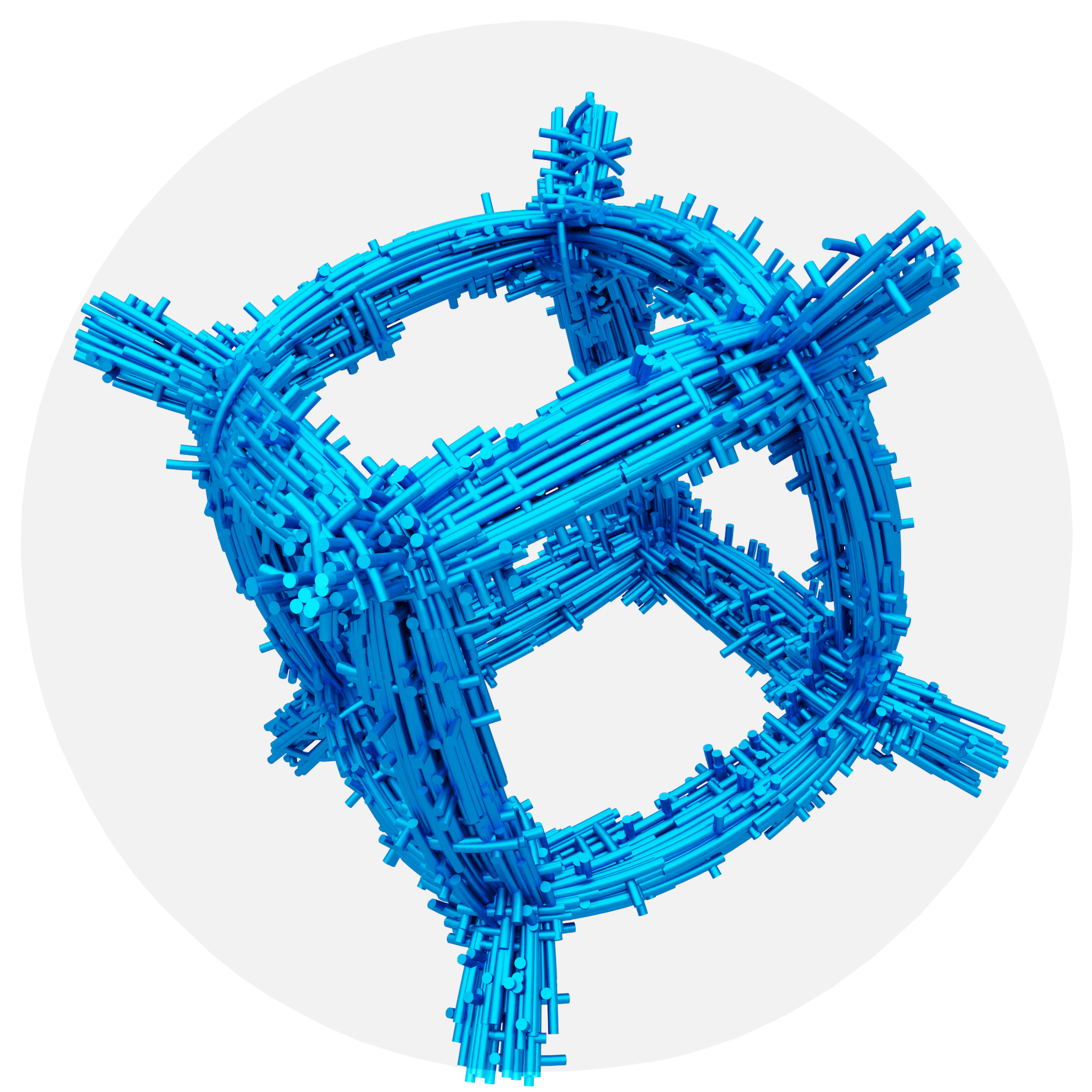}%
\hfill%
\includegraphics[width=\imagewidth]{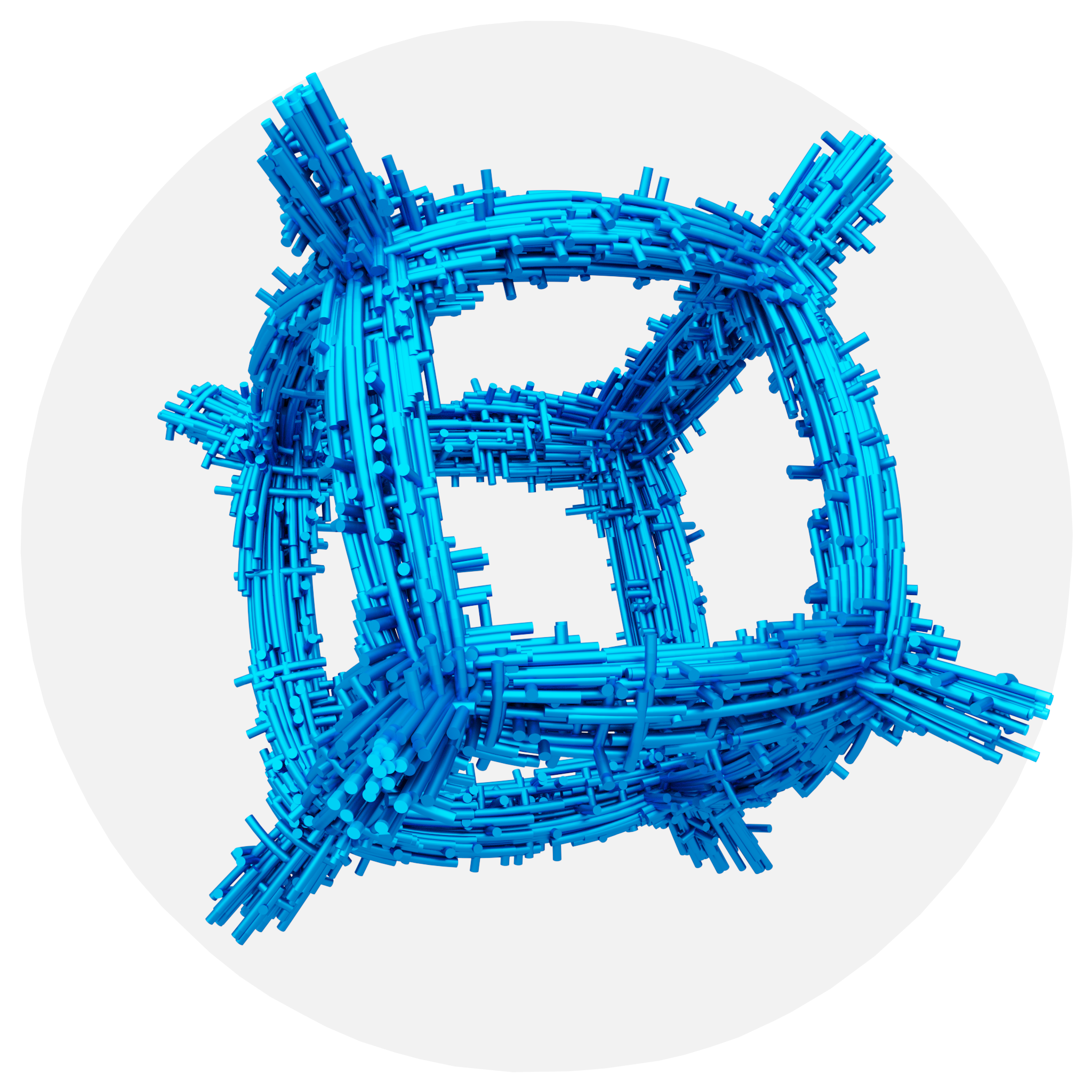} \\
\includegraphics[width=\imagewidth]{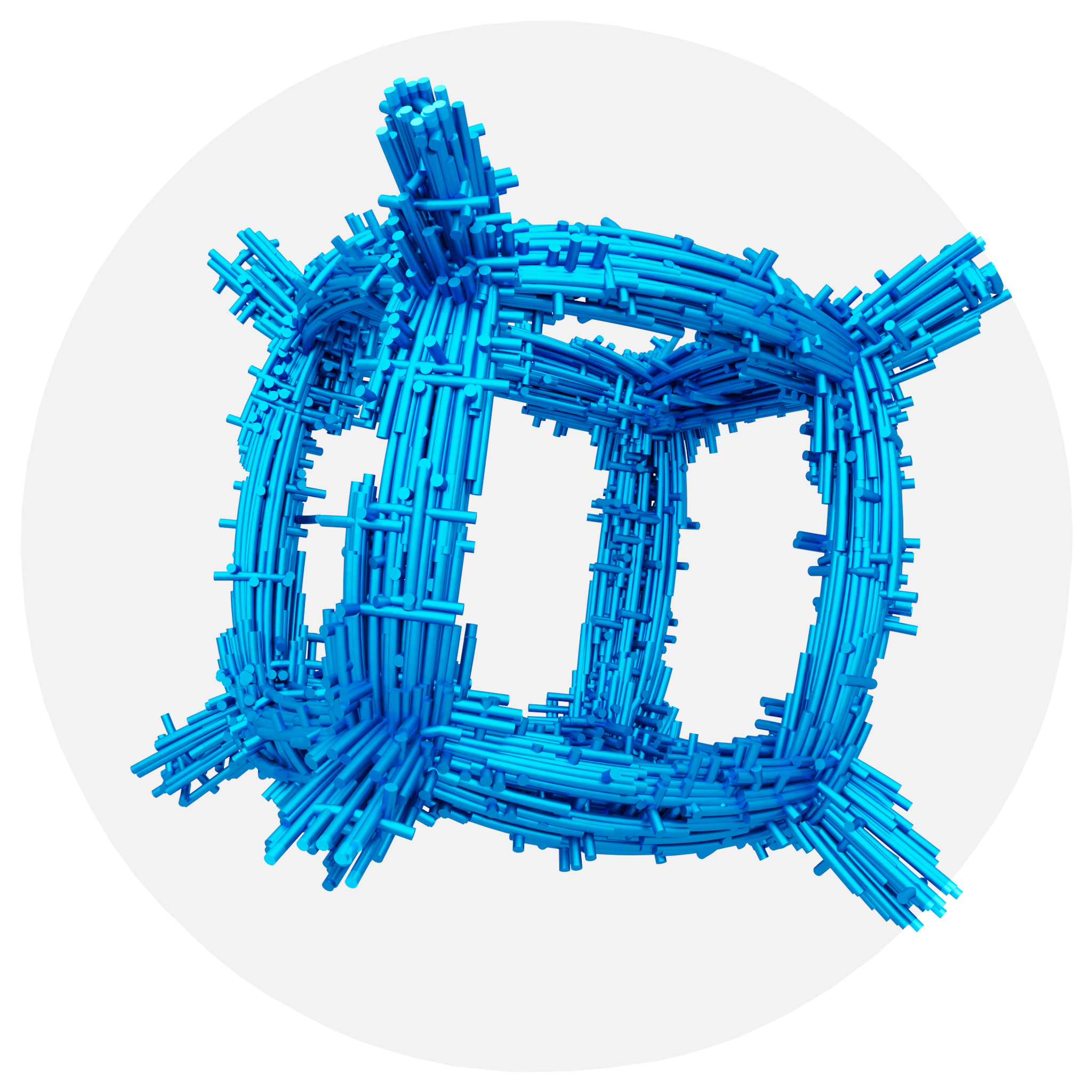}%
\hfill%
\includegraphics[width=\imagewidth]{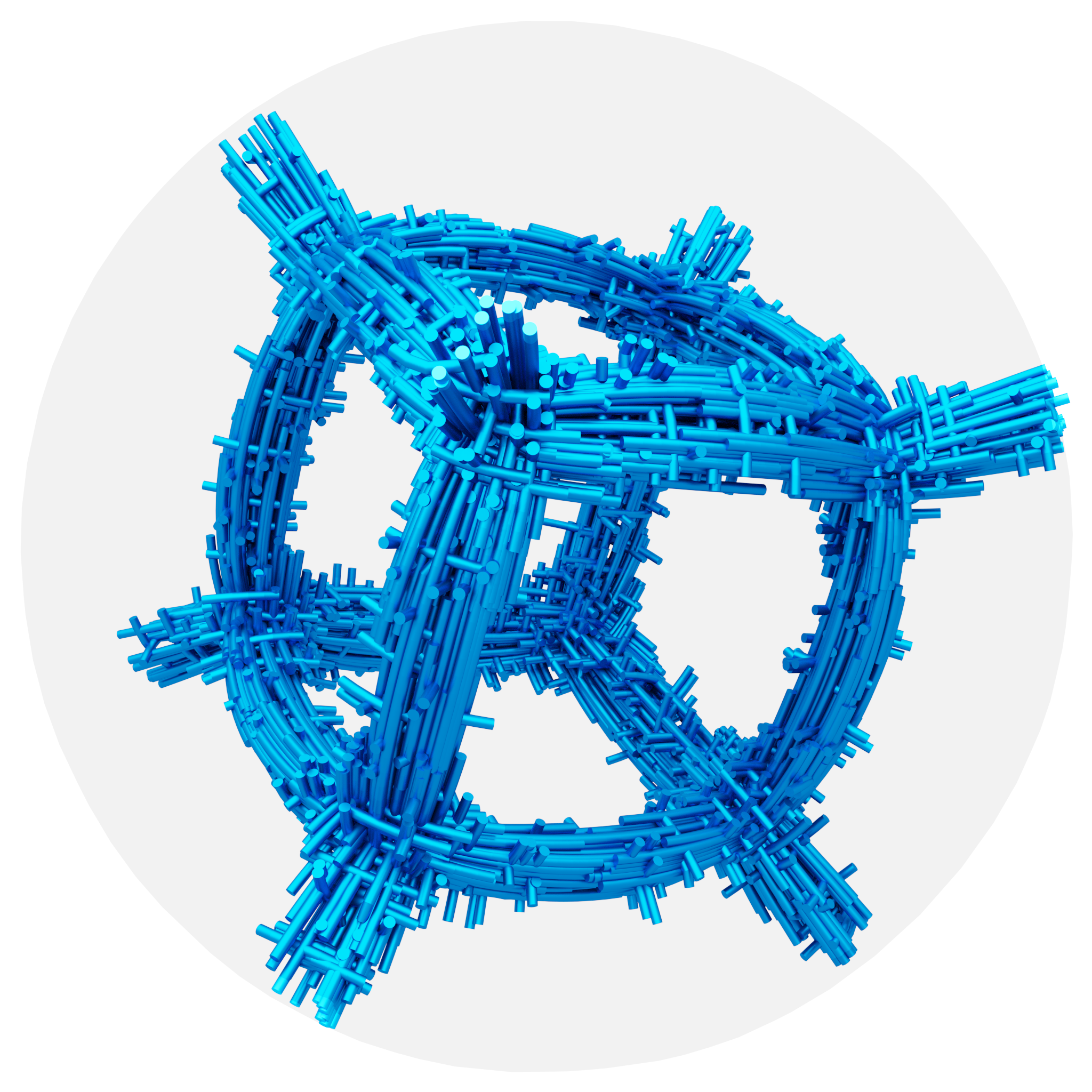}
\caption{Robustness to random initialization.
Odeco MBO transforms vertexwise random initial fields into
qualitatively identical results up to the symmetry of the sphere.
All results were computed
on a sphere with $48950$ vertices.
Integral curves are depicted in regions of rapid rotation,
i.e., in proximity to singularities.}
\label{fig.mbo_init}
\end{figure}
In Figure \ref{fig.mbo_init}, we demonstrate the robustness of MBO to random initialization.
On the sphere, global rotations of a field do not change the objective value.
Initializing odeco MBO with different fields of vertexwise random frames, we find
that it converges to randomly rotated copies of a qualitatively identical, symmetric field.

Figure \ref{fig.energy_divergence} illustrates the behavior of octahedral
and odeco fields as the density of the underlying tetrahedral mesh increases.
Due to the unit-norm constraint,
the gradient of an octahedral field becomes unbounded close to its singularities.
This leads to logarithmic divergence of the total energy as the tet
mesh becomes finer. In contrast, the additional scaling degrees of freedom
available to odeco fields allow renormalization of singularities, as shown
in Figure \ref{fig.sing_energy}. Thus, odeco field energy plateaus
as mesh density increases. Note also the much smaller variance in energy between
runs for odeco fields, quantitatively illustrating robustness to random initialization.
\begin{figure}[hb]
\centering
\begin{tikzpicture}
\begin{semilogxaxis}[
    x dir = reverse,
    minor y tick num = 4,
    xlabel={\footnotesize Tet Volume (Geometric Mean)},
    ylabel={\footnotesize Energy},
    xlabel near ticks,
    ylabel near ticks,
    every tick label/.append style = {font=\tiny},
    legend pos = north west,
    legend cell align = left,
    legend style = {font=\tiny, row sep=0.1pt},
    cycle list = {red, blue},
    mark size = 0.5pt]
\addplot+[only marks] table[x index = 0, y index = 1, col sep = comma] {figures/sphere_energy_octa.csv};
\addlegendentry{Octahedral}
\addplot+[only marks] table[x index = 0, y index = 1, col sep = comma] {figures/sphere_energy_odeco.csv};
\addlegendentry{Odeco}
\end{semilogxaxis}
\end{tikzpicture}
\caption{Energy density diverges for octahedral fields,
but plateaus for odeco fields, as mesh density increases. Also notice the higher
variance for octahedral fields.
Ten fields of each type were computed by RTR on each of thirteen
tetrahedral meshes of the unit sphere of various densities.}
\label{fig.energy_divergence}
\end{figure}
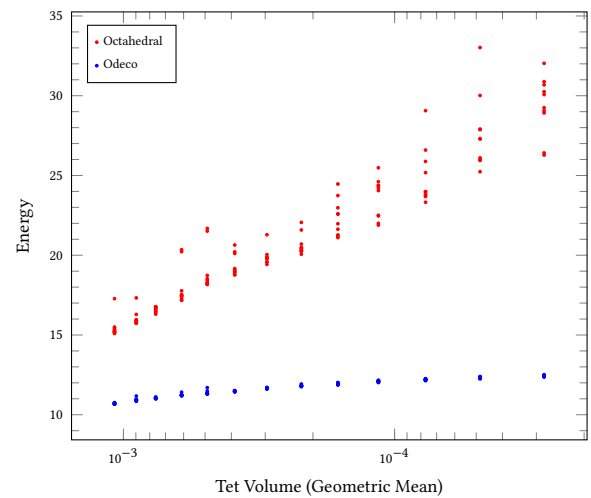%
\begin{figure}
\newcommand{\imagewidth}{0.48\columnwidth}
\centering
\includegraphics[width=\imagewidth]{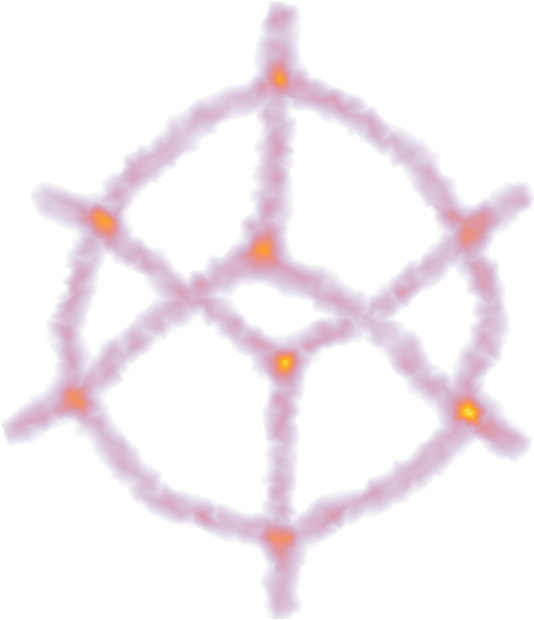}%
\hfill%
\includegraphics[width=\imagewidth]{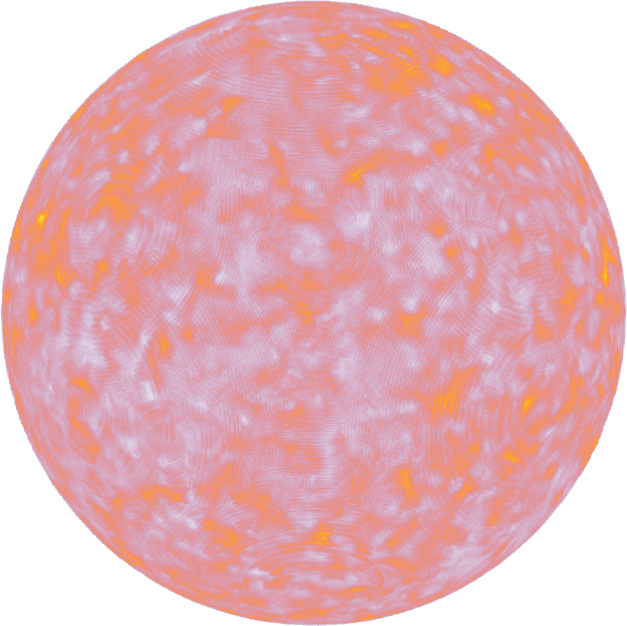}%
\caption{The energy density at singularities of an octahedral field
dominates the total energy (left). In contrast, an odeco field's energy is distributed more
uniformly (right). Results computed using MBO + RTR.}
\label{fig.sing_energy}
\end{figure}%

Table 1 (supplemental document) compares timings and energy values
for our methods and the method of \citet{RaySokolov2},
comprising $18$ types of experiments on $15$ different models, for $270$ total runs.
Direct comparison of energy values with \citet{RaySokolov2} is not possible, as their
method uses the graph Laplacian and initializes by solving a
linear system, whereas our method uses the geometric (finite element) Laplacian
and random initialization. To attempt a fair comparison, we report results of their
method alone, their method followed by RTR with the geometric Laplacian,
and their method substituted with the geometric Laplacian and random initialization.
All experiments in the table were conducted on an Ubuntu workstation with a four-core,
3.6 GHz Intel Core i7-7700 and 32 GB of RAM.
A MATLAB implementation of our algorithms accompanies this paper and also includes
our implementation of \cite{RaySokolov2}.

The results in Table 1 show that RTR converges very quickly but,
like previous work, easily gets stuck in local minima.
In contrast, mMBO takes longer to converge but produces higher-quality
fields that reflect the symmetries of the volume.
Running RTR to polish the results of mMBO produces the best energies.

Despite sometimes getting stuck in local minima (see Figure \ref{fig.proj_comparison}),
we observe
that \citeauthor{RaySokolov2}'s approximation of projection can be useful in practice.
Table 1 (supplemental document) includes experiments with MBO and mMBO
(see \S\ref{subsec.mbo}) substituting \citeauthor{RaySokolov2}'s projection
for the true projection. In most cases, the resulting energy is very similar,
suggesting that the diffusion iterations in MBO smooth out any errors resulting from
incorrect projection. This hybrid MBO can be a useful tradeoff between correctness
and speed.

\begin{figure}
\newcommand{\imagewidth}{0.28\columnwidth}
\centering
\begin{tabular}{lccc}
{\rotatebox[origin=c]{90}{\footnotesize{\cite{RaySokolov2}}}} &
\includegraphics[align=c,width=\imagewidth]{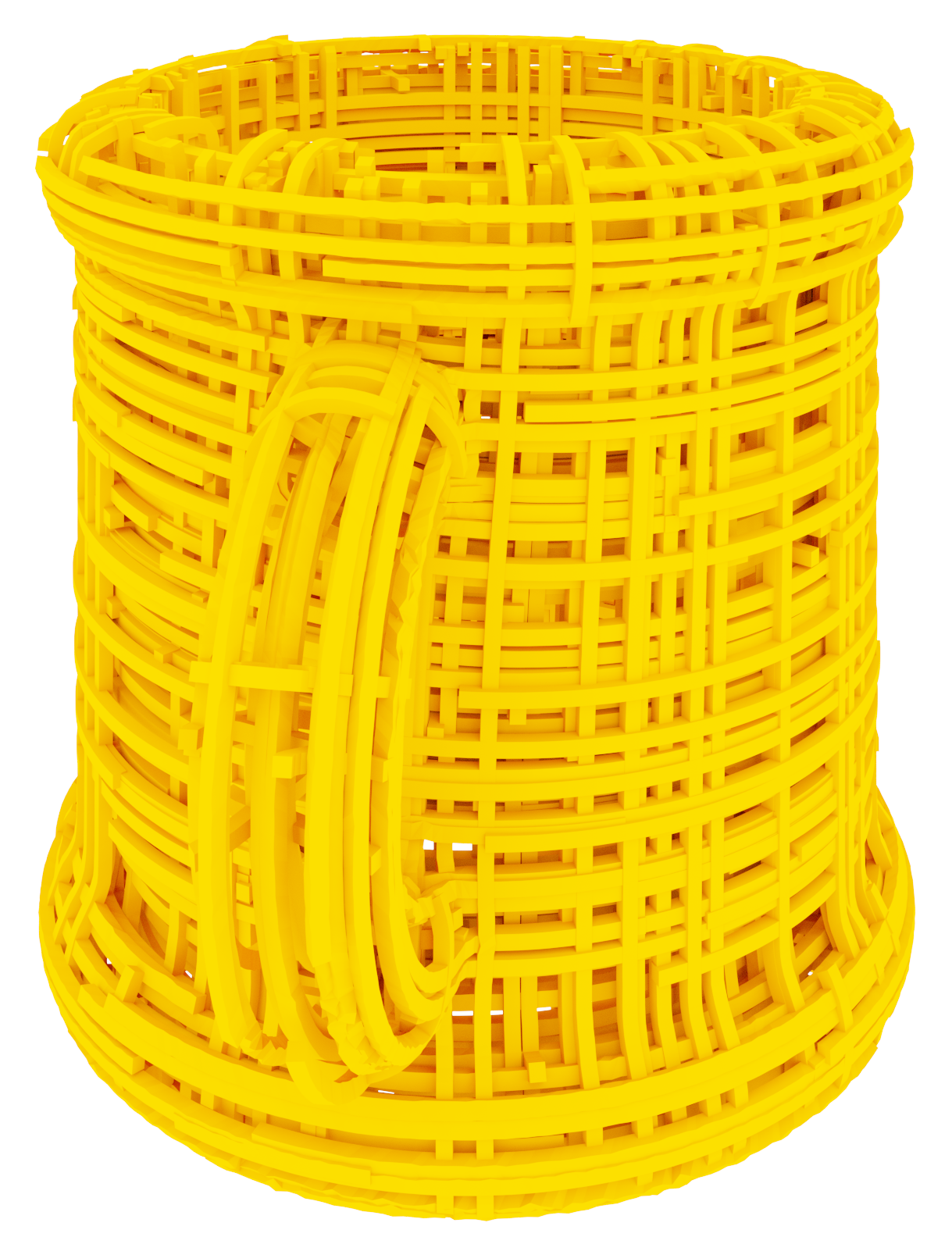} &
\includegraphics[align=c,width=\imagewidth]{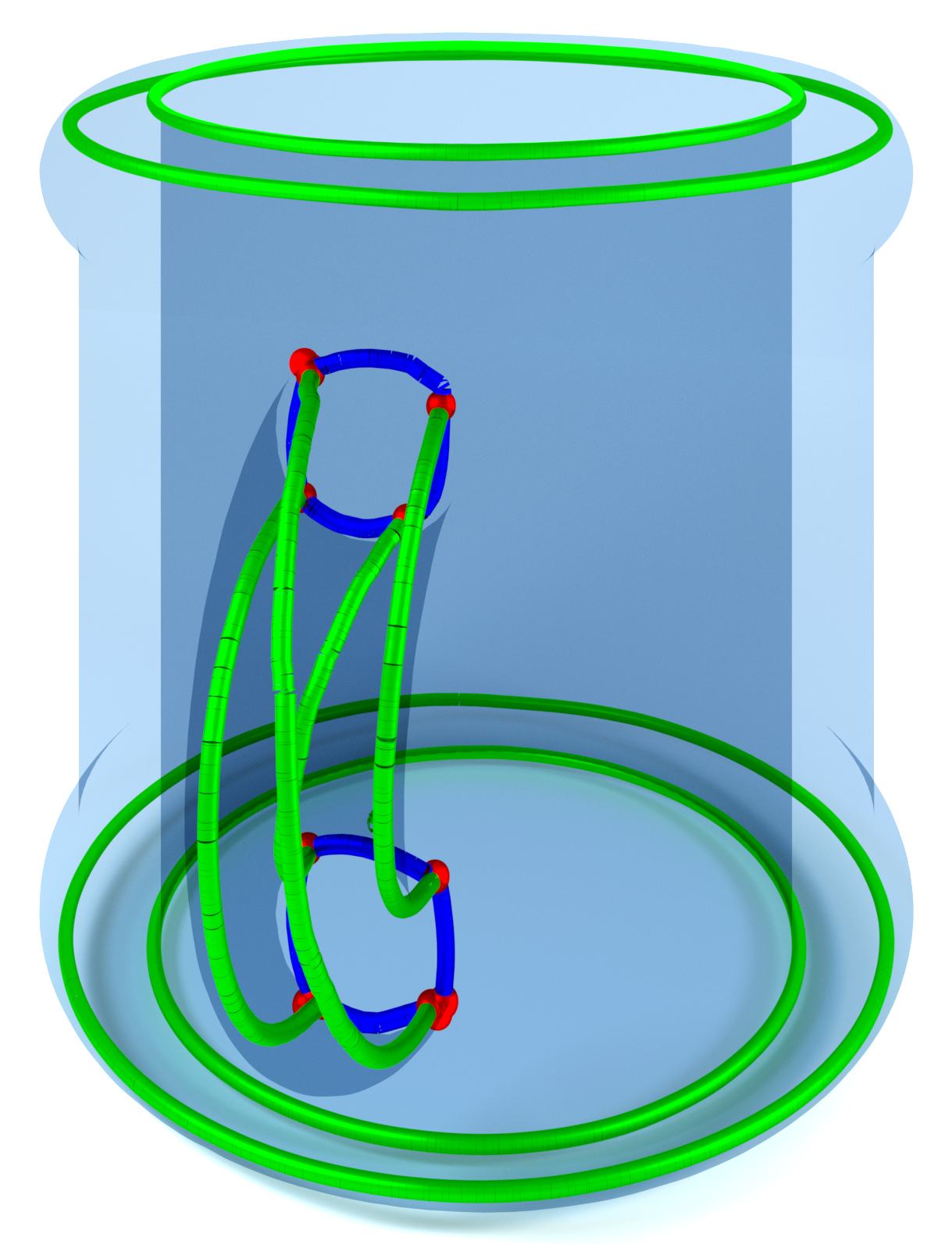} &
\includegraphics[align=c,width=\imagewidth]{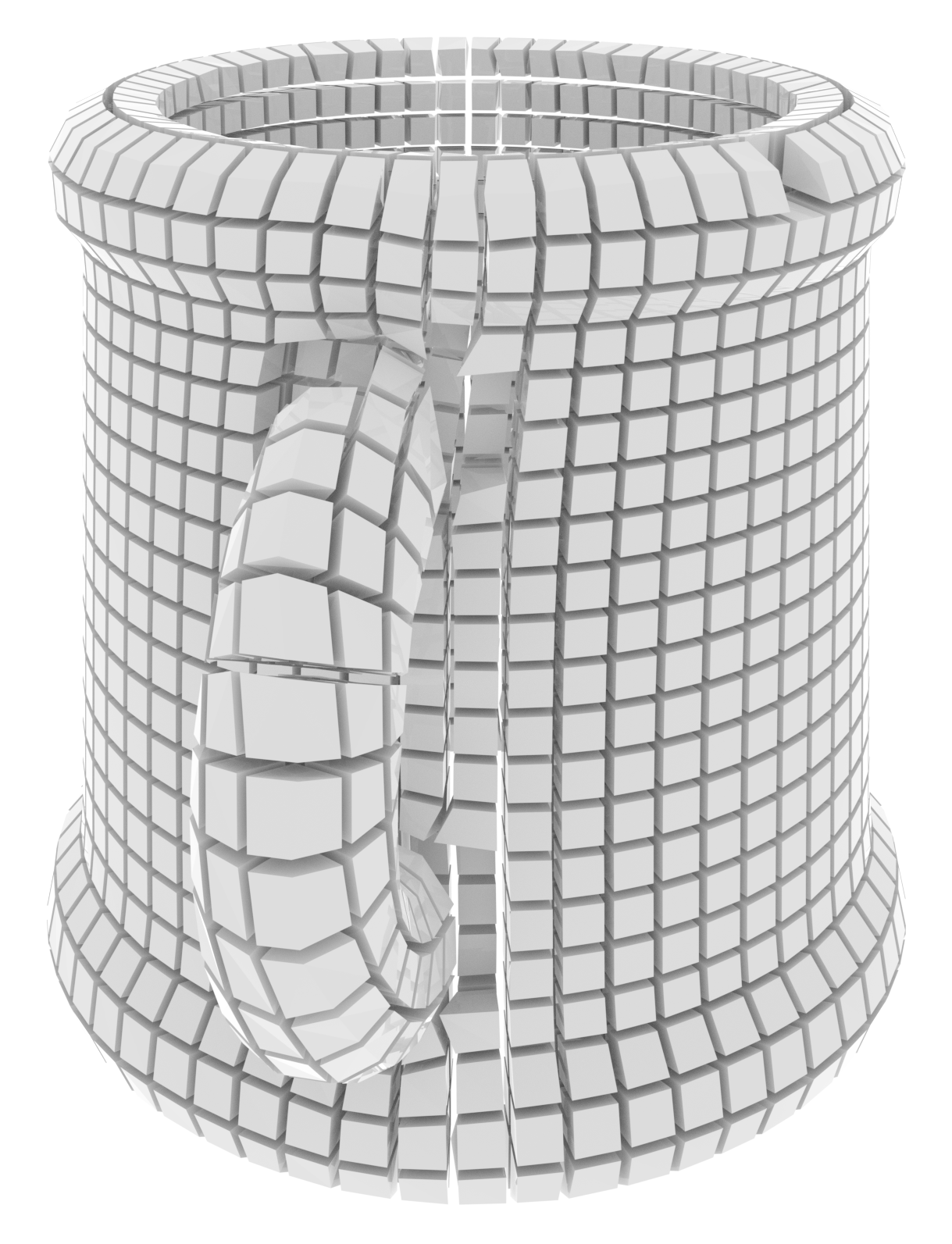} \\
{\rotatebox[origin=c]{90}{\footnotesize{mMBO + RTR}}} &
\includegraphics[align=c,width=\imagewidth]{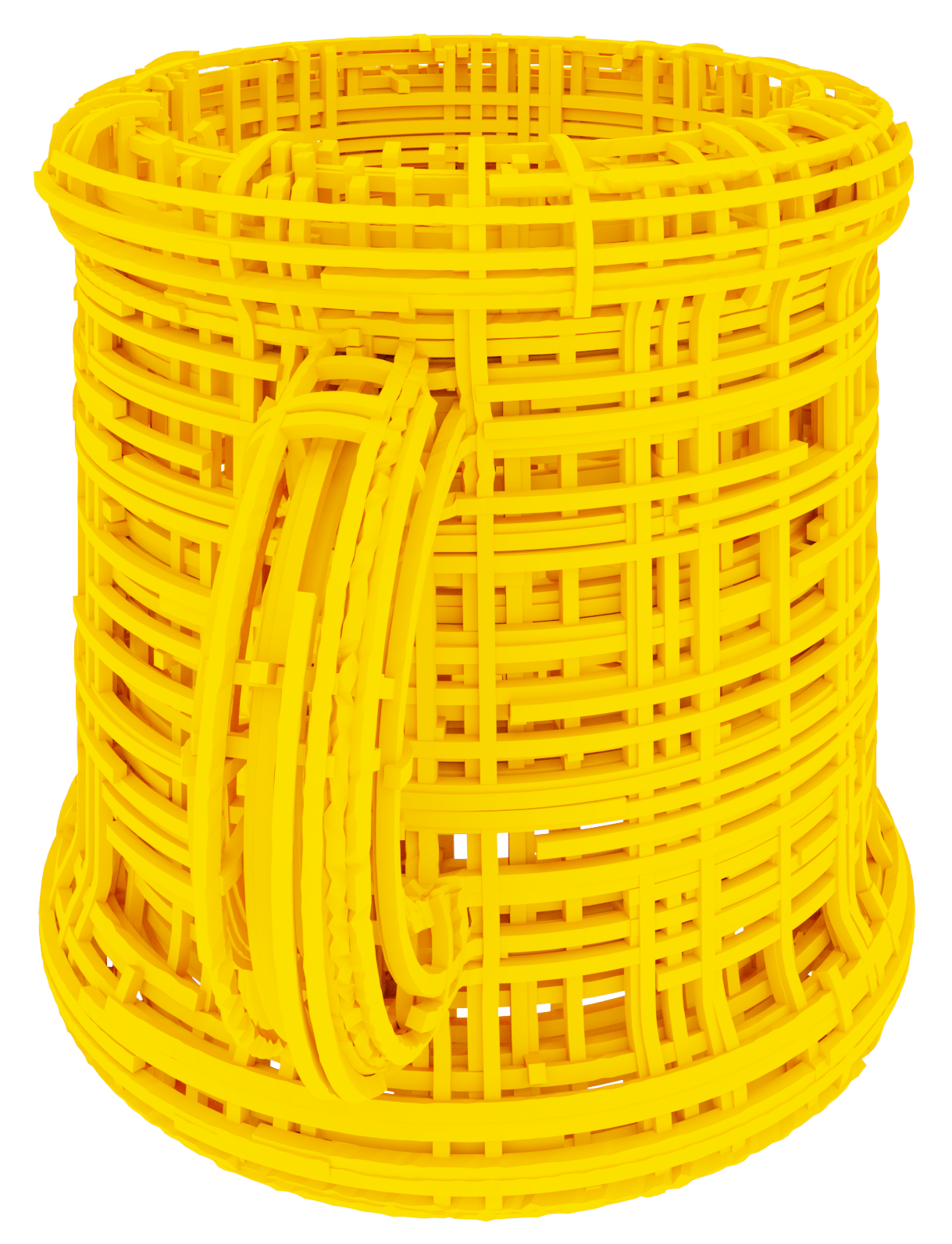} &
 \includegraphics[align=c,width=\imagewidth]{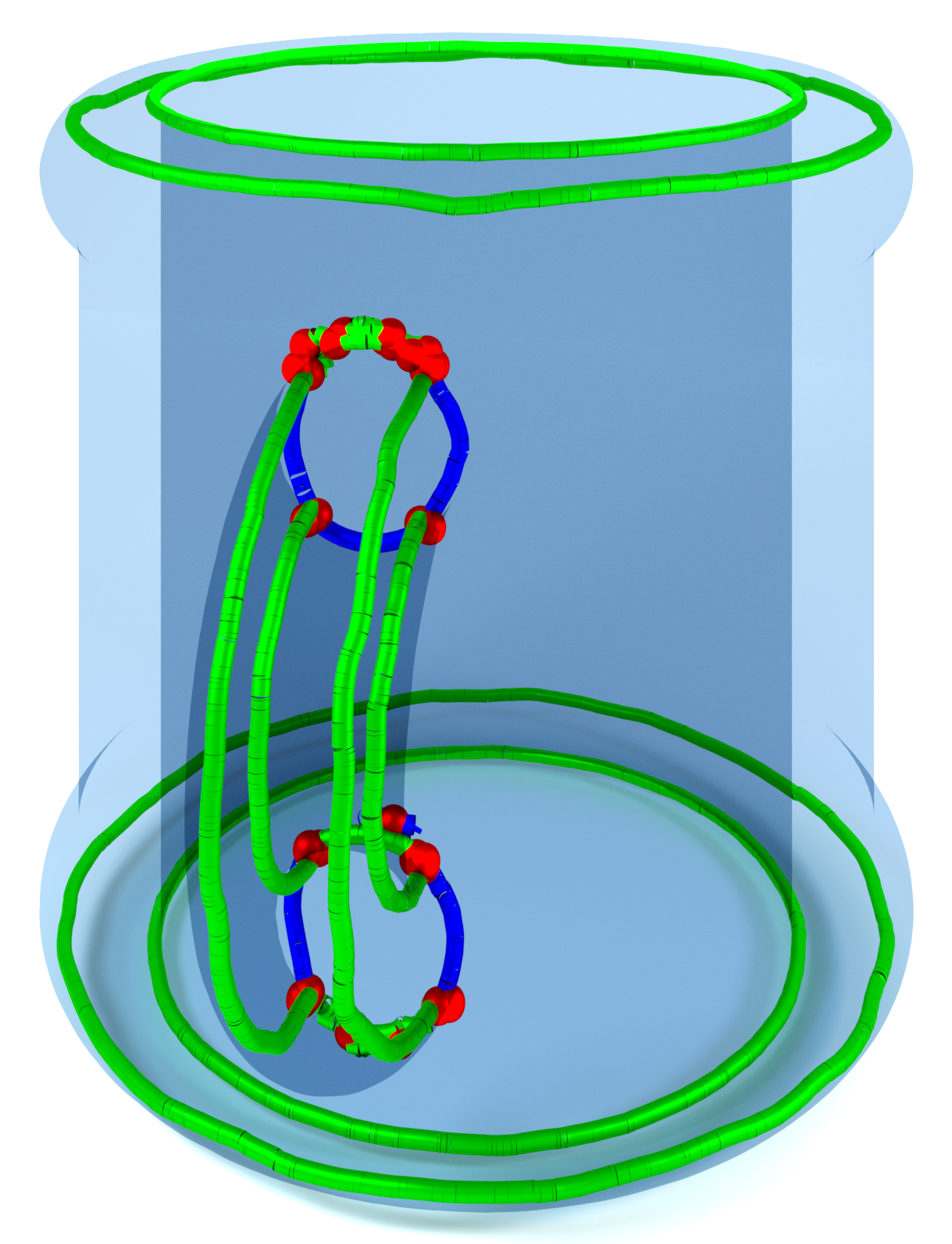} &
\includegraphics[align=c,width=\imagewidth]{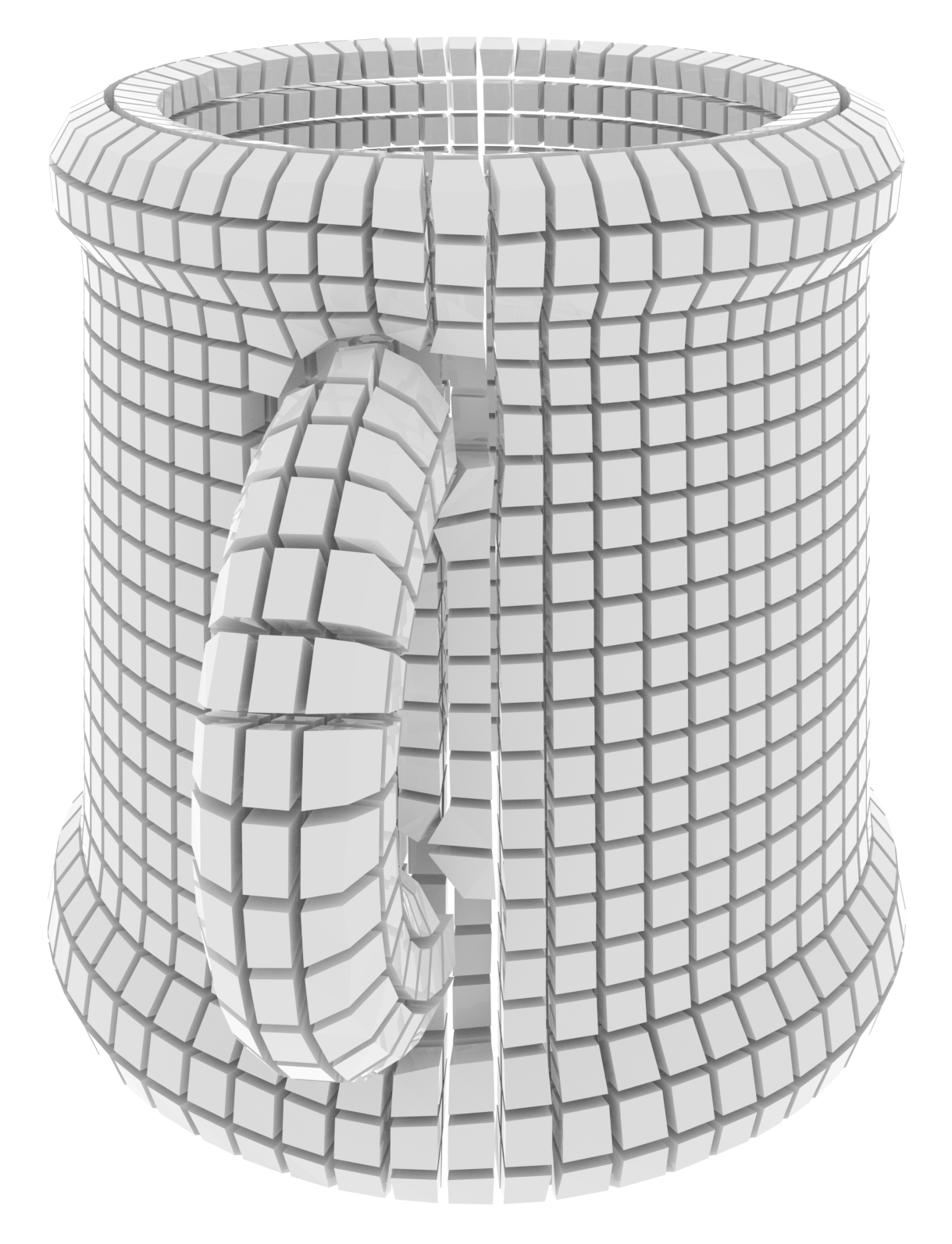}%
\end{tabular}%
\caption{Twisted singular curves on the handle of the cup
in \citeauthor{RaySokolov2}'s result lead to twisting in the resulting mesh (top).
Our field yields a mesh with a more regular structure throughout (bottom).}%
\label{fig.field_comparison.cup}
\end{figure}

\begin{figure}
\newcommand{\imagewidth}{0.48\columnwidth}
\centering
\begin{tabular}{cc}
\includegraphics[align=c,width=\imagewidth]{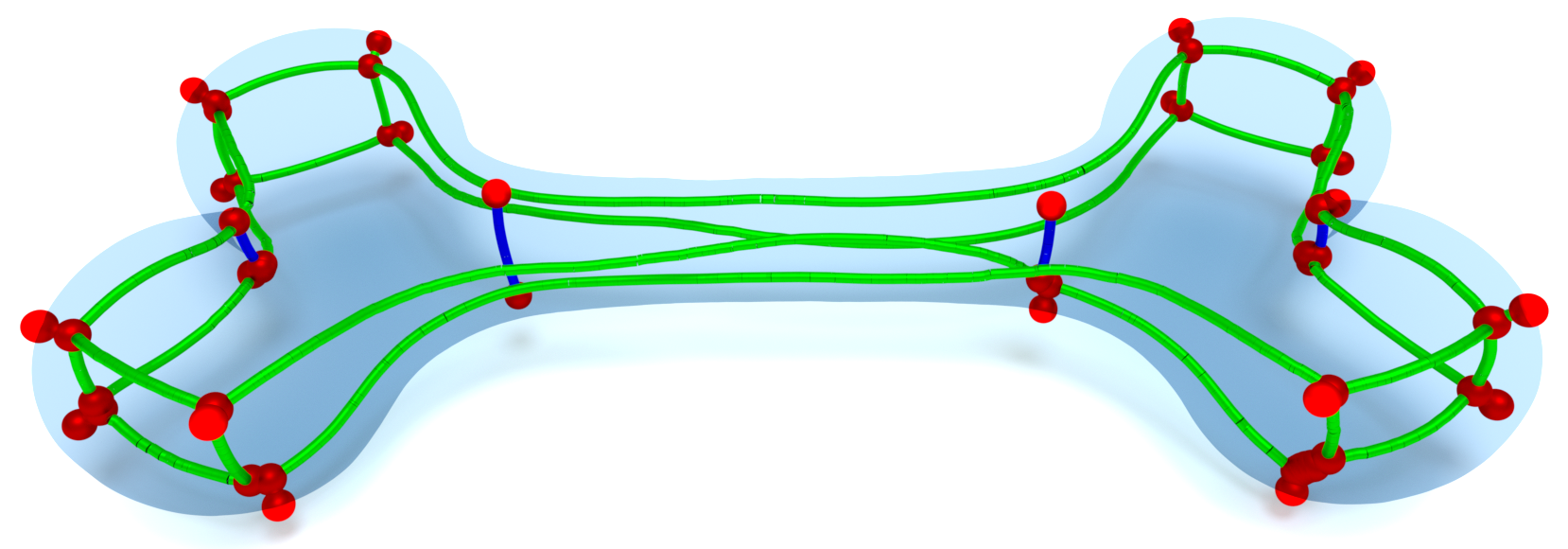} &
\includegraphics[align=c,width=\imagewidth]{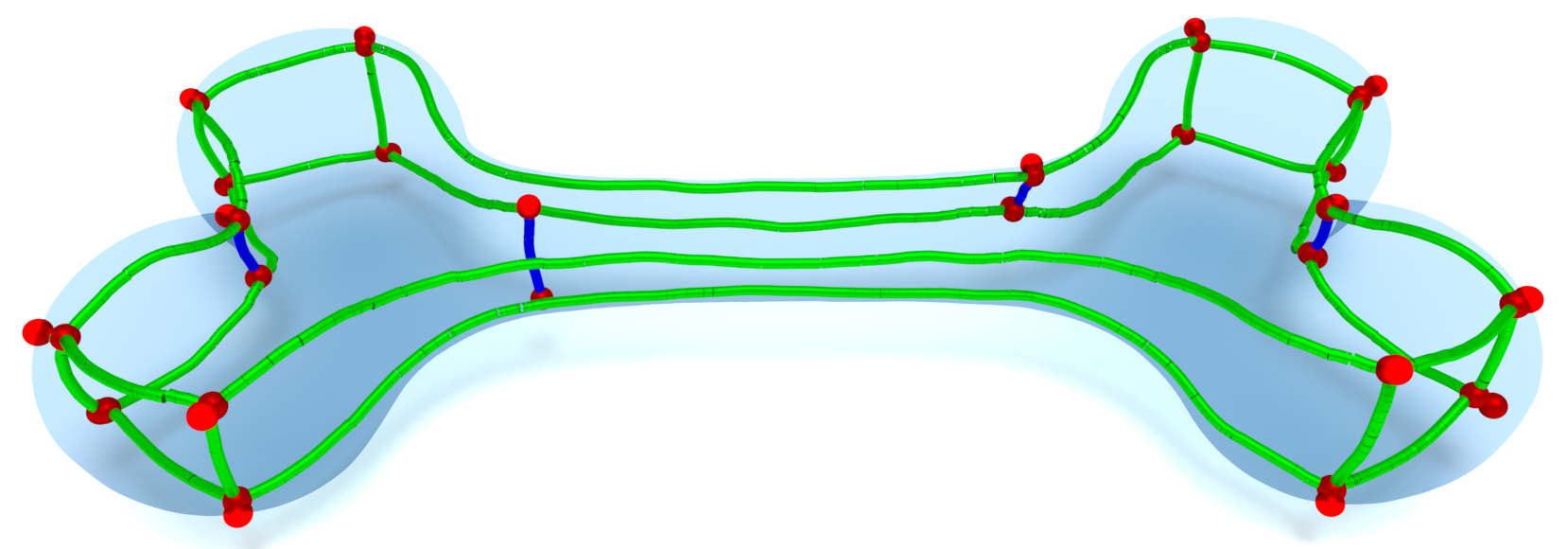} \\
\includegraphics[align=c,width=\imagewidth]{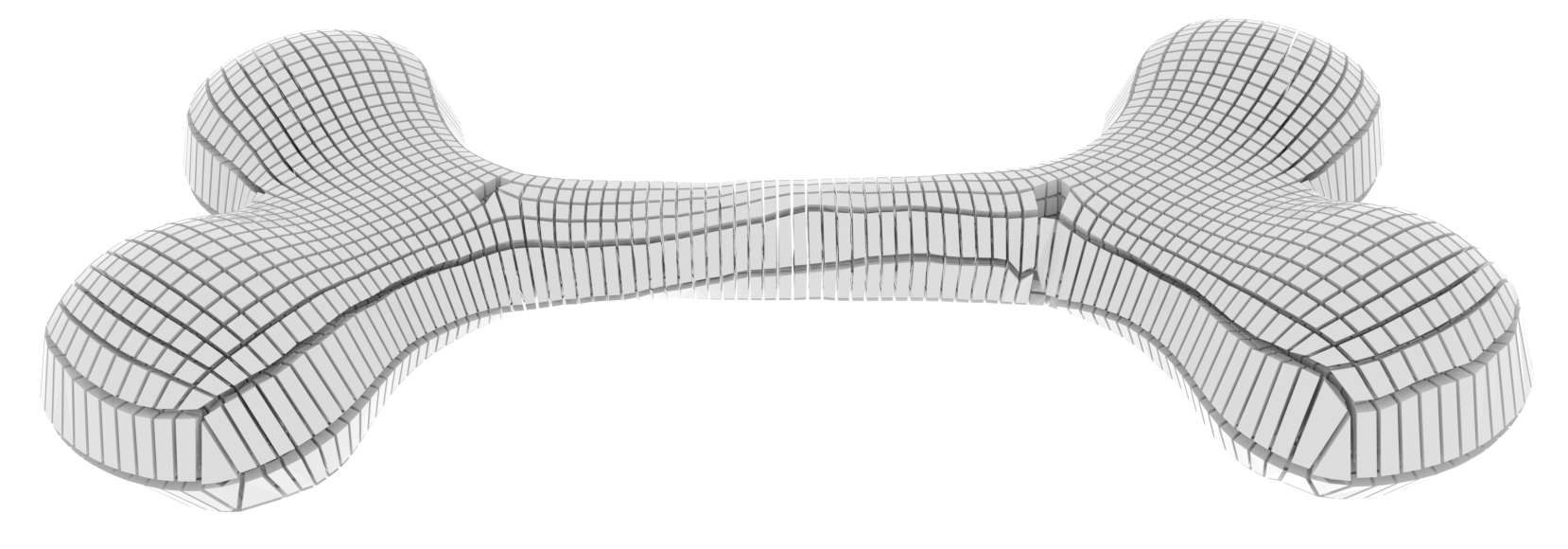} &
\includegraphics[align=c,width=\imagewidth]{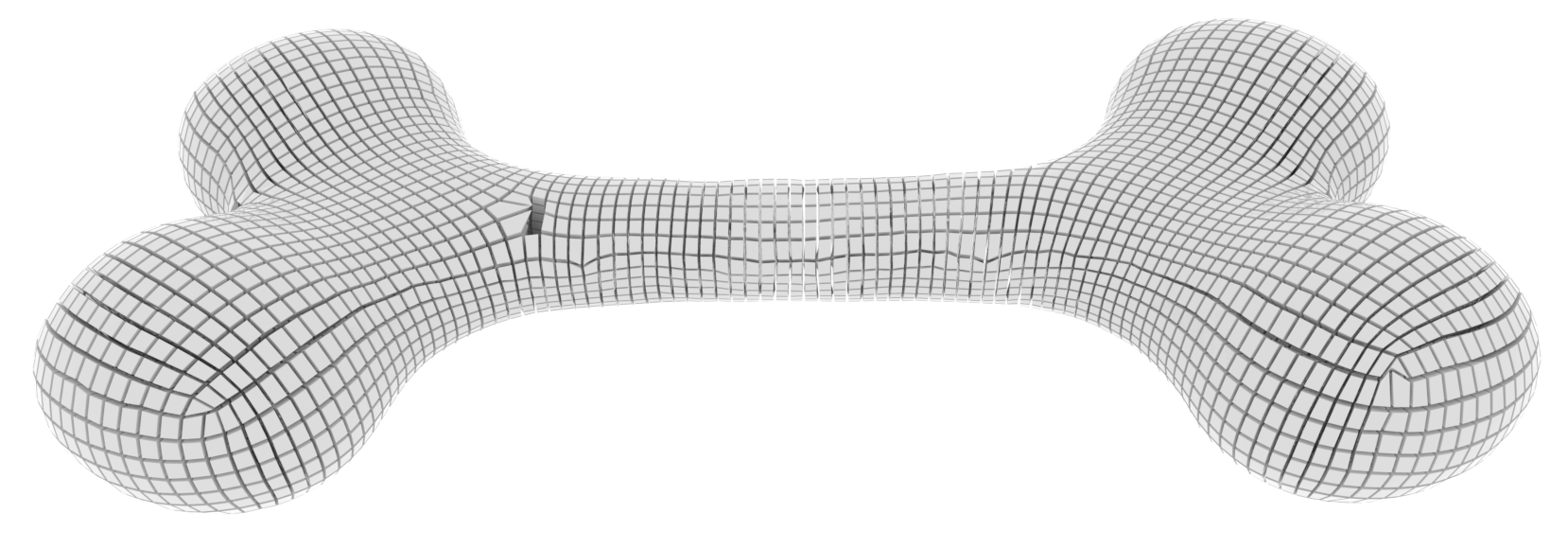} \\
\cite{RaySokolov2} & mMBO + RTR
\end{tabular}%
\caption{On the bone model, twisted singular curves in a field computed by
the method of \citet{RaySokolov2} lead to a degenerate integer grid map, producing
highly-distorted hexahedra (left). Our field yields a mesh without this degeneracy (right).}%
\label{fig.field_comparison.bone}
\end{figure}

\begin{figure}
\newcommand{\imagewidth}{0.28\columnwidth}
\centering
\begin{tabular}{lccc}
{\rotatebox[origin=c]{90}{\footnotesize{\cite{RaySokolov2}}}} &
\includegraphics[align=c,width=\imagewidth]{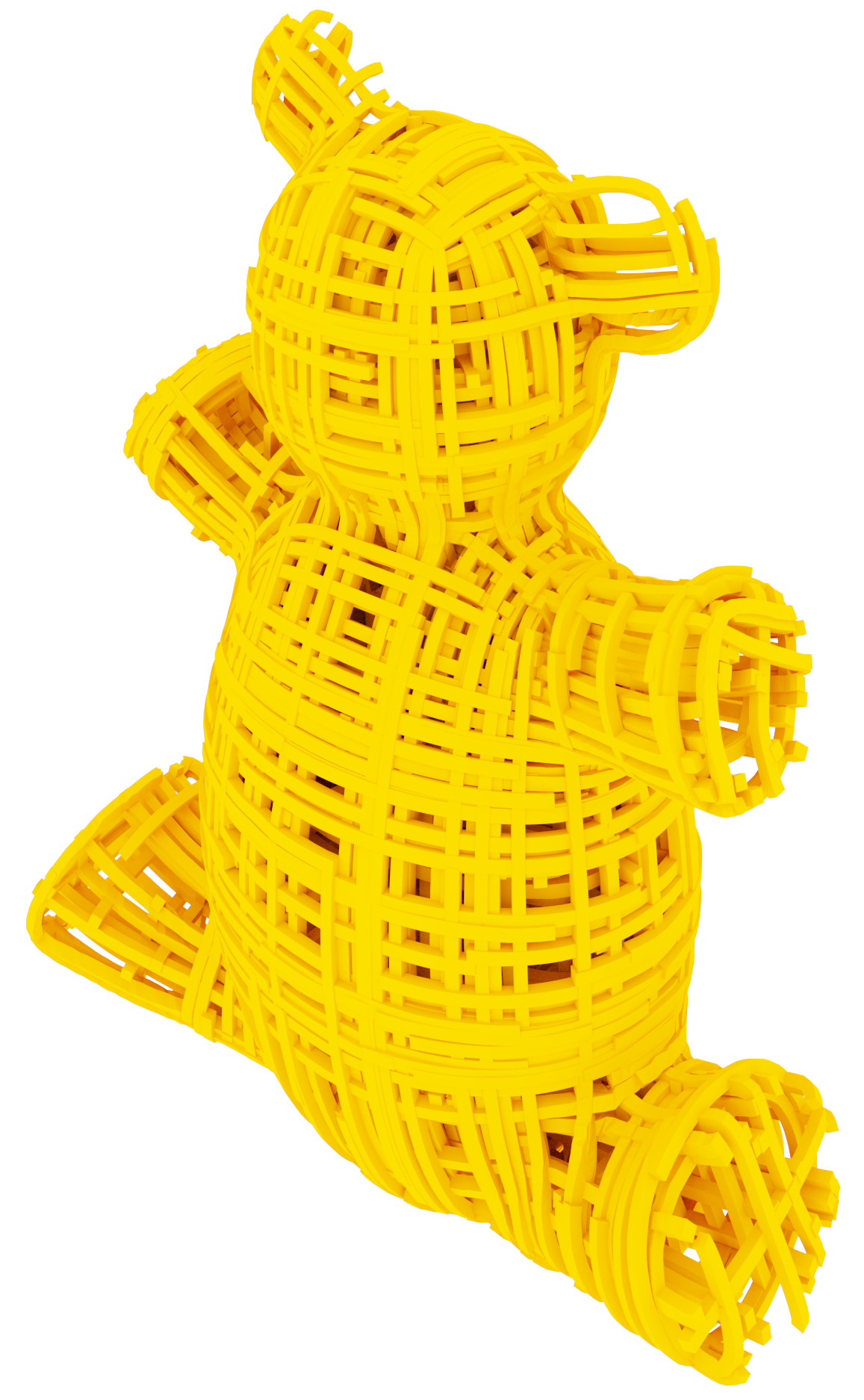} &
\includegraphics[align=c,width=\imagewidth]{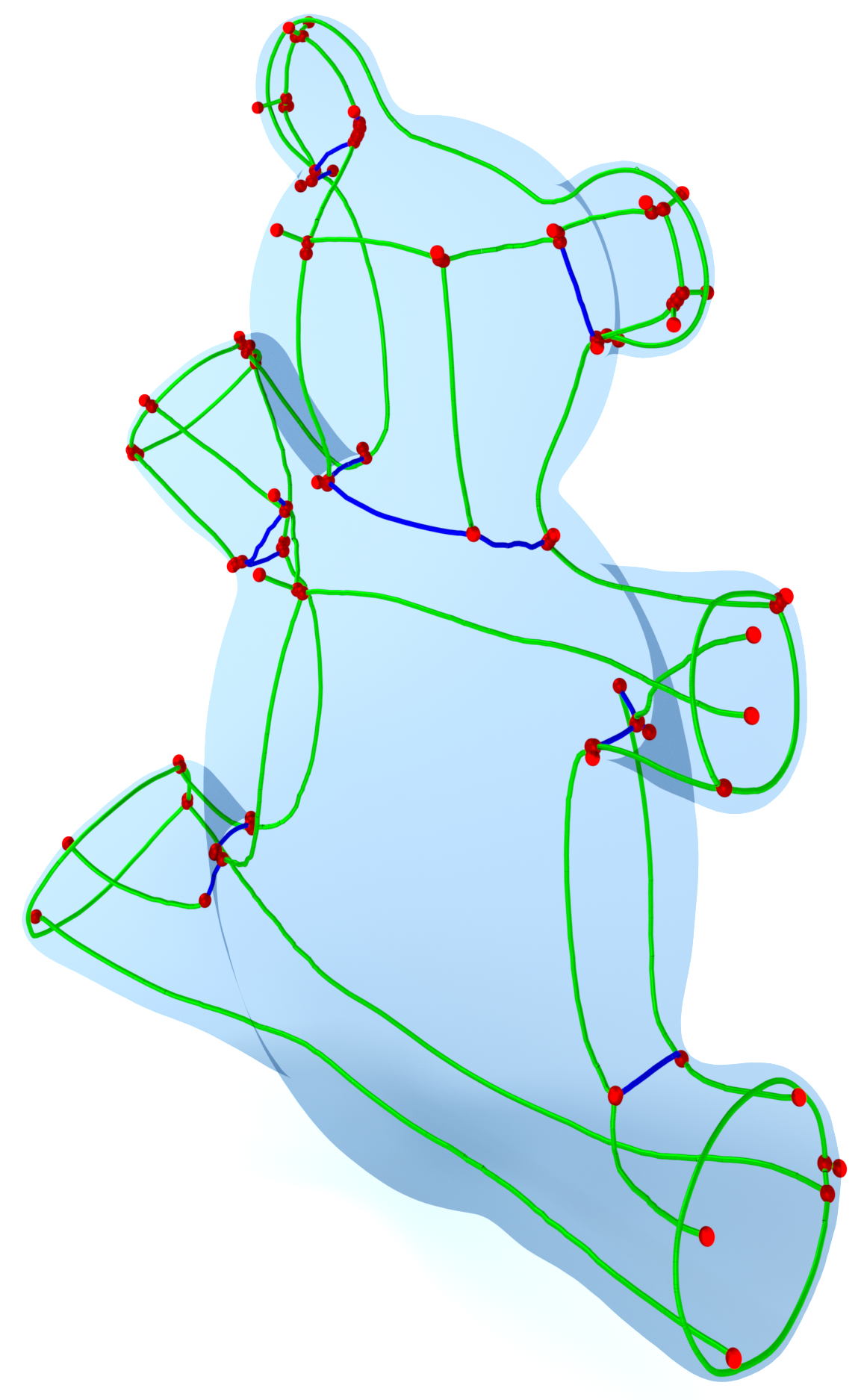} &
\includegraphics[align=c,width=\imagewidth]{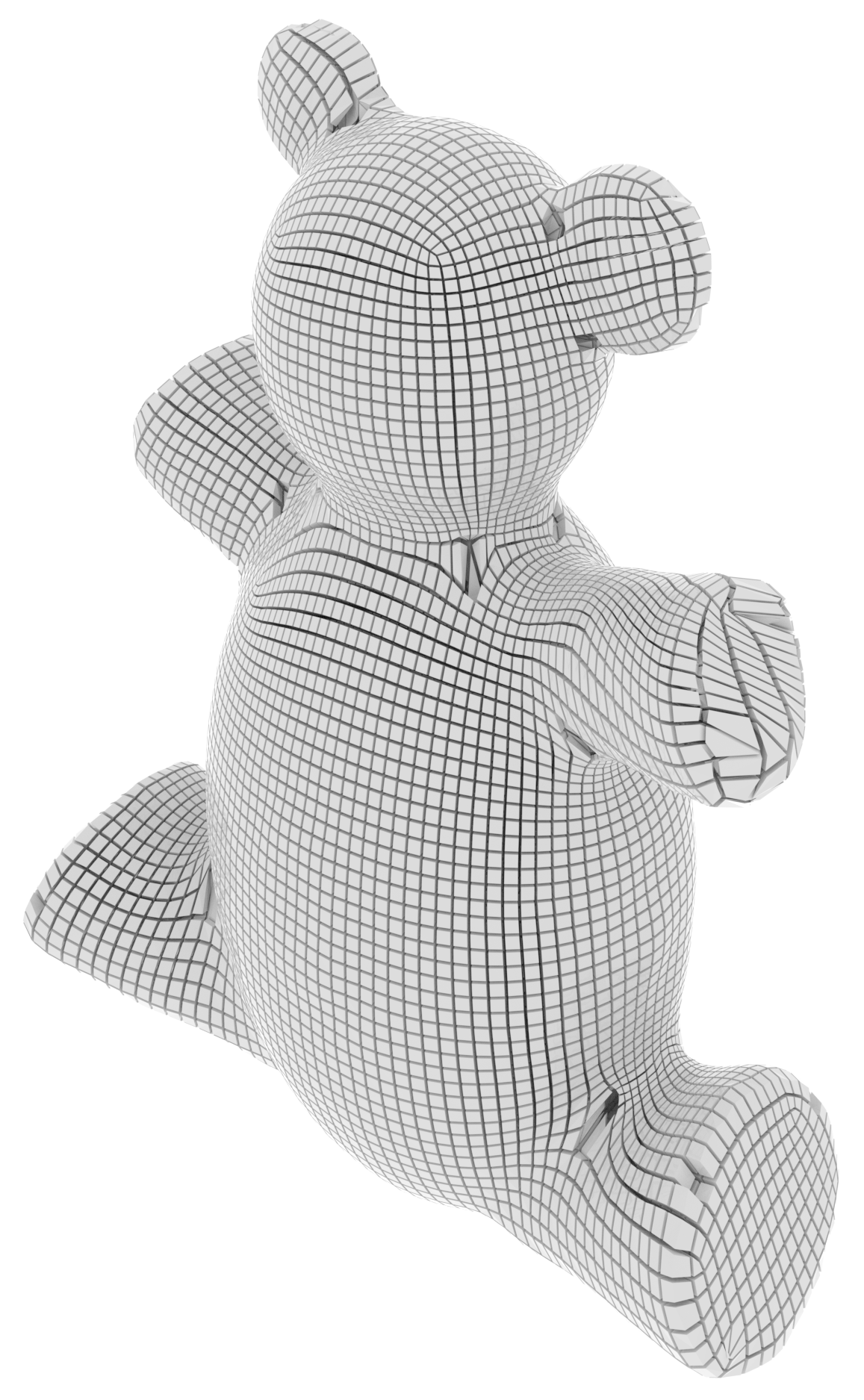} \\
{\rotatebox[origin=c]{90}{\footnotesize{mMBO + RTR}}} &
\includegraphics[align=c,width=\imagewidth]{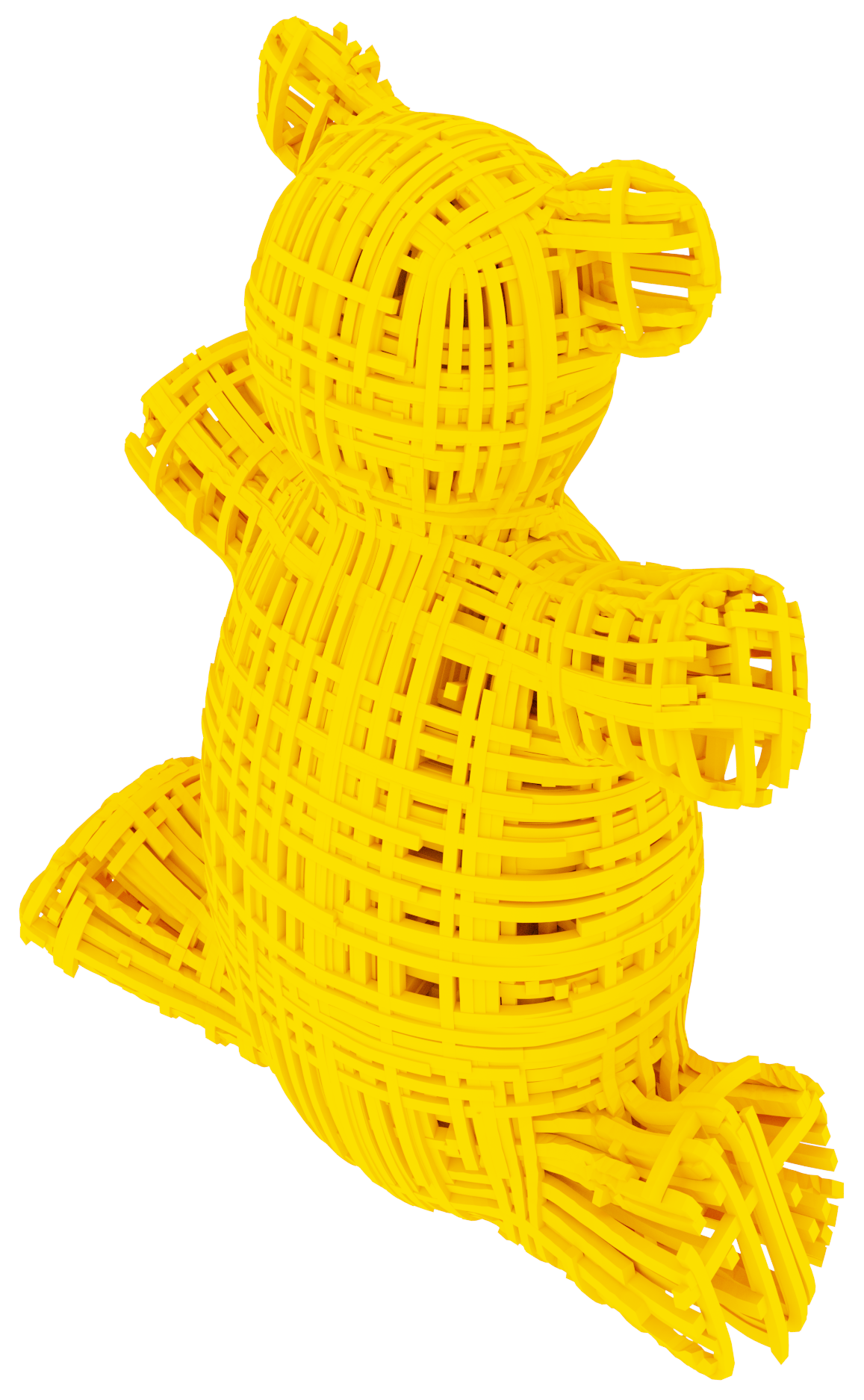} &
\includegraphics[align=c,width=\imagewidth]{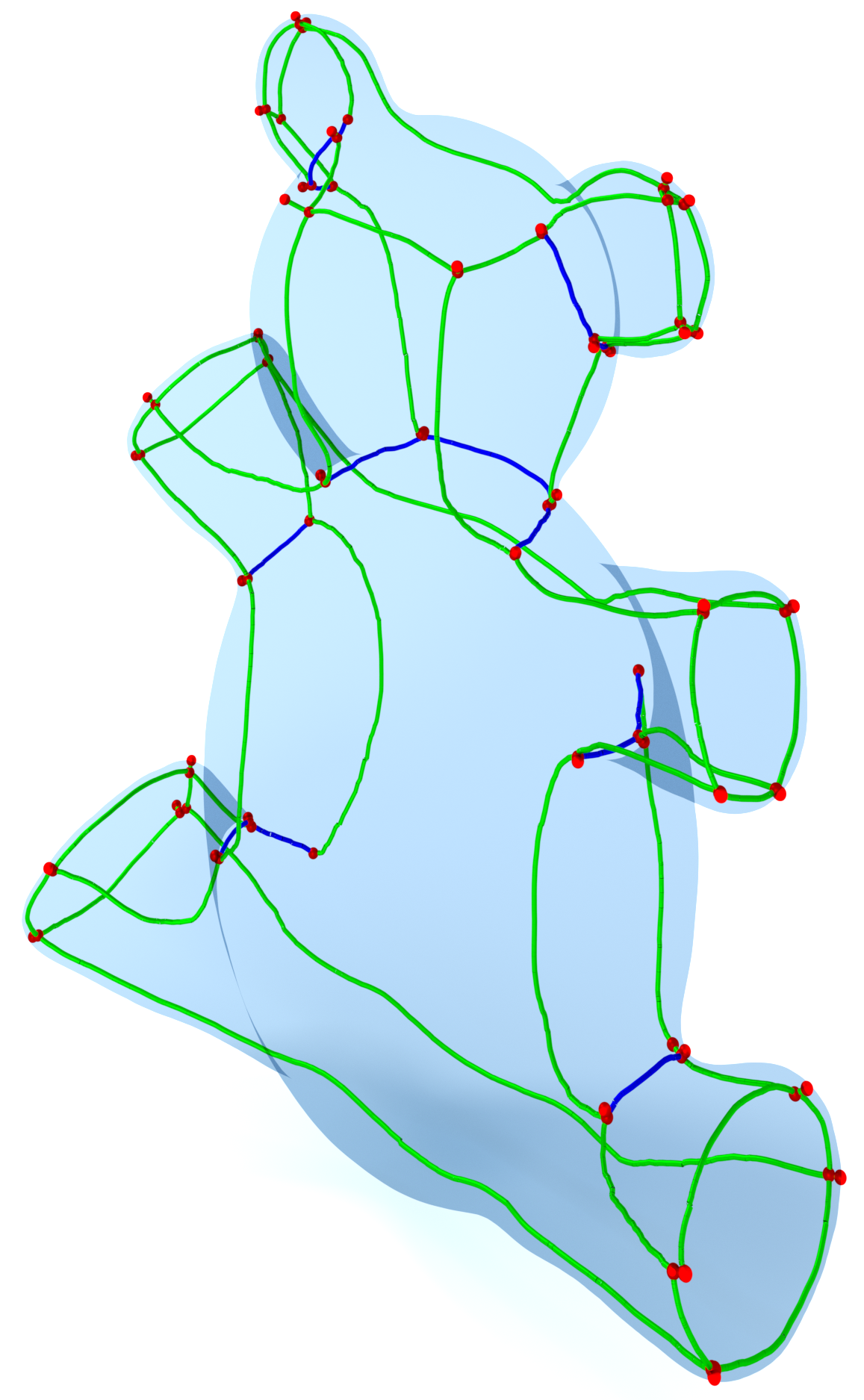} &
\includegraphics[align=c,width=\imagewidth]{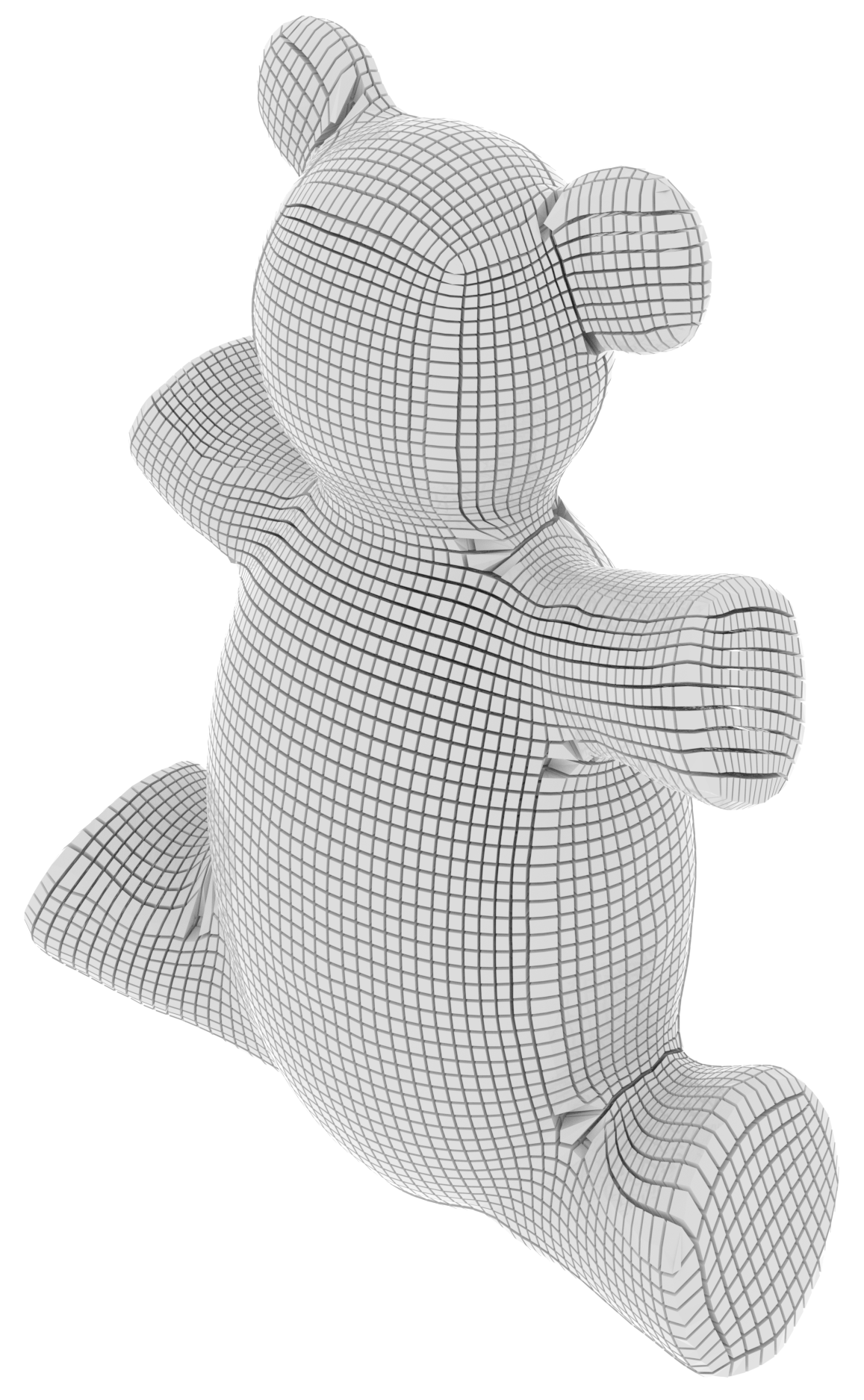}%
\end{tabular}%
\caption{
Due to the twisting of singular curves in the field generated by
\citeauthor{RaySokolov2}'s method (top), the mesh is highly distorted on the right arm.
In contrast, our field (bottom) has a regular cubic singular structure, leading to fewer mesh degeneracies.}%
\label{fig.field_comparison.teddy}
\end{figure}

\begin{figure}
\newcommand{\imagewidth}{0.28\columnwidth}
\centering
\begin{tabular}{lccc}
{\rotatebox[origin=c]{90}{\footnotesize{\cite{RaySokolov2}}}} &
\includegraphics[align=c,width=\imagewidth]{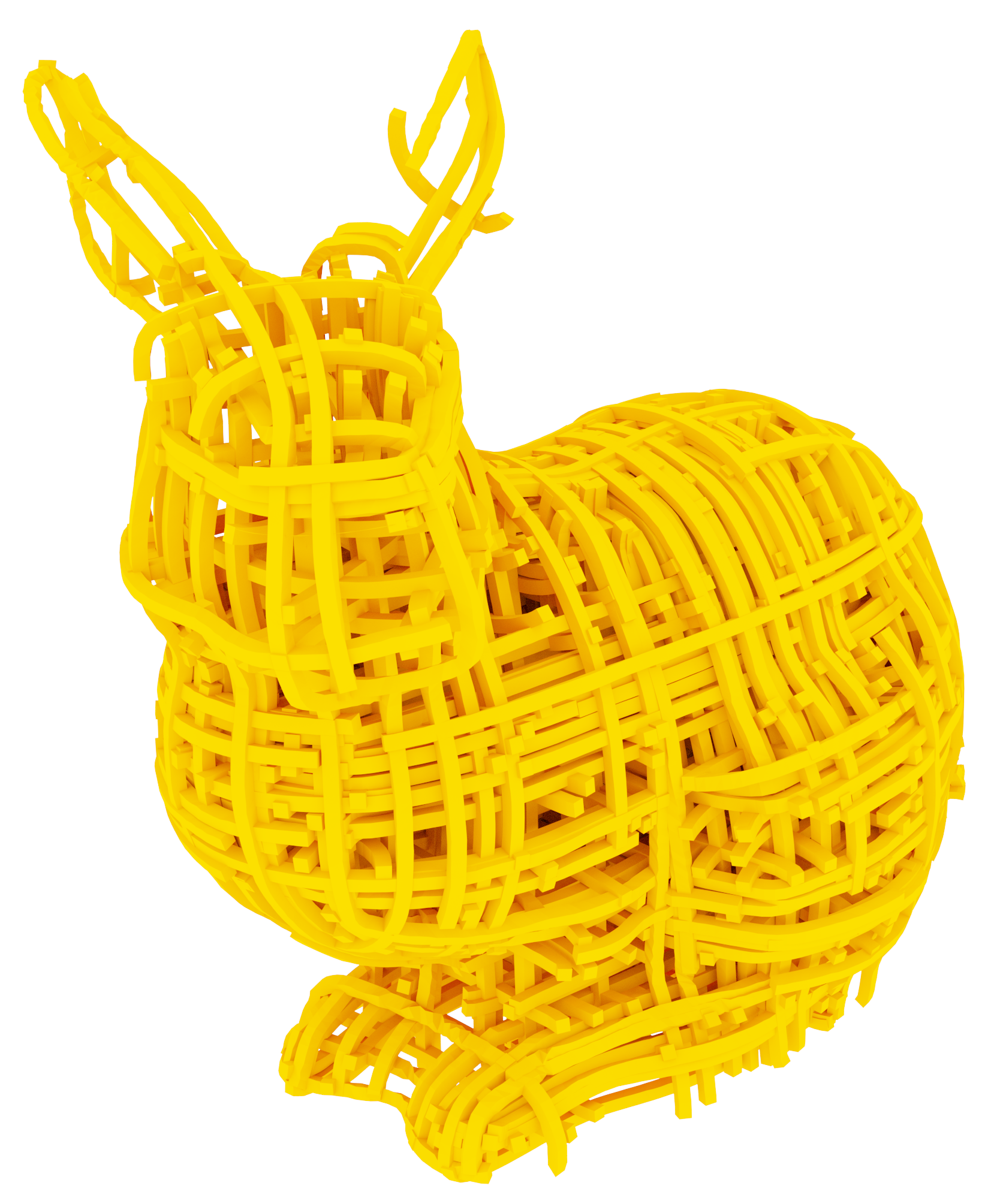} &
\includegraphics[align=c,width=\imagewidth]{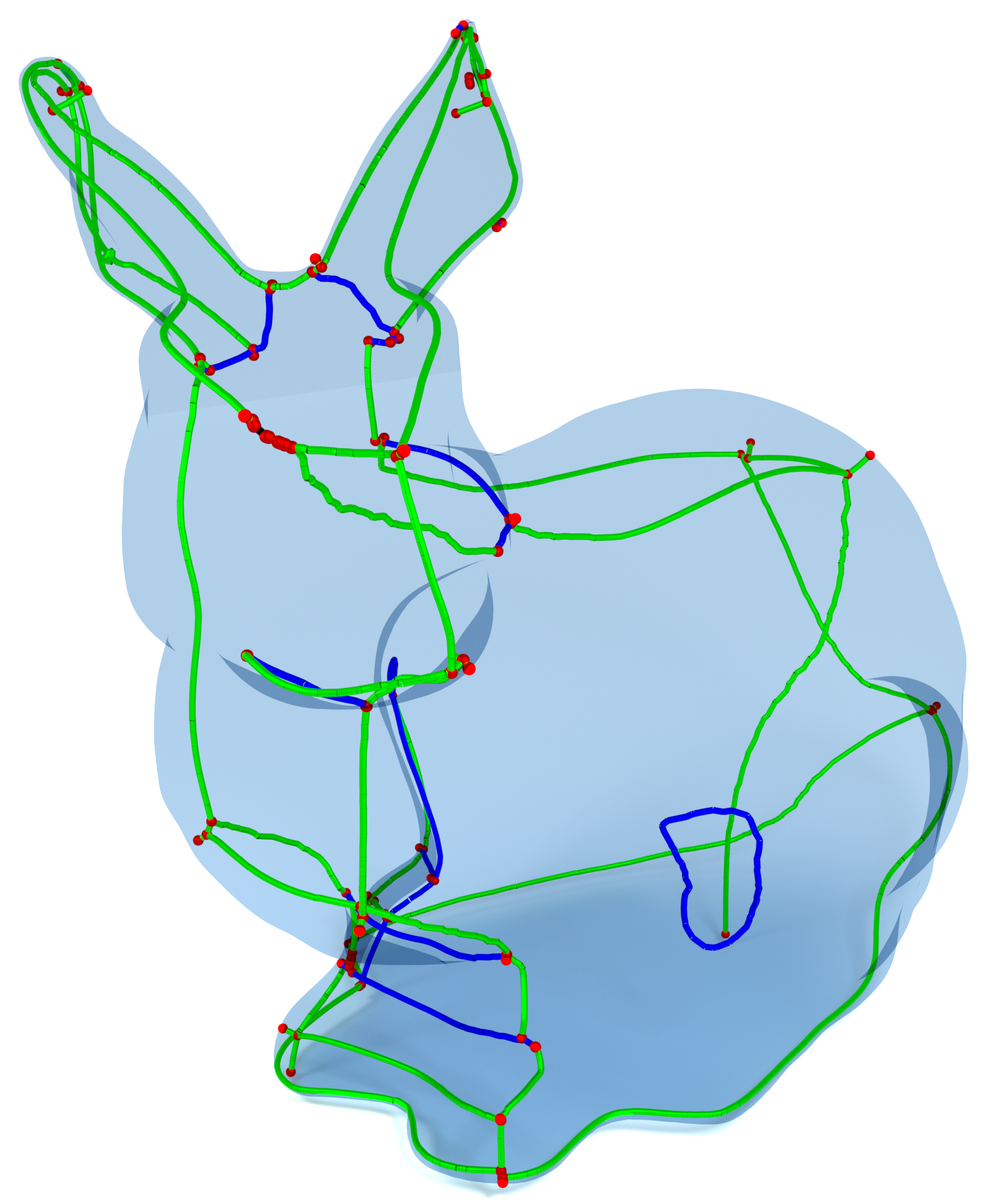} &
\includegraphics[align=c,width=\imagewidth]{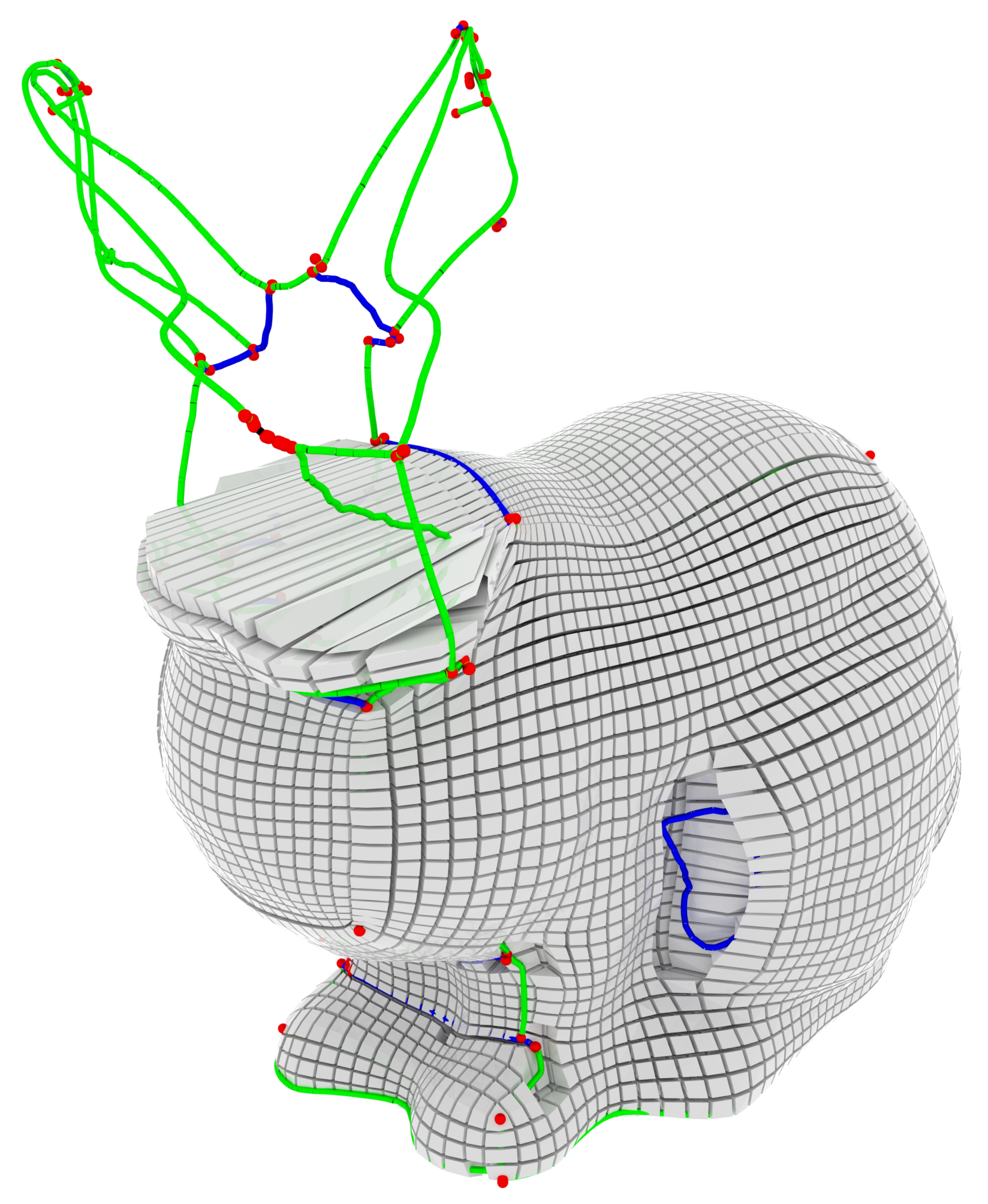} \\
{\rotatebox[origin=c]{90}{\footnotesize{mMBO + RTR}}} &
\includegraphics[align=c,width=\imagewidth]{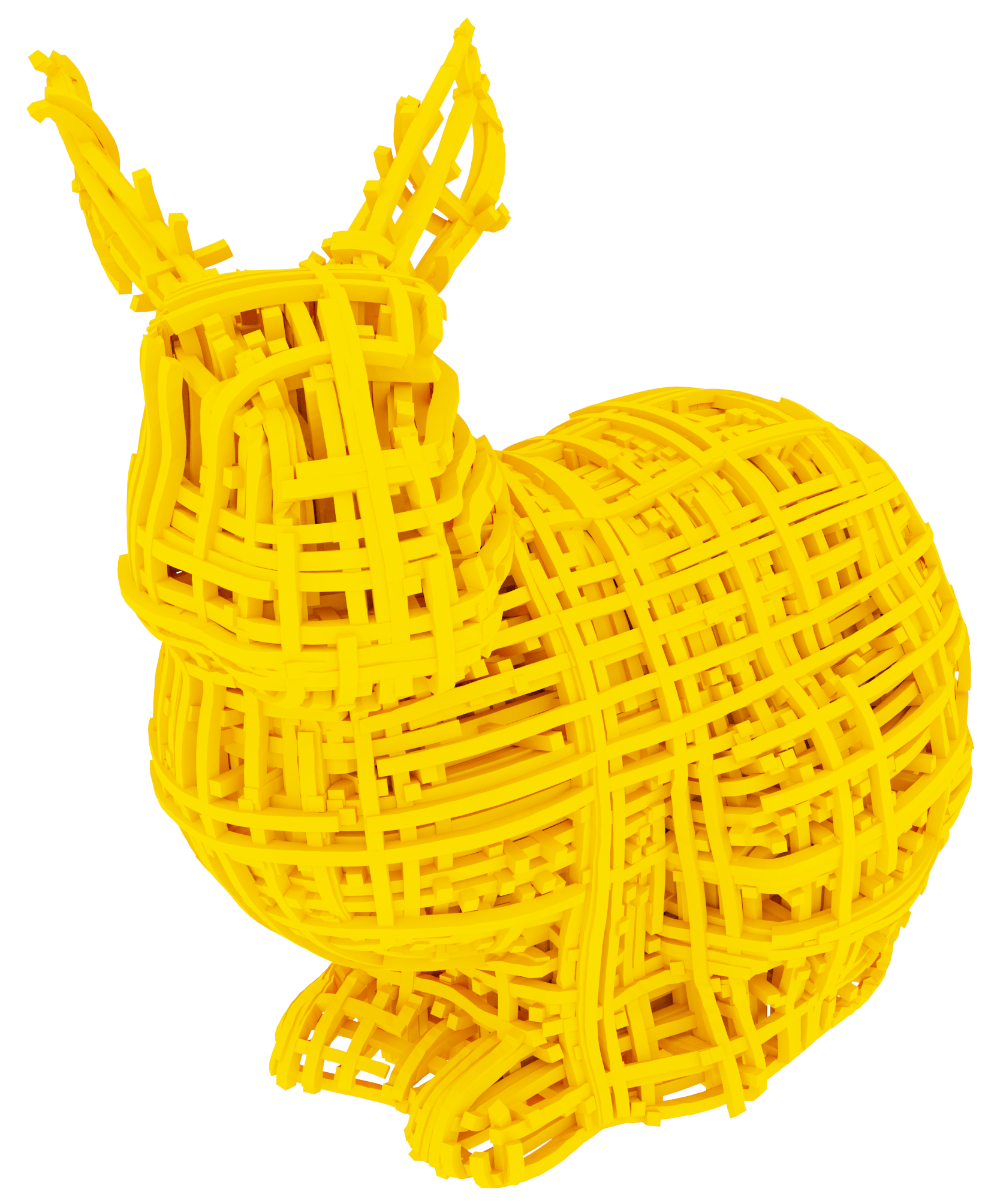} &
 \includegraphics[align=c,width=\imagewidth]{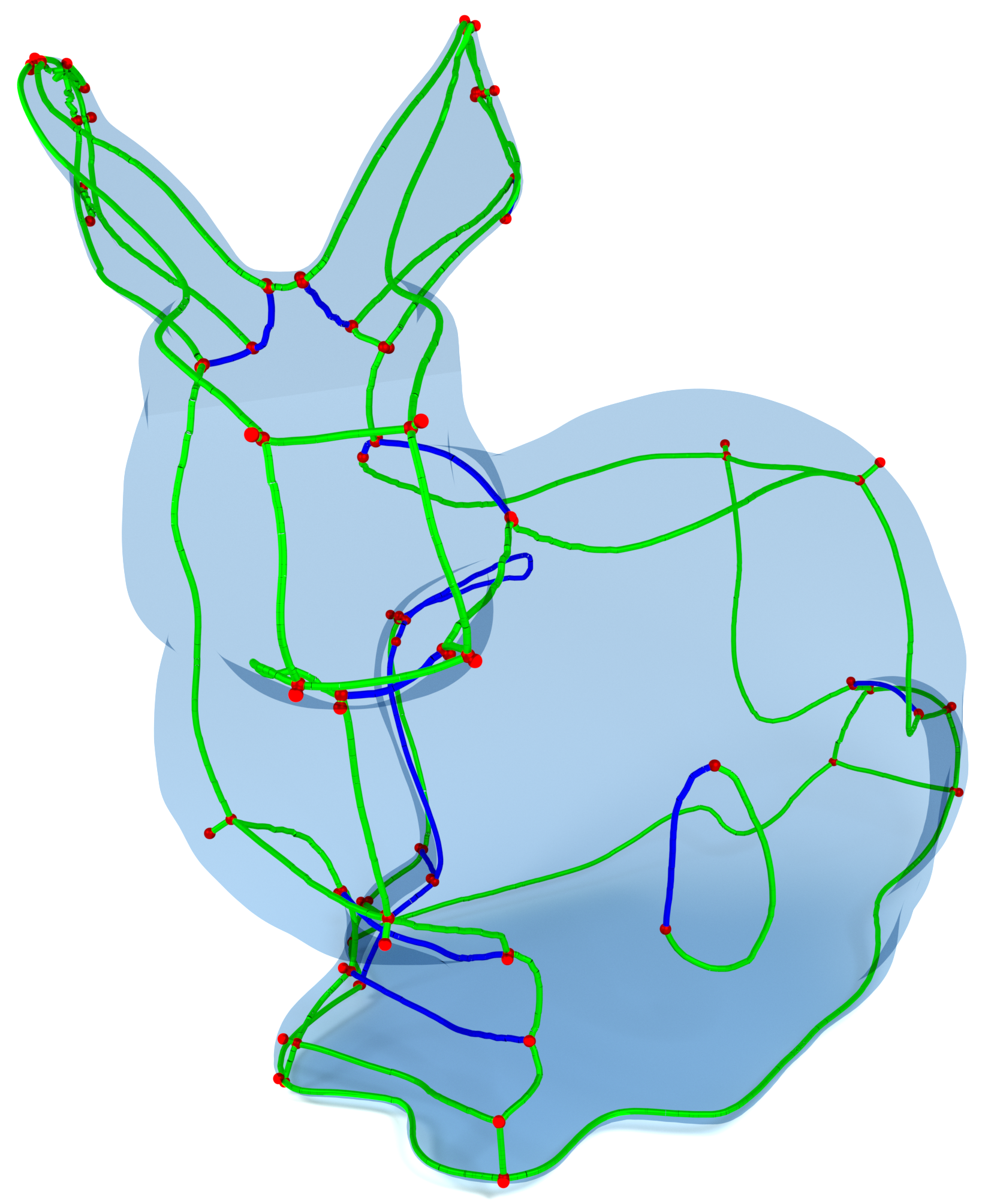} &
\includegraphics[align=c,width=\imagewidth]{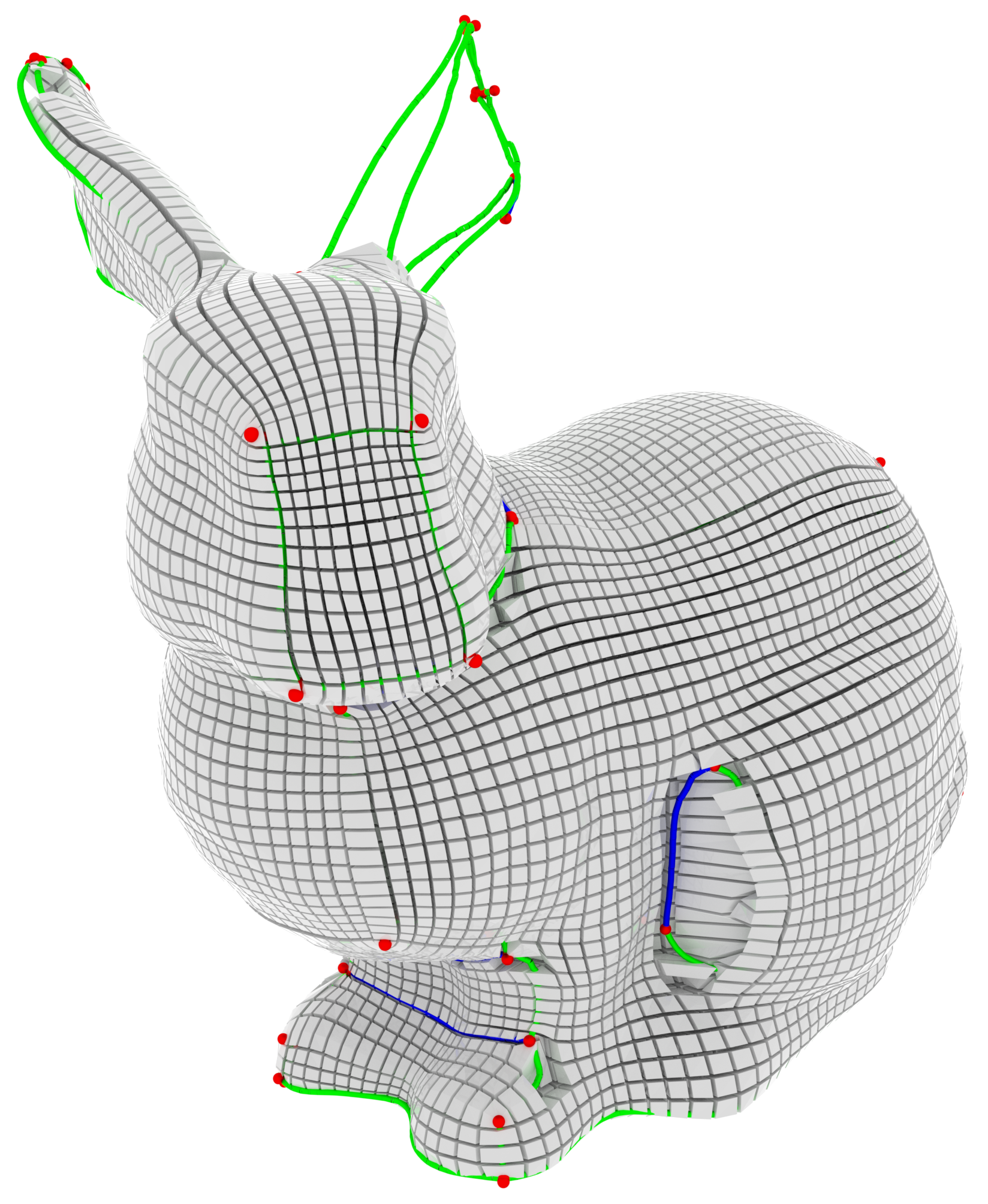}%
\end{tabular}%
\caption{Note the simpler, more regular singular curves in our result
as compared to the result of \citet{RaySokolov2}---especially
on the bunny's nose. The degeneracy in \citeauthor{RaySokolov2}'s result
leads to a collapse of the integer grid map on the head.}%
\label{fig.field_comparison.bunny}
\end{figure}

In Figures \ref{fig.field_comparison.cup}--\ref{fig.field_comparison.bunny}, we compare
fields computed by octahedral mMBO + RTR to those computed by the method of \citet{RaySokolov2}.
Our fields not only have lower Dirichlet energy, but also better singular
structures. To visualize singular structures, we use the visualization technique of \citet{liu}.
To illustrate the importance of singular structure, we have generated hexahedral meshes 
from both sets of fields, following the methods laid out in \cite{nieser2011cubecover} and \cite{Lyon:2016:HRH}.
The meshes are computed from the raw field data, with no preprocessing other than tetrahedral mesh refinement
to resolve and localize singular curves. In particular, we do not
``correct'' singularities---thus, both sets of meshes show some degeneracies resulting from collapsed
or flipped tetrahedra during the parametrization step. However, our fields yield meshes with fewer, smaller
degeneracies, less distortion, and more regular structure.

\begin{figure}
\newcommand{\imagewidth}{0.47\columnwidth}
\centering
\begin{tabular}{cc}
\includegraphics[align=c,width=\imagewidth]{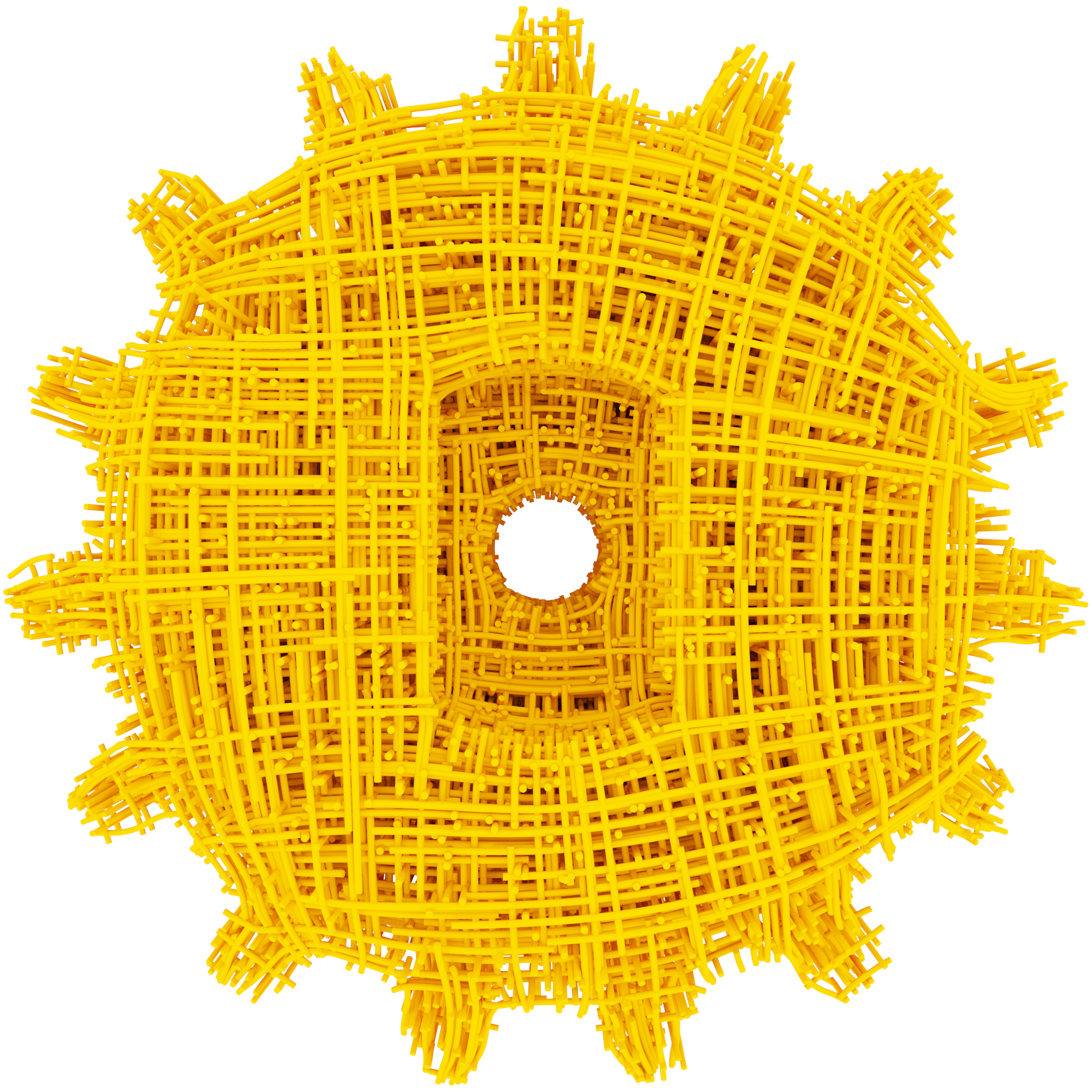} &
\includegraphics[align=c,width=\imagewidth]{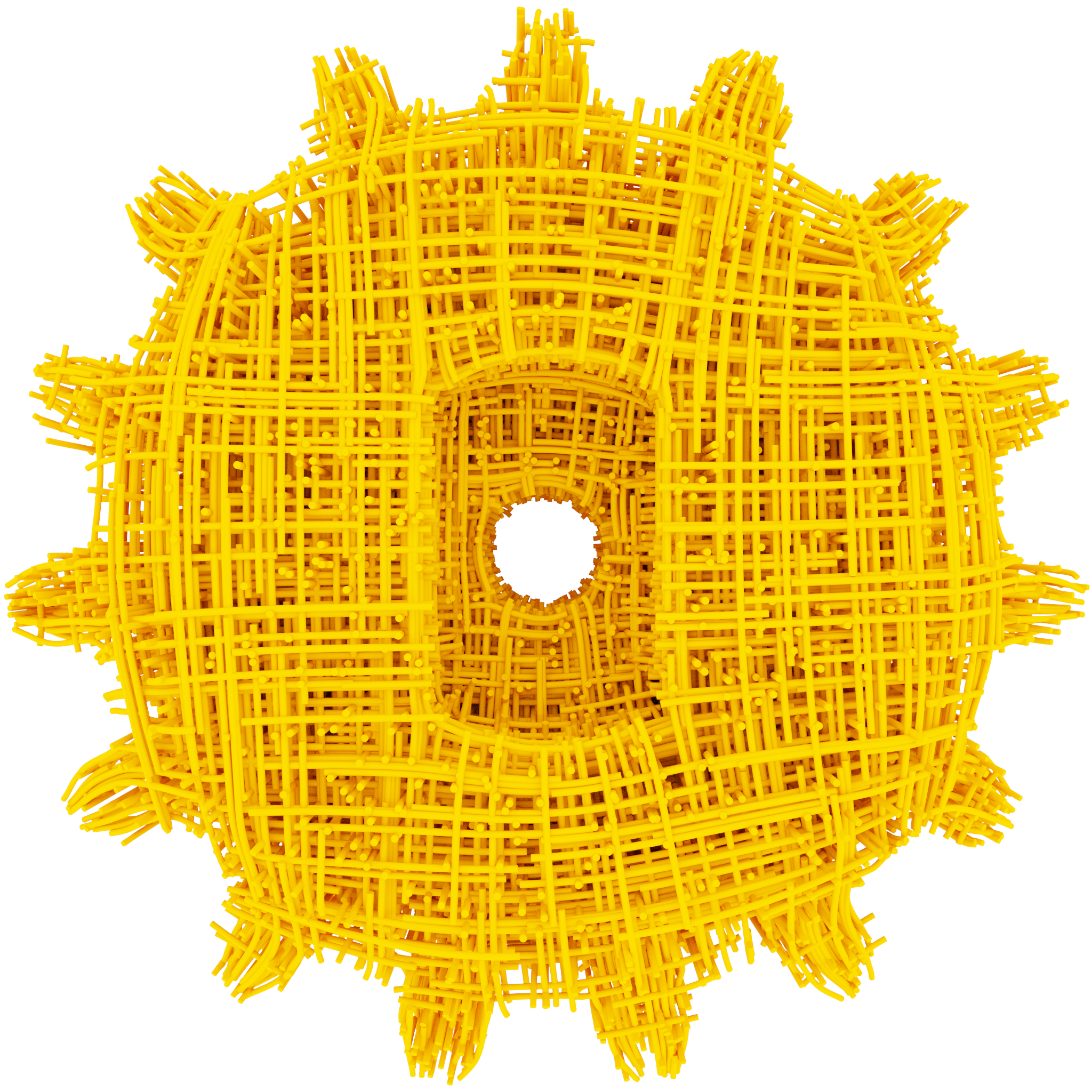} \\
$E = 10580$ & $E = 9901.4$ \\
\cite{Gao:2017:RHM} & mMBO + RTR
\end{tabular}%
\caption{Comparison of octahedral field algorithms on the gear model. A field
produced with mMBO + RTR (right) exhibits lower energy and
greater symmetry than one produced by the method of \citet{Gao:2017:RHM} (left).}%
\label{fig.field_comparison.gear}
\end{figure}

\begin{figure}
\newcommand{\imagewidth}{0.47\columnwidth}
\centering
\begin{tabular}{cc}
\includegraphics[align=c,width=\imagewidth]{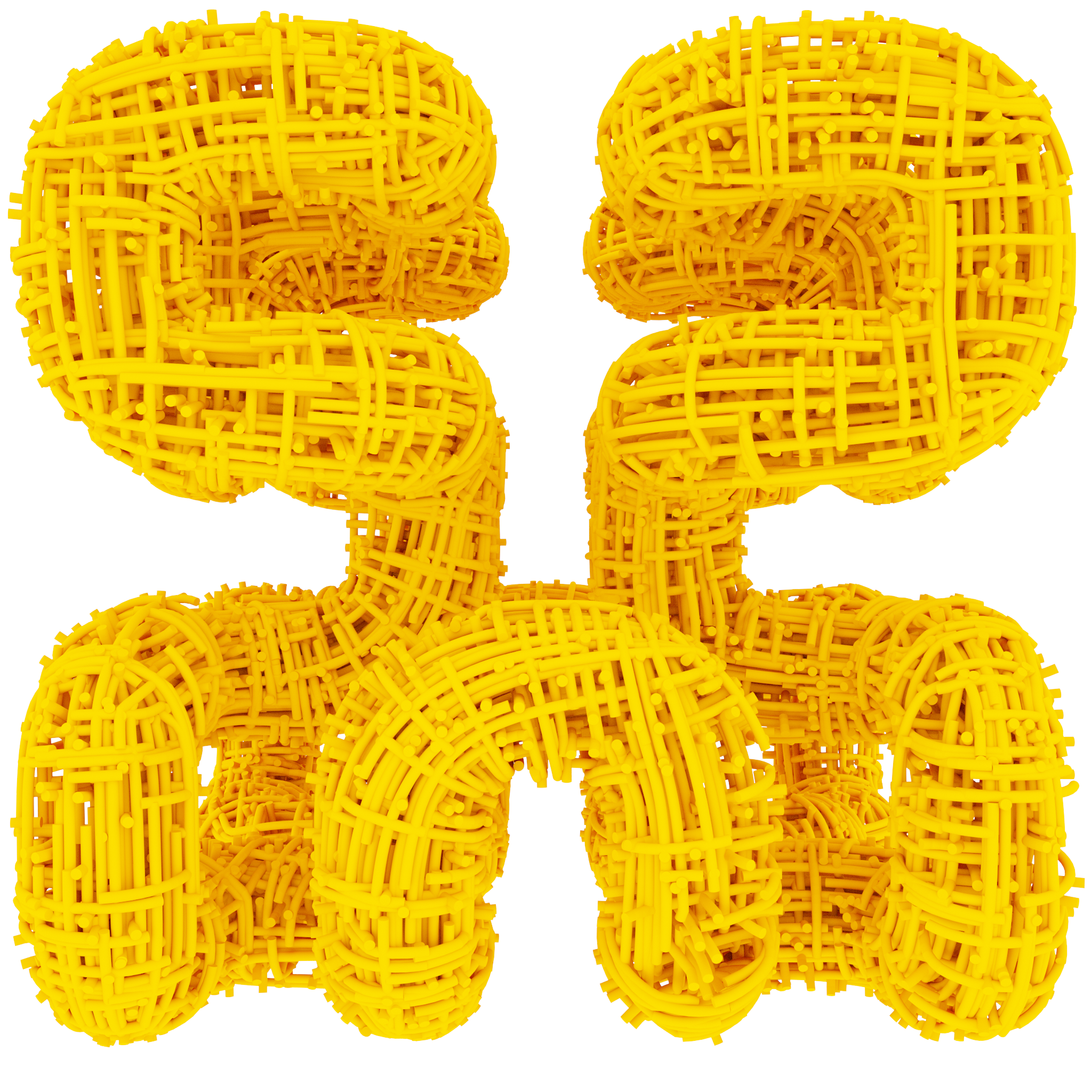} &
\includegraphics[align=c,width=\imagewidth]{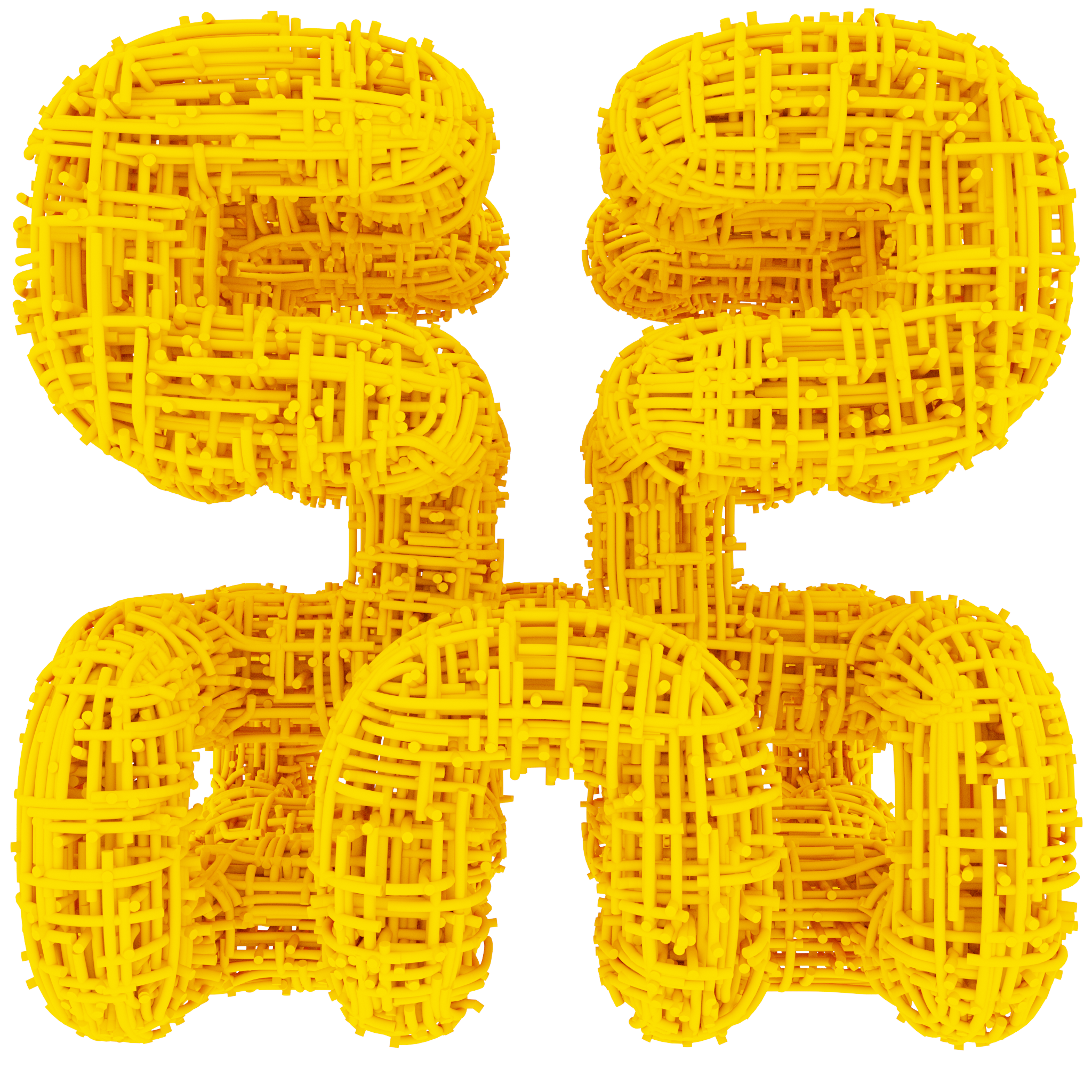} \\
$E = 8150.1$ & $E = 6319.9$ \\
\cite{Gao:2017:RHM} & mMBO + RTR
\end{tabular}%
\caption{Comparison of octahedral field algorithms on the space-filling
torus model. A field
produced with mMBO + RTR (right) exhibits lower energy and
greater symmetry than one produced by the method of \citet{Gao:2017:RHM} (left).}%
\label{fig.field_comparison.hilbert}
\end{figure}

Figures \ref{fig.field_comparison.gear} and \ref{fig.field_comparison.hilbert}
compare fields produced by mMBO + RTR to fields produced by the method of \citet{Gao:2017:RHM}.
Our fields better respect the symmetries of the underlying models.
We note that mMBO + RTR yields a higher-quality result
without requiring a multiscale method like the one \citet{Gao:2017:RHM} employ.

\section{Discussion and Future Work}
\label{sec.concl}

A stronger understanding of the unknowns in the volumetric frame field problem enables both theoretical and practical developments.  On the theoretical side, our reexamination of the typical representation of octahedral frames not only yields useful geodesic formulas and projection operators but also suggests generalization to odeco frames.  For both frame representations, optimization algorithms designed explicitly to traverse the manifold of unknowns yield gains in efficiency and in the quality of computed results.

Our work is intended not only to improve on existing frame field computation algorithms but also to inspire additional research into the structure of spaces of frames and to highlight the significance of this structure to practical aspects of field computation and meshing. Most immediately, a few open questions are left from our discussion. 
On the theoretical side, conjectures \ref{conj.octa_exact} and
\ref{conj.odeco_pos} remain to be proven. We anticipate that both
can be embedded in a more general framework explaining when
relaxations of projection problems are exact.
On the algorithmic side, RTR seems to get stuck in local minima
much more often than MBO, especially on denser meshes.
We hypothesize that this is because RTR strictly moves along the manifold,
while MBO is free to tunnel through the ambient space. On the other hand,
RTR converges much faster and yields high-quality fields on coarse meshes.
Given these observations, we anticipate that it may be possible to incorporate RTR into a multiscale method that leverages its efficiency while avoiding local minima that appear at the finest scales.

\begin{figure}
\newcommand{\imagewidth}{0.4\columnwidth}
\centering
\begin{tabular}{lcc}
{\rotatebox[origin=c]{90}{\footnotesize{\cite{RaySokolov2}}}} &
\includegraphics[align=c,width=\imagewidth]{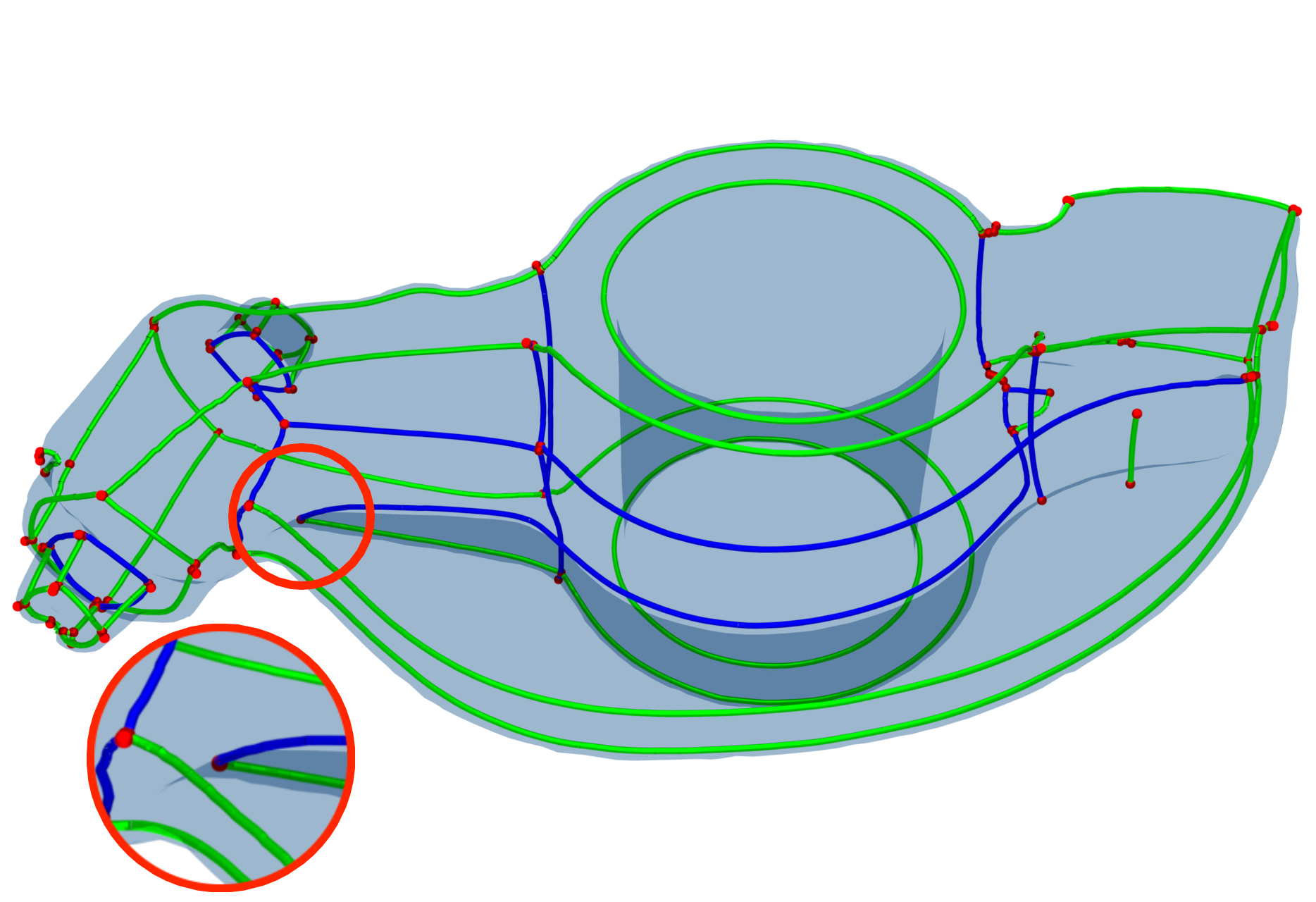} &
\includegraphics[align=c,width=\imagewidth]{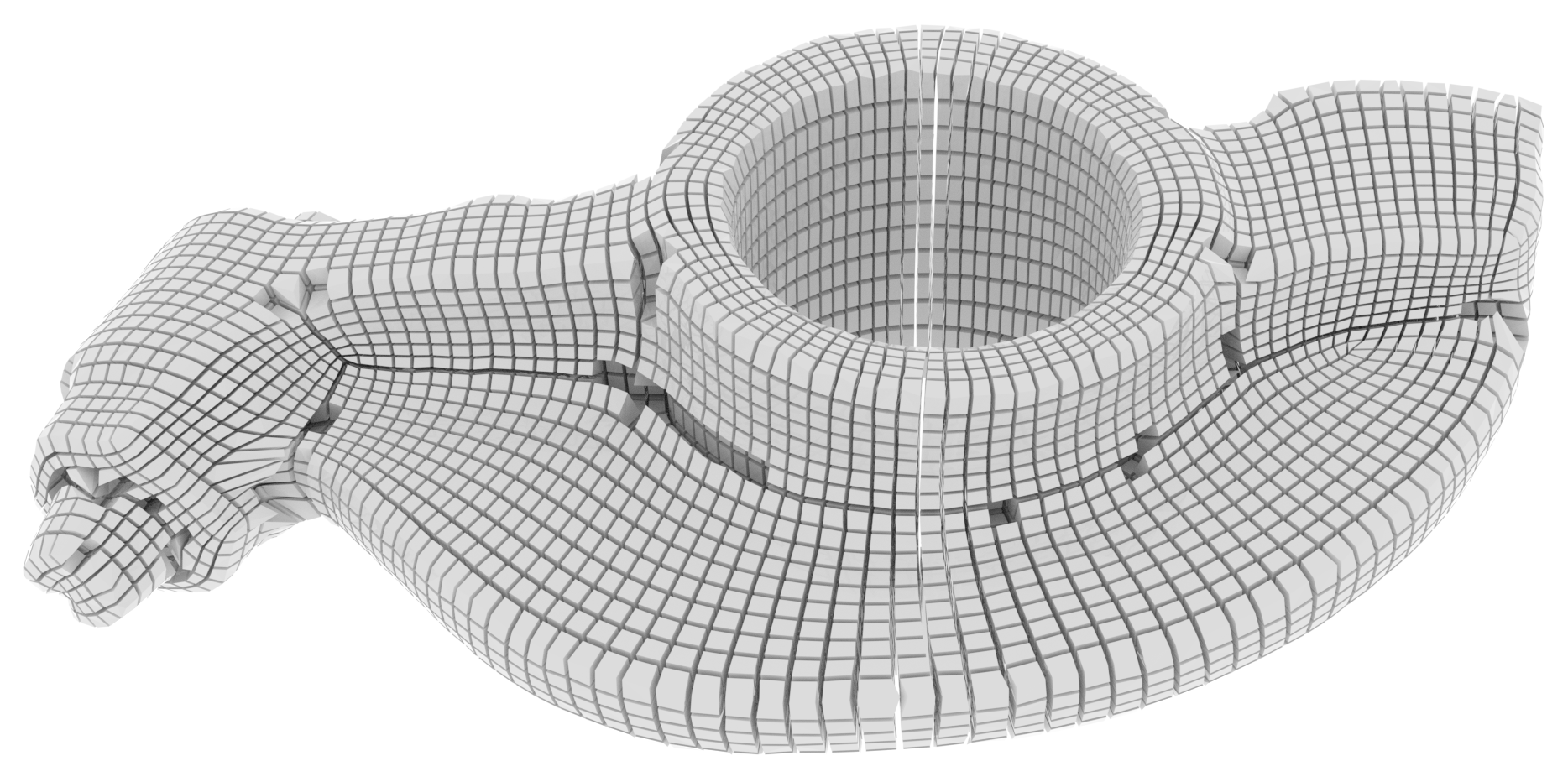} \\
{\rotatebox[origin=c]{90}{\footnotesize{mMBO + RTR}}} &
\includegraphics[align=c,width=\imagewidth]{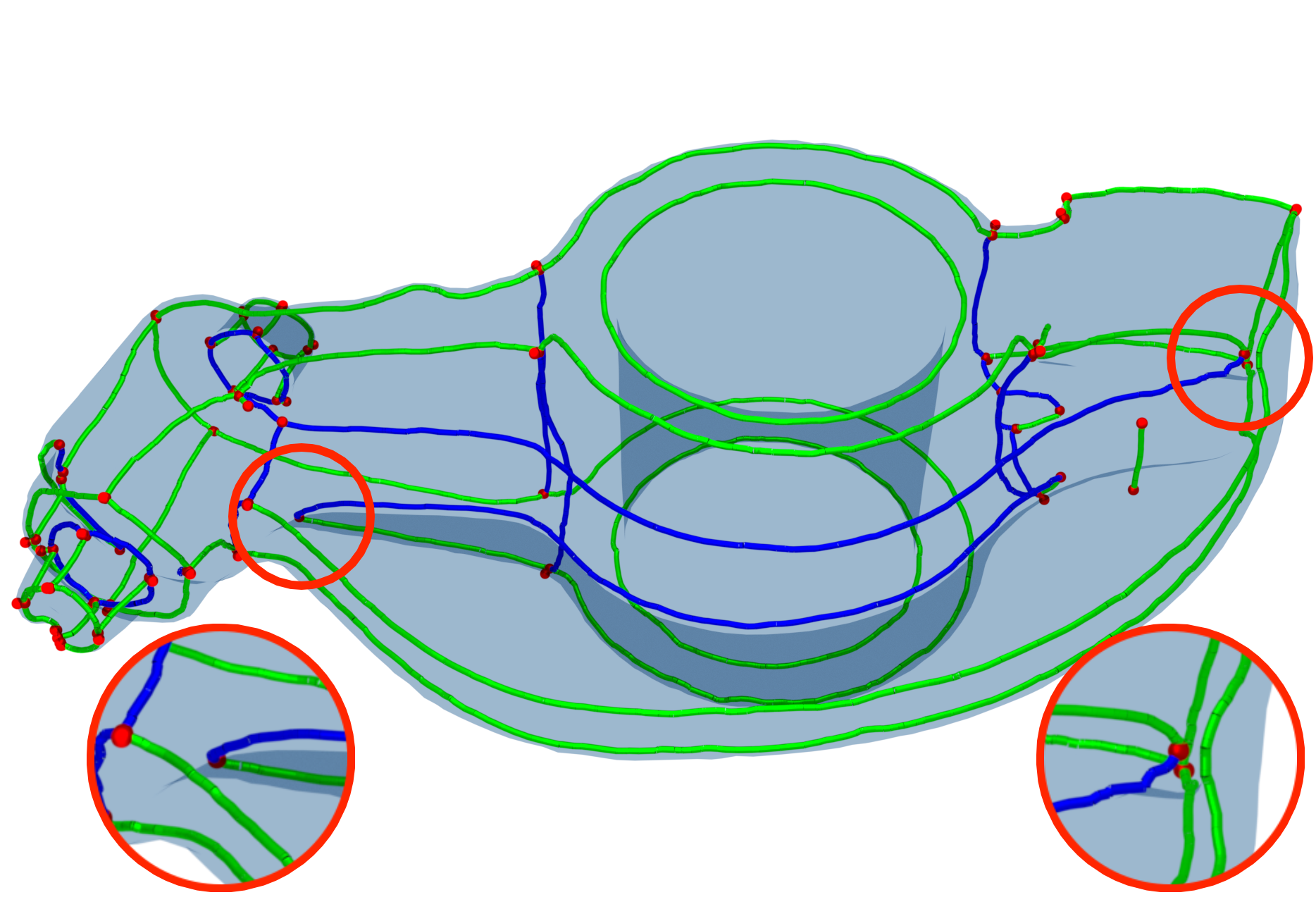} &
\includegraphics[align=c,width=\imagewidth]{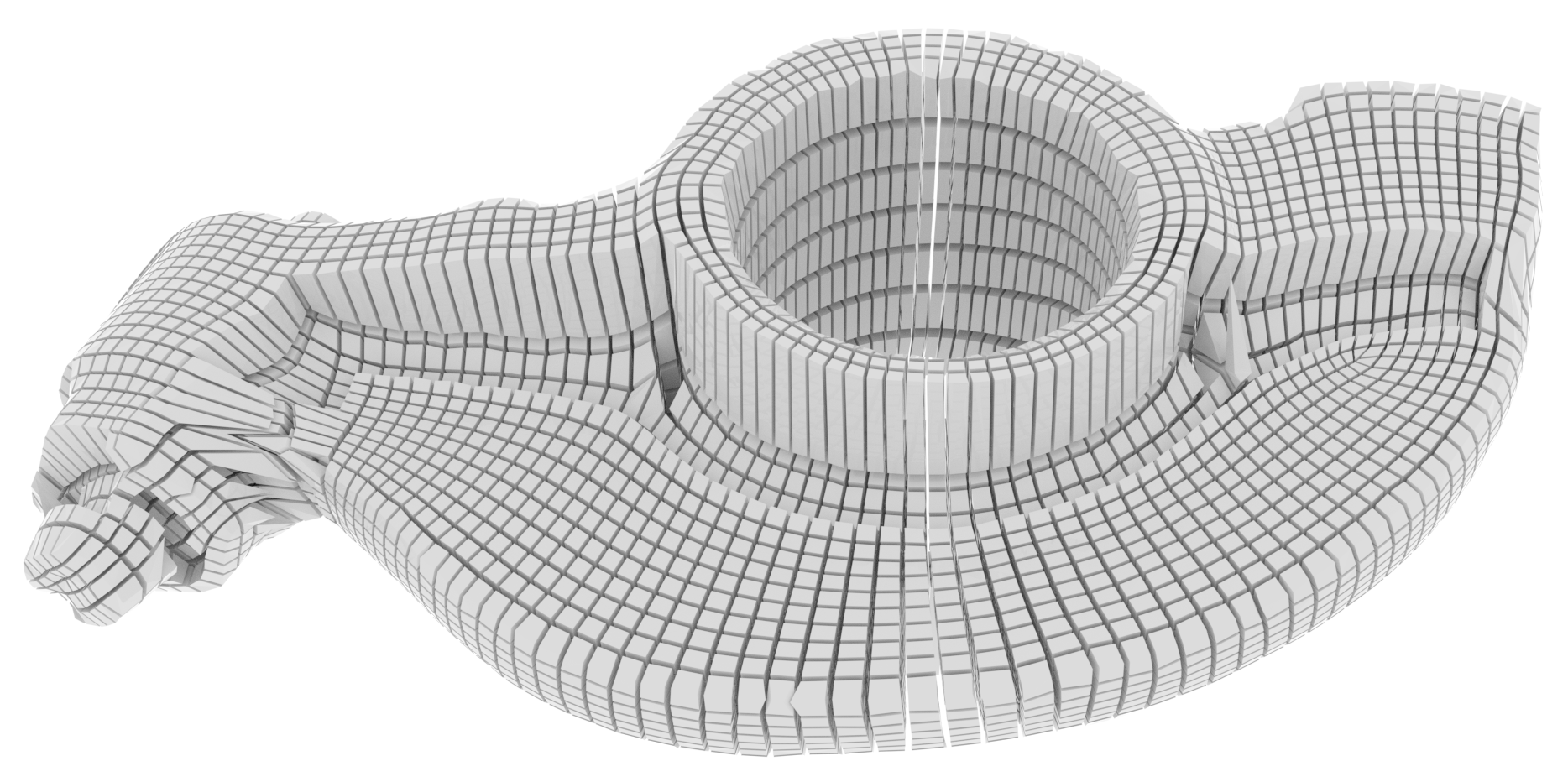} %
\end{tabular}%
\caption{A field of lower Dirichlet energy (bottom) may
still result in more mesh degeneracies than a field of higher Dirichlet energy (top)
due to topological impediments to meshability,
namely the presence of an additional valence $3$--$5$ junction.}%
\label{fig.field_comparison.rockerarm}
\end{figure}

As with most existing frame field computation algorithms,
even when mMBO+RTR converges to a smooth field, the topology of the field is not always
hex-meshable. Figure \ref{fig.field_comparison.rockerarm} shows a case where
the method of \citet{RaySokolov2} yields a field of higher
Dirichlet energy, but which leads to fewer degeneracies than our field. In
particular, the presence of a singular curve whose type changes from valence $3$
to $5$ leads to a degeneracy in the integer-grid map.
While our method succeeds at finding fields of lower
Dirichlet energy---the objective function of our method and that of
\citet{RaySokolov2}---minimizing this energy is an incomplete proxy
for our ultimate goal, namely to obtain
smooth, meshable fields. We would like to investigate additional
metrics, e.g., integrability of frame fields, that might make it possible
to express meshability constraints rigorously.

To define such additional metrics, it might be fruitful
to consider further alternative frame field representations. For example,
given our use of Lie algebra representations, a logical next step might be
to introduce an $\SO(3)$--principal bundle and work with connections on that
bundle as variables. In this theory, quantities such as curvature, torsion,
or the Chern-Simons functional might encode important features of frame fields.
The discretization of connections in \cite{Corman2019} represents a promising first
step.

Finally, while our algorithms do not explicitly take account of symmetry, we
find that fields computed by MBO consistently reproduce the symmetries of the
volume. It would be interesting to develop a better theoretical understanding
of this behavior and to develop machinery for explicitly promoting conformation to near-symmetry in deformed domains.

These and other challenges appear when extending well-known machinery from geometry processing on surfaces to volumetric domains, as demanded by applications in engineering, simulation, and other disciplines.  Over the longer term, careful consideration of problems like the ones we consider in this paper will lead to a versatile and generalizable approach to geometry processing.

\appendix

\section{Mathematical Background}
\label{app.background}

Representation theory and algebraic geometry provide a concise
language for the discussion of the octahedral and odeco varieties above.
We begin with definitions of a few terms used throughout the paper.
While a detailed introduction to either of these subjects
is outside the scope of our discussion,
we refer the reader to \cite{stillwell2008naive} for an introduction to matrix
Lie groups and to \cite{blekherman2012semidefinite} for a primer on real
algebraic geometry and optimization.

\subsection{Lie Groups and Representations}

A \textbf{Lie group} is a group $G$ that is also a smooth manifold, and
such that the multiplication $\cdot : G \times G \to G$ and inversion $\,^{-1} : G \to G$
are smooth maps.
A \textbf{matrix Lie group} is a subgroup of the invertible matrices $\GL(n) \subset \R^{n \times n}$
that is simultaneously a submanifold.
Examples include the orthogonal groups $\mathrm{O}(n)$ and the special
orthogonal groups $\SO(n)$. The primary Lie group of interest to us 
in this paper will be $\SO(3)$.

For a matrix Lie group $G \subset \R^{n\times n}$,
the \textbf{Lie algebra} is a linear subspace of
$\R^{n\times n}$ identified with the tangent space to $G$ at the identity $I$, denoted
$\mathfrak{g} = T_I G$. The \textbf{inner automorphism} $\operatorname{Ad}_g : G \to G$
taking $h \mapsto ghg^{-1}$ preserves the identity, and its derivative at the
identity induces the \textbf{Lie bracket}
$[\cdot, \cdot] : \mathfrak{g} \times \mathfrak{g} \to \mathfrak{g}$.
In the case of a matrix Lie group, this is just the matrix commutator.

As left translation $\Ltrans_g : h \mapsto gh$ is a diffeomorphism,
every element of $\mathfrak{g}$ is also associated to a left-invariant vector
field on $G$. The exponential map $\exp: \mathfrak{g} \to G$ is given by
starting at $I$ and integrating this vector field for one unit of time. For a matrix Lie group,
$\exp$ is the ordinary matrix exponential.

A \textbf{representation} of a matrix Lie group $G$ is a smooth group homomorphism
$\rho: G \to \GL(n)$ for some $n$. A representation of $G$ induces a representation
of each subgroup $H \subset G$. A Lie group representation comes with a
\textbf{Lie algebra representation} $(D\rho)_I : \mathfrak{g} \to \R^{n \times n}$, which
preserves the Lie bracket. Moreover, representations commute with exponentials:
\begin{equation} \label{eq.exp_rep}
	\rho \circ \exp = \exp \circ (D\rho)_I.
\end{equation}
An \textbf{irreducible} representation is one that cannot be decomposed
as a direct sum of sub-representations. The irreducible representations of
\emph{compact} Lie groups such as $\SO(3)$ are finite-dimensional.
The \textbf{adjoint representation} is the irreducible representation of $G$ on its own Lie algebra given by conjugation:
\begin{equation}
\begin{aligned}
	\operatorname{Ad} : G &\to \operatorname{Aut}(\mathfrak{g}) \\
	\operatorname{Ad}(g) a &\coloneqq g^{-1} a g.
\end{aligned}
\end{equation}
\begin{lemma} \label{lem.adj_comm}
	$\operatorname{Ad}$ commutes with representations of $G$:
	\begin{equation}
		(D\rho)_I(g^{-1} a g) = \rho(g)^{-1} ((D\rho)_I a) \rho(g).
	\end{equation}
\end{lemma}
\begin{proof}
	Using \eqref{eq.exp_rep} and conjugating by $g$, we obtain
	\begin{equation} \rho(g^{-1} \exp(t a) g) = \rho(g)^{-1} \exp(t (D\rho)_I a) \rho(g). \end{equation}
	Differentiating in $t$ at $t = 0$ yields the result.
\end{proof}
The \textbf{stabilizer} $\stab(v)$ of a vector $v$ is the subgroup consisting
of all elements $g \in G$ that preserve $v$---that is, such that $\rho(g)v = v$.

The Lie algebra $\mathfrak{so}(3)$ associated to $\SO(3)$ consists of the
skew-symmetric $3 \times 3$ matrices. $\mathfrak{so}(3)$ has a basis
consisting of infinitesimal rotations about the three coordinate axes:
\[ l_1 = \begin{pmatrix}0 & 0 & 0 \\ 0 & 0 & 1 \\ 0 & -1 & 0 \end{pmatrix} \quad
   l_2 = \begin{pmatrix}0 & 0 & 1 \\ 0 & 0 & 0 \\ -1 & 0 & 0 \end{pmatrix} \quad
   l_3 = \begin{pmatrix}0 & 1 & 0 \\ -1 & 0 & 0 \\ 0 & 0 & 0 \end{pmatrix}. \]
Their Lie brackets are
\begin{equation} \label{so3.commutators}
[l_i, l_i] = 0 \quad [l_1, l_2] = l_3 \quad [l_2, l_3] = l_1 \quad [l_3, l_1] = l_2.
\end{equation}
These generators might be familiar as the angular momentum operators
from quantum mechanics. Indeed, $\SO(3)$ acts on functions on the sphere
by rotating them, and its Lie algebra elements act as rotational derivatives.
This action decomposes into irreducible representations.
A basis for each irreducible representation of $\SO(3)$ is provided
by \textbf{spherical harmonics}; each irreducible
representation is referred to as a \emph{band} of spherical harmonics.
The real representations are odd-dimensional vector spaces, and the
representation matrices $\rho(\cdot)$ are also called Wigner $D$-Matrices.

$\SO(3)$ admits a \textbf{bi-invariant Riemannian metric}
whose value at the identity is given by
\begin{equation}
	\langle u, v \rangle = \frac{1}{2} \sum_{i, j} u_{ij} v_{ij}.
\end{equation}
Note that the generators $l_i$ form an orthonormal basis under this metric.
Bi-invariance means that
\begin{equation} \label{eq.bi_invariant}
\langle (D\Ltrans_g)_I v, (D\Ltrans_g)_I w \rangle = \langle v, w \rangle =
\langle (D\Rtrans_g)_I v, (D\Rtrans_g)_I w \rangle,
\end{equation}
where $v, w \in \mathfrak{so}(3)$ and $\Rtrans_g$ is right translation by $g \in SO(3)$,
defined analogously to left translation. \eqref{eq.bi_invariant} completely
describes the Riemannian metric on $\SO(3)$. Under this metric, the matrix
exponential is also the Riemannian exponential map at the identity, giving rise to geodesics.

The \textbf{octahedral group} $\Oct$ is a discrete subgroup of $\SO(3)$ generated by
right-angle rotations about the three axes,
\[ \left\{r_i = \exp\left(\frac{\pi}{2}l_i\right)\right\}_{i=1}^3. \]
$\Oct$ has $24$ elements comprising
all the symmetries of the cube, or equivalently, of its dual, the octahedron.

\subsection{Semidefinite Relaxations}
\label{subsec.relaxations}

A \textbf{(real) algebraic variety} is the set of solutions of a system of polynomial equations
over $n$ real variables:
\[ V(p_1, \dots, p_k) = \{ x \in \R^n \mid p_i(x) = 0, i = 1, \dots, k \}, \]
where $p_1, \dots, p_k$ are polynomials.

If $p_1, \dots, p_k$ are quadratic in $x$, they may be written
\[ p_i(x)
= \begin{pmatrix} 1 & x\tran \end{pmatrix} A_i \begin{pmatrix} 1 \\ x \end{pmatrix}
= \langle A_i, X \rangle, \]
where $A_i$ are symmetric matrices, $\langle, \rangle$ denotes the usual inner
product on matrices inducing the Frobenius norm, and
\[ X = \begin{pmatrix} 1 \\ x \end{pmatrix} \begin{pmatrix} 1 & x\tran \end{pmatrix}
= \begin{pmatrix} 1 & x\tran \\ x & x x\tran \end{pmatrix}. \]

A quadratically-constrained quadratic program (\textbf{QCQP}) is of the form
\[ \argmin \{f(x) \mid x \in V(p_1, \dots, p_k)\}, \]
where $f, p_1, \dots, p_k$ are all quadratic. A QCQP may be
rewritten in the more illuminating form
\begin{equation}
\label{relax.prototype}
\begin{alignedat}{4}
	&\argmin_X \; &\langle C, X \rangle \\
	&\text{subject to} &X_{11} &= 1 \\
		&&\langle A_i, X \rangle &= 0, &\; i = 1, \dots, k \\
		&&X &\succeq 0 \\
		&&\rank(X) &= 1.
\end{alignedat}
\end{equation}
Absent the rank constraint, \eqref{relax.prototype} would be a semidefinite
program (\textbf{SDP}). Semidefinite programs may be solved in polynomial time by
interior point methods, implemented in common optimization software packages
like Mosek \shortcite{mosek}.
Thus it is natural to ignore the rank constraint and solve the associated
SDP; this technique is called \textbf{semidefinite relaxation}.
The optimal objective value of the relaxed problem is a lower bound for the globally optimal
value of the QCQP \eqref{relax.prototype},
and if the recovered solution to the SDP has rank one,
the solution is said to be \textbf{exact}, as it is the globally optimal solution
of \eqref{relax.prototype}.

\begin{acks}

The authors would like to thank Elina Robeva for helping us understand her work on the odeco varieties, Pablo Parrilo and Diego Cifuentes for sharing their insights into semidefinite relaxation, and Nicolas Boumal for helping us get started with manifold optimization. We thank Heng Liu for providing tetrahedral refinement, parametrization, and hexahedral mesh extraction code. We are grateful to Xifeng Gao and Daniele Panozzo for guiding us in using their code from \cite{Gao:2017:RHM}. Finally, we thank Mirela Ben-Chen, Amit Bermano, Edward Chien, Etienne Corman, Keenan Crane, Fernando de Goes, Steven Gortler, Leif Kobbelt, Braxton Osting, Nir Sharon, Nir Sochen, Amir Vaxman, Josh Vekhter, and Paul Zhang for valuable discussions.

David Palmer appreciates the generous support of the Fannie and John Hertz Foundation
through the Hertz Graduate Fellowship.
David Bommes appreciates the generous support of the \grantsponsor{ERC}{European Research Council (ERC)}{https://erc.europa.eu} under the European Union’s Horizon 2020 research and innovation program (AlgoHex, grant agreement no.~\grantnum{ERC}{853343}).
Justin Solomon acknowledges the generous support of \grantsponsor{ARO}{Army Research Office}{https://www.arl.army.mil/who-we-are/aro/} grant \grantnum{ARO}{W911NF-12-R-0011}, \grantsponsor{NSF}{National Science Foundation}{https://www.nsf.gov} grant \grantnum{NSF}{IIS-1838071}, \grantsponsor{AFOSR}{Air Force Office of Scientific Research}{https://www.wpafb.af.mil/afrl/afosr/} award \grantnum{AFOSR}{FA9550-19-1-0319}, and a gift from Adobe Systems.  Any opinions, findings, and conclusions or recommendations expressed in this material are those of the authors and do not necessarily reflect the views of these organizations.

\end{acks}

\bibliographystyle{ACM-Reference-Format}
\bibliography{bibliography}

\end{document}


\title{Supplementary Material: Algebraic Representations for Volumetric Frame Fields}

\author{David Palmer}
\orcid{0000-0002-1931-5673}
\affiliation{%
  \institution{Massachusetts Institute of Technology}
  \streetaddress{77 Massachusetts Avenue}
  \city{Cambridge}
  \state{MA}
  \postcode{02139}
  \country{USA}}
\email{drp@mit.edu}

\author{David Bommes}
\affiliation{%
  \institution{University of Bern}
  \streetaddress{Hochschulstrasse 6}
  \city{Bern}
  \postcode{3012}
  \country{Switzerland}}
\email{david.bommes@inf.unibe.ch}

\author{Justin Solomon}
\orcid{0000-0002-7701-7586}
\affiliation{%
  \institution{Massachusetts Institute of Technology}
  \streetaddress{77 Massachusetts Avenue}
  \city{Cambridge}
  \state{MA}
  \postcode{02139}
  \country{USA}}
\email{jsolomon@mit.edu}

\maketitle

\section{Energy and Timing Results}

In the following table, ``X+RTR'' indicates that RTR was used to polish the result of algorithm X on the previous line, and the corresponding runtimes and iteration counts are the totals for X and RTR. Ray is the algorithm of \citet{RaySokolov2} with their initialization and the combinatorial Laplacian. Ray2 indicates their algorithm but with random initialization and geometric Laplacian (to provide a fairer comparison against our results). RayMBO and RayMMBO are MBO and mMBO, respectively, but using Ray et al.'s approximate projection approach rather than SDP-based projection. For consistency, all energy values are computed by converting to odeco frames and using the geometric Laplacian.

\tablecaption{Energy and computation time comparison on various models. \label{tbl.1}}
\tablefirsthead{%
\toprule Mesh & {Vertices} & Type  & Method & {Energy} & {Time (\si{\second})} & {Iterations} \\ \midrule}
\tablehead{%
\toprule Mesh & {Vertices} & Type  & Method & {Energy} & {Time (\si{\second})} & {Iterations} \\ \midrule}
\tabletail{\midrule}
\tablelasttail{\bottomrule}
\begin{center}
\begin{xtabular}{lS[table-format=6.0]llS[table-format=6.2]S[table-format=4.2]r}
37881\_gear.2                    & 55337    & Octa  & Ray         & 1627.7881976314  & 90.886845   & 354        \\
37881\_gear.2                    & 55337    & Octa  & Ray+RTR     & 1387.84483806352 & 100.502334  & 426        \\
37881\_gear.2                    & 55337    & Octa  & Ray2        & 1418.57420822576 & 1164.663405 & 4876       \\
37881\_gear.2                    & 55337    & Octa  & Ray2+RTR    & 1418.57419221369 & 1167.082352 & 4879       \\
37881\_gear.2                    & 55337    & Octa  & RayMBO      & 1830.1853713872  & 6.2545      & 14         \\
37881\_gear.2                    & 55337    & Octa  & RayMBO+RTR  & 1407.08649047954 & 19.72424    & 111        \\
37881\_gear.2                    & 55337    & Octa  & RayMMBO     & 1338.28078152288 & 31.995769   & 89         \\
37881\_gear.2                    & 55337    & Octa  & RayMMBO+RTR & 1334.91798361169 & 40.659427   & 117        \\
\end{xtabular}
\end{center}

\section{Geodesics in the Octahedral Variety}
The \textbf{angular momentum operators} $L_i$ are the images under a Lie algebra representation of $\mathfrak{so}(3)$ of the Lie algebra generators $l_i$. For the nine-dimensional real irreducible representation of $\mathrm{SO}(3)$, we use
\[
L_1 = %
\]

In the main text (\S4.1) we give an explicit formula for geodesics in the octahedral variety
in terms of the following two ingredients:
\[ \exp(\theta L_3) = \left(
\begin{array}{ccccccccc}
 \cos (4 \theta ) & 0 & 0 & 0 & 0 & 0 & 0 & 0 & \sin (4 \theta ) \\
 0 & \cos (3 \theta ) & 0 & 0 & 0 & 0 & 0 & \sin (3 \theta ) & 0 \\
 0 & 0 & \cos (2 \theta ) & 0 & 0 & 0 & \sin (2 \theta ) & 0 & 0 \\
 0 & 0 & 0 & \cos (\theta ) & 0 & \sin (\theta ) & 0 & 0 & 0 \\
 0 & 0 & 0 & 0 & 1 & 0 & 0 & 0 & 0 \\
 0 & 0 & 0 & -\sin (\theta ) & 0 & \cos (\theta ) & 0 & 0 & 0 \\
 0 & 0 & -\sin (2 \theta ) & 0 & 0 & 0 & \cos (2 \theta ) & 0 & 0 \\
 0 & -\sin (3 \theta ) & 0 & 0 & 0 & 0 & 0 & \cos (3 \theta ) & 0 \\
 -\sin (4 \theta ) & 0 & 0 & 0 & 0 & 0 & 0 & 0 & \cos (4 \theta ) \\
\end{array}
\right) \]
and
\[ R_{23} = \exp\left(\frac{\pi}{2}L_1\right) =
\left(
\begin{array}{ccccccccc}
 0 & 0 & 0 & 0 & 0 & \frac{\sqrt{\frac{7}{2}}}{2} & 0 & -\frac{1}{2 \sqrt{2}} & 0 \\
 0 & -\frac{3}{4} & 0 & \frac{\sqrt{7}}{4} & 0 & 0 & 0 & 0 & 0 \\
 0 & 0 & 0 & 0 & 0 & \frac{1}{2 \sqrt{2}} & 0 & \frac{\sqrt{\frac{7}{2}}}{2} & 0 \\
 0 & \frac{\sqrt{7}}{4} & 0 & \frac{3}{4} & 0 & 0 & 0 & 0 & 0 \\
 0 & 0 & 0 & 0 & \frac{3}{8} & 0 & \frac{\sqrt{5}}{4} & 0 & \frac{\sqrt{35}}{8} \\
 -\frac{\sqrt{\frac{7}{2}}}{2} & 0 & -\frac{1}{2 \sqrt{2}} & 0 & 0 & 0 & 0 & 0 & 0 \\
 0 & 0 & 0 & 0 & \frac{\sqrt{5}}{4} & 0 & \frac{1}{2} & 0 & -\frac{\sqrt{7}}{4} \\
 \frac{1}{2 \sqrt{2}} & 0 & -\frac{\sqrt{\frac{7}{2}}}{2} & 0 & 0 & 0 & 0 & 0 & 0 \\
 0 & 0 & 0 & 0 & \frac{\sqrt{35}}{8} & 0 & -\frac{\sqrt{7}}{4} & 0 & \frac{1}{8} \\
\end{array}
\right) \]

\section{Defining Equations of the Odeco Variety}
For odeco frames aligned to the vertical $\boldsymbol{z}$ (\emph{cf.}
\S5.1.2), $s_{\boldsymbol{z}}$
satisfies three homogeneous quadratic equations defined by the following symmetric
matrices:
\begin{gather*}
	\left(
\begin{array}{ccccc}
 -4 & 0 & -3 \sqrt{2} & 0 & 0 \\
 0 & 0 & 0 & 18 & 0 \\
 -3 \sqrt{2} & 0 & 0 & 0 & 18 \\
 0 & 18 & 0 & 72 & 0 \\
 0 & 0 & 18 & 0 & 72 \\
\end{array}
\right) \\
\left(
\begin{array}{ccccc}
 0 & 6 \sqrt{2} & 0 & 0 & 0 \\
 6 \sqrt{2} & 0 & 0 & 0 & 36 \\
 0 & 0 & 0 & -36 & 0 \\
 0 & 0 & -36 & 0 & 0 \\
 0 & 36 & 0 & 0 & 0 \\
\end{array}
\right) \\
\left(
\begin{array}{ccccc}
 -4 & 0 & 3 \sqrt{2} & 0 & 0 \\
 0 & 0 & 0 & -18 & 0 \\
 3 \sqrt{2} & 0 & 0 & 0 & -18 \\
 0 & -18 & 0 & 72 & 0 \\
 0 & 0 & -18 & 0 & 72 \\
\end{array}
\right)
\end{gather*}

Below, we reproduce explicitly the $27$ symmetric $15 \times 15$ matrices $A_i$ in the equations
(8) cutting out the odeco variety, expressed in the spherical harmonics
basis.

{\tiny%
\[\left(
\begin{array}{ccccccccccccccc}
 \frac{64}{\pi } & 0 & 0 & \frac{52 \sqrt{5}}{\pi } & 0 & \frac{4 \sqrt{15}}{\pi } & 0 & 0 & 0 & 0 & \frac{72}{\pi } & 0 & \frac{36 \sqrt{5}}{\pi
} & 0 & 0 \\
 0 & -\frac{30}{\pi } & 0 & 0 & 0 & 0 & 0 & 0 & -\frac{90 \sqrt{3}}{\pi } & 0 & 0 & 0 & 0 & 0 & 0 \\
 0 & 0 & \frac{90}{\pi } & 0 & 0 & 0 & 0 & \frac{135 \sqrt{\frac{21}{2}}}{2 \pi } & 0 & \frac{45 \sqrt{\frac{3}{2}}}{2 \pi } & 0 & 0 & 0 & 0 & 0
\\
 \frac{52 \sqrt{5}}{\pi } & 0 & 0 & \frac{110}{\pi } & 0 & \frac{50 \sqrt{3}}{\pi } & 0 & 0 & 0 & 0 & -\frac{144 \sqrt{5}}{\pi } & 0 & \frac{450}{\pi
} & 0 & 0 \\
 0 & 0 & 0 & 0 & 0 & 0 & 0 & 0 & 0 & 0 & 0 & -\frac{945 \sqrt{\frac{3}{2}}}{2 \pi } & 0 & \frac{135 \sqrt{\frac{21}{2}}}{2 \pi } & 0 \\
 \frac{4 \sqrt{15}}{\pi } & 0 & 0 & \frac{50 \sqrt{3}}{\pi } & 0 & -\frac{30}{\pi } & 0 & 0 & 0 & 0 & \frac{72 \sqrt{15}}{\pi } & 0 & -\frac{90 \sqrt{3}}{\pi
} & 0 & 0 \\
 0 & 0 & 0 & 0 & 0 & 0 & 0 & 0 & 0 & 0 & 0 & 0 & 0 & 0 & 0 \\
 0 & 0 & \frac{135 \sqrt{\frac{21}{2}}}{2 \pi } & 0 & 0 & 0 & 0 & 0 & 0 & \frac{405 \sqrt{7}}{2 \pi } & 0 & 0 & 0 & 0 & 0 \\
\end{array}
\right)\]

\[\left(
\begin{array}{ccccccccccccccc}
 0 & 0 & 0 & 0 & \frac{48 \sqrt{15}}{\pi } & 0 & 0 & 0 & 0 & 0 & 0 & \frac{18 \sqrt{10}}{\pi } & 0 & \frac{18 \sqrt{70}}{\pi } & 0 \\
 0 & 0 & \frac{120}{\pi } & 0 & 0 & 0 & 0 & -\frac{45 \sqrt{\frac{21}{2}}}{2 \pi } & 0 & \frac{585 \sqrt{\frac{3}{2}}}{2 \pi } & 0 & 0 & 0 & 0 &
0 \\
 0 & \frac{120}{\pi } & 0 & 0 & 0 & 0 & \frac{135 \sqrt{21}}{2 \pi } & 0 & -\frac{225 \sqrt{3}}{2 \pi } & 0 & 0 & 0 & 0 & 0 & 0 \\
 0 & 0 & 0 & 0 & \frac{60 \sqrt{3}}{\pi } & 0 & 0 & 0 & 0 & 0 & 0 & -\frac{2745}{2 \sqrt{2} \pi } & 0 & \frac{495 \sqrt{\frac{7}{2}}}{2 \pi } & 0
\\
 \frac{48 \sqrt{15}}{\pi } & 0 & 0 & \frac{60 \sqrt{3}}{\pi } & 0 & \frac{180}{\pi } & 0 & 0 & 0 & 0 & \frac{513 \sqrt{15}}{2 \pi } & 0 & -\frac{405
\sqrt{3}}{\pi } & 0 & \frac{135 \sqrt{21}}{2 \pi } \\
 0 & 0 & 0 & 0 & \frac{180}{\pi } & 0 & 0 & 0 & 0 & 0 & 0 & \frac{1035 \sqrt{\frac{3}{2}}}{2 \pi } & 0 & -\frac{45 \sqrt{\frac{21}{2}}}{2 \pi } &
0 \\
 0 & 0 & \frac{135 \sqrt{21}}{2 \pi } & 0 & 0 & 0 & 0 & 0 & 0 & \frac{405 \sqrt{\frac{7}{2}}}{\pi } & 0 & 0 & 0 & 0 & 0 \\
 0 & -\frac{45 \sqrt{\frac{21}{2}}}{2 \pi } & 0 & 0 & 0 & 0 & 0 & 0 & -\frac{405 \sqrt{\frac{7}{2}}}{2 \pi } & 0 & 0 & 0 & 0 & 0 & 0 \\
\end{array}
\right)\]

\[\left(
\begin{array}{ccccccccccccccc}
 0 & \frac{4 \sqrt{15}}{\pi } & 0 & 0 & 0 & 0 & 0 & 0 & \frac{36 \sqrt{5}}{\pi } & 0 & 0 & 0 & 0 & 0 & 0 \\
 \frac{4 \sqrt{15}}{\pi } & 0 & 0 & \frac{50 \sqrt{3}}{\pi } & 0 & 0 & 0 & 0 & 0 & 0 & \frac{72 \sqrt{15}}{\pi } & 0 & 0 & 0 & 0 \\
 0 & 0 & 0 & 0 & -\frac{45}{\pi } & 0 & 0 & 0 & 0 & 0 & 0 & -\frac{495 \sqrt{\frac{3}{2}}}{2 \pi } & 0 & -\frac{135 \sqrt{\frac{21}{2}}}{2 \pi }
& 0 \\
 0 & \frac{50 \sqrt{3}}{\pi } & 0 & 0 & 0 & 0 & 0 & 0 & \frac{450}{\pi } & 0 & 0 & 0 & 0 & 0 & 0 \\
 0 & 0 & -\frac{45}{\pi } & 0 & 0 & 0 & 0 & \frac{135 \sqrt{\frac{21}{2}}}{2 \pi } & 0 & -\frac{495 \sqrt{\frac{3}{2}}}{2 \pi } & 0 & 0 & 0 & 0 &
0 \\
 0 & 0 & 0 & 0 & 0 & 0 & 0 & 0 & 0 & 0 & 0 & 0 & 0 & 0 & 0 \\
 0 & 0 & 0 & 0 & 0 & 0 & 0 & 0 & 0 & 0 & 0 & 0 & 0 & 0 & 0 \\
 0 & 0 & 0 & 0 & \frac{135 \sqrt{\frac{21}{2}}}{2 \pi } & 0 & 0 & 0 & 0 & 0 & 0 & \frac{405 \sqrt{7}}{2 \pi } & 0 & 0 & 0 \\
 \frac{36 \sqrt{5}}{\pi } & 0 & 0 & \frac{450}{\pi } & 0 & 0 & 0 & 0 & 0 & 0 & \frac{648 \sqrt{5}}{\pi } & 0 & 0 & 0 & 0 \\
 0 & 0 & 0 & 0 & -\frac{495 \sqrt{\frac{3}{2}}}{2 \pi } & 0 & 0 & 0 & 0 & 0 & 0 & -\frac{1215}{\pi } & 0 & -\frac{405 \sqrt{7}}{2 \pi } & 0 \\
 0 & \frac{72 \sqrt{15}}{\pi } & 0 & 0 & 0 & 0 & 0 & 0 & \frac{648 \sqrt{5}}{\pi } & 0 & 0 & 0 & 0 & 0 & 0 \\
 0 & 0 & -\frac{495 \sqrt{\frac{3}{2}}}{2 \pi } & 0 & 0 & 0 & 0 & \frac{405 \sqrt{7}}{2 \pi } & 0 & -\frac{1215}{\pi } & 0 & 0 & 0 & 0 & 0 \\
 0 & 0 & 0 & 0 & 0 & 0 & 0 & 0 & 0 & 0 & 0 & 0 & 0 & 0 & 0 \\
 0 & 0 & -\frac{135 \sqrt{\frac{21}{2}}}{2 \pi } & 0 & 0 & 0 & 0 & 0 & 0 & -\frac{405 \sqrt{7}}{2 \pi } & 0 & 0 & 0 & 0 & 0 \\
 0 & 0 & 0 & 0 & 0 & 0 & 0 & 0 & 0 & 0 & 0 & 0 & 0 & 0 & 0 \\
\end{array}
\right)\]

\[\left(
\begin{array}{ccccccccccccccc}
 0 & 0 & \frac{20 \sqrt{15}}{\pi } & 0 & 0 & 0 & 0 & \frac{18 \sqrt{70}}{\pi } & 0 & \frac{18 \sqrt{10}}{\pi } & 0 & 0 & 0 & 0 & 0 \\
 0 & 0 & 0 & 0 & \frac{75}{\pi } & 0 & 0 & 0 & 0 & 0 & 0 & \frac{405 \sqrt{\frac{3}{2}}}{2 \pi } & 0 & \frac{45 \sqrt{\frac{21}{2}}}{2 \pi } & 0
\\
 \frac{20 \sqrt{15}}{\pi } & 0 & 0 & \frac{25 \sqrt{3}}{\pi } & 0 & -\frac{15}{\pi } & 0 & 0 & 0 & 0 & \frac{135 \sqrt{15}}{2 \pi } & 0 & -\frac{45
\sqrt{3}}{\pi } & 0 & -\frac{135 \sqrt{21}}{2 \pi } \\
 0 & 0 & \frac{25 \sqrt{3}}{\pi } & 0 & 0 & 0 & 0 & \frac{495 \sqrt{\frac{7}{2}}}{2 \pi } & 0 & -\frac{855}{2 \sqrt{2} \pi } & 0 & 0 & 0 & 0 & 0
\\
 0 & \frac{75}{\pi } & 0 & 0 & 0 & 0 & \frac{135 \sqrt{21}}{2 \pi } & 0 & -\frac{495 \sqrt{3}}{2 \pi } & 0 & 0 & 0 & 0 & 0 & 0 \\
 0 & 0 & -\frac{15}{\pi } & 0 & 0 & 0 & 0 & -\frac{45 \sqrt{\frac{21}{2}}}{2 \pi } & 0 & \frac{45 \sqrt{\frac{3}{2}}}{2 \pi } & 0 & 0 & 0 & 0 & 0
\\
 0 & 0 & 0 & 0 & \frac{135 \sqrt{21}}{2 \pi } & 0 & 0 & 0 & 0 & 0 & 0 & \frac{405 \sqrt{\frac{7}{2}}}{\pi } & 0 & 0 & 0 \\
 \frac{18 \sqrt{70}}{\pi } & 0 & 0 & \frac{495 \sqrt{\frac{7}{2}}}{2 \pi } & 0 & -\frac{45 \sqrt{\frac{21}{2}}}{2 \pi } & 0 & 0 & 0 & 0 & \frac{243
\sqrt{\frac{35}{2}}}{\pi } & 0 & -\frac{405 \sqrt{\frac{7}{2}}}{2 \pi } & 0 & 0 \\
\end{array}
\right)\]

\[\left(
\begin{array}{ccccccccccccccc}
 0 & 0 & \frac{8 \sqrt{15}}{\pi } & 0 & 0 & 0 & 0 & \frac{18 \sqrt{70}}{\pi } & 0 & -\frac{18 \sqrt{10}}{\pi } & 0 & 0 & 0 & 0 & 0 \\
 0 & 0 & 0 & 0 & \frac{90}{\pi } & 0 & 0 & 0 & 0 & 0 & 0 & \frac{675 \sqrt{\frac{3}{2}}}{2 \pi } & 0 & -\frac{45 \sqrt{\frac{21}{2}}}{2 \pi } & 0
\\
 \frac{8 \sqrt{15}}{\pi } & 0 & 0 & \frac{10 \sqrt{3}}{\pi } & 0 & -\frac{30}{\pi } & 0 & 0 & 0 & 0 & \frac{243 \sqrt{15}}{2 \pi } & 0 & -\frac{90
\sqrt{3}}{\pi } & 0 & -\frac{135 \sqrt{21}}{2 \pi } \\
 0 & 0 & \frac{10 \sqrt{3}}{\pi } & 0 & 0 & 0 & 0 & \frac{495 \sqrt{\frac{7}{2}}}{2 \pi } & 0 & -\frac{1035}{2 \sqrt{2} \pi } & 0 & 0 & 0 & 0 & 0
\\
 0 & \frac{90}{\pi } & 0 & 0 & 0 & 0 & \frac{135 \sqrt{21}}{2 \pi } & 0 & -\frac{405 \sqrt{3}}{2 \pi } & 0 & 0 & 0 & 0 & 0 & 0 \\
 0 & 0 & -\frac{30}{\pi } & 0 & 0 & 0 & 0 & \frac{45 \sqrt{\frac{21}{2}}}{2 \pi } & 0 & -\frac{225 \sqrt{\frac{3}{2}}}{2 \pi } & 0 & 0 & 0 & 0 &
0 \\
 0 & 0 & 0 & 0 & \frac{135 \sqrt{21}}{2 \pi } & 0 & 0 & 0 & 0 & 0 & 0 & \frac{405 \sqrt{\frac{7}{2}}}{\pi } & 0 & 0 & 0 \\
 \frac{18 \sqrt{70}}{\pi } & 0 & 0 & \frac{495 \sqrt{\frac{7}{2}}}{2 \pi } & 0 & \frac{45 \sqrt{\frac{21}{2}}}{2 \pi } & 0 & 0 & 0 & 0 & \frac{243
\sqrt{\frac{35}{2}}}{\pi } & 0 & \frac{405 \sqrt{\frac{7}{2}}}{2 \pi } & 0 & 0 \\
\end{array}
\right)\]

\[\left(
\begin{array}{ccccccccccccccc}
 \frac{64}{\pi } & 0 & 0 & \frac{52 \sqrt{5}}{\pi } & 0 & -\frac{4 \sqrt{15}}{\pi } & 0 & 0 & 0 & 0 & \frac{72}{\pi } & 0 & -\frac{36 \sqrt{5}}{\pi
} & 0 & 0 \\
 0 & -\frac{30}{\pi } & 0 & 0 & 0 & 0 & 0 & 0 & -\frac{90 \sqrt{3}}{\pi } & 0 & 0 & 0 & 0 & 0 & 0 \\
 0 & 0 & 0 & 0 & 0 & 0 & 0 & -\frac{135 \sqrt{\frac{21}{2}}}{2 \pi } & 0 & -\frac{945 \sqrt{\frac{3}{2}}}{2 \pi } & 0 & 0 & 0 & 0 & 0 \\
 \frac{52 \sqrt{5}}{\pi } & 0 & 0 & \frac{110}{\pi } & 0 & -\frac{50 \sqrt{3}}{\pi } & 0 & 0 & 0 & 0 & -\frac{144 \sqrt{5}}{\pi } & 0 & -\frac{450}{\pi
} & 0 & 0 \\
 0 & 0 & 0 & 0 & \frac{90}{\pi } & 0 & 0 & 0 & 0 & 0 & 0 & \frac{45 \sqrt{\frac{3}{2}}}{2 \pi } & 0 & -\frac{135 \sqrt{\frac{21}{2}}}{2 \pi } & 0
\\
 -\frac{4 \sqrt{15}}{\pi } & 0 & 0 & -\frac{50 \sqrt{3}}{\pi } & 0 & -\frac{30}{\pi } & 0 & 0 & 0 & 0 & -\frac{72 \sqrt{15}}{\pi } & 0 & -\frac{90
\sqrt{3}}{\pi } & 0 & 0 \\
 0 & 0 & 0 & 0 & 0 & 0 & 0 & 0 & 0 & 0 & 0 & 0 & 0 & 0 & 0 \\
 0 & 0 & -\frac{135 \sqrt{\frac{21}{2}}}{2 \pi } & 0 & 0 & 0 & 0 & 0 & 0 & -\frac{405 \sqrt{7}}{2 \pi } & 0 & 0 & 0 & 0 & 0 \\
\end{array}
\right)\]

\[\left(
\begin{array}{ccccccccccccccc}
 0 & 0 & 0 & 0 & \frac{8 \sqrt{15}}{\pi } & 0 & 0 & 0 & 0 & 0 & 0 & -\frac{18 \sqrt{10}}{\pi } & 0 & -\frac{18 \sqrt{70}}{\pi } & 0 \\
 0 & 0 & \frac{90}{\pi } & 0 & 0 & 0 & 0 & \frac{45 \sqrt{\frac{21}{2}}}{2 \pi } & 0 & \frac{675 \sqrt{\frac{3}{2}}}{2 \pi } & 0 & 0 & 0 & 0 & 0
\\
 0 & \frac{90}{\pi } & 0 & 0 & 0 & 0 & -\frac{135 \sqrt{21}}{2 \pi } & 0 & -\frac{405 \sqrt{3}}{2 \pi } & 0 & 0 & 0 & 0 & 0 & 0 \\
 0 & 0 & 0 & 0 & \frac{10 \sqrt{3}}{\pi } & 0 & 0 & 0 & 0 & 0 & 0 & -\frac{1035}{2 \sqrt{2} \pi } & 0 & -\frac{495 \sqrt{\frac{7}{2}}}{2 \pi } &
0 \\
 \frac{8 \sqrt{15}}{\pi } & 0 & 0 & \frac{10 \sqrt{3}}{\pi } & 0 & \frac{30}{\pi } & 0 & 0 & 0 & 0 & \frac{243 \sqrt{15}}{2 \pi } & 0 & \frac{90
\sqrt{3}}{\pi } & 0 & -\frac{135 \sqrt{21}}{2 \pi } \\
 0 & 0 & 0 & 0 & \frac{30}{\pi } & 0 & 0 & 0 & 0 & 0 & 0 & \frac{225 \sqrt{\frac{3}{2}}}{2 \pi } & 0 & \frac{45 \sqrt{\frac{21}{2}}}{2 \pi } & 0
\\
 0 & 0 & -\frac{135 \sqrt{21}}{2 \pi } & 0 & 0 & 0 & 0 & 0 & 0 & -\frac{405 \sqrt{\frac{7}{2}}}{\pi } & 0 & 0 & 0 & 0 & 0 \\
 0 & \frac{45 \sqrt{\frac{21}{2}}}{2 \pi } & 0 & 0 & 0 & 0 & 0 & 0 & \frac{405 \sqrt{\frac{7}{2}}}{2 \pi } & 0 & 0 & 0 & 0 & 0 & 0 \\
\end{array}
\right)\]

\[\left(
\begin{array}{ccccccccccccccc}
 0 & 0 & 0 & 0 & \frac{20 \sqrt{15}}{\pi } & 0 & 0 & 0 & 0 & 0 & 0 & \frac{18 \sqrt{10}}{\pi } & 0 & -\frac{18 \sqrt{70}}{\pi } & 0 \\
 0 & 0 & \frac{75}{\pi } & 0 & 0 & 0 & 0 & -\frac{45 \sqrt{\frac{21}{2}}}{2 \pi } & 0 & \frac{405 \sqrt{\frac{3}{2}}}{2 \pi } & 0 & 0 & 0 & 0 & 0
\\
 0 & \frac{75}{\pi } & 0 & 0 & 0 & 0 & -\frac{135 \sqrt{21}}{2 \pi } & 0 & -\frac{495 \sqrt{3}}{2 \pi } & 0 & 0 & 0 & 0 & 0 & 0 \\
 0 & 0 & 0 & 0 & \frac{25 \sqrt{3}}{\pi } & 0 & 0 & 0 & 0 & 0 & 0 & -\frac{855}{2 \sqrt{2} \pi } & 0 & -\frac{495 \sqrt{\frac{7}{2}}}{2 \pi } & 0
\\
 \frac{20 \sqrt{15}}{\pi } & 0 & 0 & \frac{25 \sqrt{3}}{\pi } & 0 & \frac{15}{\pi } & 0 & 0 & 0 & 0 & \frac{135 \sqrt{15}}{2 \pi } & 0 & \frac{45
\sqrt{3}}{\pi } & 0 & -\frac{135 \sqrt{21}}{2 \pi } \\
 0 & 0 & 0 & 0 & \frac{15}{\pi } & 0 & 0 & 0 & 0 & 0 & 0 & -\frac{45 \sqrt{\frac{3}{2}}}{2 \pi } & 0 & -\frac{45 \sqrt{\frac{21}{2}}}{2 \pi } & 0
\\
 0 & 0 & -\frac{135 \sqrt{21}}{2 \pi } & 0 & 0 & 0 & 0 & 0 & 0 & -\frac{405 \sqrt{\frac{7}{2}}}{\pi } & 0 & 0 & 0 & 0 & 0 \\
 0 & -\frac{45 \sqrt{\frac{21}{2}}}{2 \pi } & 0 & 0 & 0 & 0 & 0 & 0 & -\frac{405 \sqrt{\frac{7}{2}}}{2 \pi } & 0 & 0 & 0 & 0 & 0 & 0 \\
\end{array}
\right)\]

\[\left(
\begin{array}{ccccccccccccccc}
 0 & 0 & \frac{48 \sqrt{15}}{\pi } & 0 & 0 & 0 & 0 & -\frac{18 \sqrt{70}}{\pi } & 0 & \frac{18 \sqrt{10}}{\pi } & 0 & 0 & 0 & 0 & 0 \\
 0 & 0 & 0 & 0 & \frac{120}{\pi } & 0 & 0 & 0 & 0 & 0 & 0 & \frac{585 \sqrt{\frac{3}{2}}}{2 \pi } & 0 & \frac{45 \sqrt{\frac{21}{2}}}{2 \pi } & 0
\\
 \frac{48 \sqrt{15}}{\pi } & 0 & 0 & \frac{60 \sqrt{3}}{\pi } & 0 & -\frac{180}{\pi } & 0 & 0 & 0 & 0 & \frac{513 \sqrt{15}}{2 \pi } & 0 & \frac{405
\sqrt{3}}{\pi } & 0 & \frac{135 \sqrt{21}}{2 \pi } \\
 0 & 0 & \frac{60 \sqrt{3}}{\pi } & 0 & 0 & 0 & 0 & -\frac{495 \sqrt{\frac{7}{2}}}{2 \pi } & 0 & -\frac{2745}{2 \sqrt{2} \pi } & 0 & 0 & 0 & 0 &
0 \\
 0 & \frac{120}{\pi } & 0 & 0 & 0 & 0 & -\frac{135 \sqrt{21}}{2 \pi } & 0 & -\frac{225 \sqrt{3}}{2 \pi } & 0 & 0 & 0 & 0 & 0 & 0 \\
 0 & 0 & -\frac{180}{\pi } & 0 & 0 & 0 & 0 & -\frac{45 \sqrt{\frac{21}{2}}}{2 \pi } & 0 & -\frac{1035 \sqrt{\frac{3}{2}}}{2 \pi } & 0 & 0 & 0 & 0
& 0 \\
 0 & 0 & 0 & 0 & -\frac{135 \sqrt{21}}{2 \pi } & 0 & 0 & 0 & 0 & 0 & 0 & -\frac{405 \sqrt{\frac{7}{2}}}{\pi } & 0 & 0 & 0 \\
 -\frac{18 \sqrt{70}}{\pi } & 0 & 0 & -\frac{495 \sqrt{\frac{7}{2}}}{2 \pi } & 0 & -\frac{45 \sqrt{\frac{21}{2}}}{2 \pi } & 0 & 0 & 0 & 0 & -\frac{243
\sqrt{\frac{35}{2}}}{\pi } & 0 & -\frac{405 \sqrt{\frac{7}{2}}}{2 \pi } & 0 & 0 \\
\end{array}
\right)\]

\[\left(
\begin{array}{ccccccccccccccc}
 \frac{64}{\pi } & 0 & 0 & -\frac{20 \sqrt{5}}{\pi } & 0 & \frac{28 \sqrt{15}}{\pi } & 0 & 0 & 0 & 0 & -\frac{18}{\pi } & 0 & 0 & 0 & \frac{18 \sqrt{35}}{\pi
} \\
 0 & \frac{90}{\pi } & 0 & 0 & 0 & 0 & \frac{45 \sqrt{21}}{2 \pi } & 0 & \frac{225 \sqrt{3}}{2 \pi } & 0 & 0 & 0 & 0 & 0 & 0 \\
 0 & 0 & -\frac{30}{\pi } & 0 & 0 & 0 & 0 & -\frac{45 \sqrt{\frac{21}{2}}}{\pi } & 0 & \frac{45 \sqrt{\frac{3}{2}}}{\pi } & 0 & 0 & 0 & 0 & 0 \\
 -\frac{20 \sqrt{5}}{\pi } & 0 & 0 & -\frac{70}{\pi } & 0 & -\frac{10 \sqrt{3}}{\pi } & 0 & 0 & 0 & 0 & \frac{315 \sqrt{5}}{2 \pi } & 0 & -\frac{405}{\pi
} & 0 & \frac{45 \sqrt{7}}{2 \pi } \\
 0 & 0 & 0 & 0 & 0 & 0 & 0 & 0 & 0 & 0 & 0 & \frac{945 \sqrt{\frac{3}{2}}}{2 \pi } & 0 & -\frac{135 \sqrt{\frac{21}{2}}}{2 \pi } & 0 \\
 \frac{28 \sqrt{15}}{\pi } & 0 & 0 & -\frac{10 \sqrt{3}}{\pi } & 0 & \frac{150}{\pi } & 0 & 0 & 0 & 0 & -\frac{117 \sqrt{15}}{2 \pi } & 0 & \frac{135
\sqrt{3}}{\pi } & 0 & \frac{45 \sqrt{21}}{2 \pi } \\
 0 & \frac{45 \sqrt{21}}{2 \pi } & 0 & 0 & 0 & 0 & 0 & 0 & \frac{405 \sqrt{7}}{2 \pi } & 0 & 0 & 0 & 0 & 0 & 0 \\
 0 & 0 & -\frac{45 \sqrt{\frac{21}{2}}}{\pi } & 0 & 0 & 0 & 0 & -\frac{2835}{4 \pi } & 0 & \frac{405 \sqrt{7}}{4 \pi } & 0 & 0 & 0 & 0 & 0 \\
\end{array}
\right)\]

\[\left(
\begin{array}{ccccccccccccccc}
 0 & \frac{12 \sqrt{15}}{\pi } & 0 & 0 & 0 & 0 & \frac{18 \sqrt{35}}{\pi } & 0 & -\frac{18 \sqrt{5}}{\pi } & 0 & 0 & 0 & 0 & 0 & 0 \\
 \frac{12 \sqrt{15}}{\pi } & 0 & 0 & 0 & 0 & \frac{30}{\pi } & 0 & 0 & 0 & 0 & -\frac{63 \sqrt{15}}{2 \pi } & 0 & \frac{90 \sqrt{3}}{\pi } & 0 &
-\frac{45 \sqrt{21}}{2 \pi } \\
 0 & 0 & 0 & 0 & -\frac{15}{\pi } & 0 & 0 & 0 & 0 & 0 & 0 & \frac{90 \sqrt{6}}{\pi } & 0 & 0 & 0 \\
 0 & 0 & 0 & 0 & 0 & 0 & \frac{45 \sqrt{7}}{2 \pi } & 0 & -\frac{315}{2 \pi } & 0 & 0 & 0 & 0 & 0 & 0 \\
 0 & 0 & -\frac{15}{\pi } & 0 & 0 & 0 & 0 & -\frac{45 \sqrt{\frac{21}{2}}}{2 \pi } & 0 & \frac{45 \sqrt{\frac{3}{2}}}{2 \pi } & 0 & 0 & 0 & 0 & 0
\\
 0 & \frac{30}{\pi } & 0 & 0 & 0 & 0 & \frac{45 \sqrt{21}}{2 \pi } & 0 & -\frac{135 \sqrt{3}}{2 \pi } & 0 & 0 & 0 & 0 & 0 & 0 \\
 \frac{18 \sqrt{35}}{\pi } & 0 & 0 & \frac{45 \sqrt{7}}{2 \pi } & 0 & \frac{45 \sqrt{21}}{2 \pi } & 0 & 0 & 0 & 0 & -\frac{81 \sqrt{35}}{\pi } &
0 & \frac{405 \sqrt{7}}{2 \pi } & 0 & 0 \\
 0 & 0 & 0 & 0 & -\frac{45 \sqrt{\frac{21}{2}}}{2 \pi } & 0 & 0 & 0 & 0 & 0 & 0 & \frac{405 \sqrt{7}}{\pi } & 0 & 0 & 0 \\
\end{array}
\right)\]

\[\left(
\begin{array}{ccccccccccccccc}
 0 & 0 & 0 & -\frac{12 \sqrt{5}}{\pi } & 0 & \frac{4 \sqrt{15}}{\pi } & 0 & 0 & 0 & 0 & \frac{90}{\pi } & 0 & \frac{36 \sqrt{5}}{\pi } & 0 & -\frac{18
\sqrt{35}}{\pi } \\
 0 & \frac{90}{\pi } & 0 & 0 & 0 & 0 & -\frac{45 \sqrt{21}}{2 \pi } & 0 & \frac{225 \sqrt{3}}{2 \pi } & 0 & 0 & 0 & 0 & 0 & 0 \\
 0 & 0 & -\frac{90}{\pi } & 0 & 0 & 0 & 0 & -\frac{45 \sqrt{\frac{21}{2}}}{\pi } & 0 & \frac{135 \sqrt{\frac{3}{2}}}{\pi } & 0 & 0 & 0 & 0 & 0 \\
 -\frac{12 \sqrt{5}}{\pi } & 0 & 0 & -\frac{30}{\pi } & 0 & -\frac{10 \sqrt{3}}{\pi } & 0 & 0 & 0 & 0 & \frac{153 \sqrt{5}}{2 \pi } & 0 & -\frac{90}{\pi
} & 0 & -\frac{45 \sqrt{7}}{2 \pi } \\
 0 & 0 & 0 & 0 & 0 & 0 & 0 & 0 & 0 & 0 & 0 & \frac{315 \sqrt{\frac{3}{2}}}{2 \pi } & 0 & -\frac{45 \sqrt{\frac{21}{2}}}{2 \pi } & 0 \\
 \frac{4 \sqrt{15}}{\pi } & 0 & 0 & -\frac{10 \sqrt{3}}{\pi } & 0 & \frac{30}{\pi } & 0 & 0 & 0 & 0 & \frac{9 \sqrt{15}}{2 \pi } & 0 & \frac{90 \sqrt{3}}{\pi
} & 0 & -\frac{45 \sqrt{21}}{2 \pi } \\
 0 & -\frac{45 \sqrt{21}}{2 \pi } & 0 & 0 & 0 & 0 & 0 & 0 & -\frac{405 \sqrt{7}}{2 \pi } & 0 & 0 & 0 & 0 & 0 & 0 \\
 0 & 0 & -\frac{45 \sqrt{\frac{21}{2}}}{\pi } & 0 & 0 & 0 & 0 & \frac{2835}{4 \pi } & 0 & \frac{405 \sqrt{7}}{4 \pi } & 0 & 0 & 0 & 0 & 0 \\
\end{array}
\right)\]

\[\left(
\begin{array}{ccccccccccccccc}
 \frac{32}{\pi } & 0 & 0 & -\frac{4 \sqrt{5}}{\pi } & 0 & \frac{4 \sqrt{15}}{\pi } & 0 & 0 & 0 & 0 & -\frac{54}{\pi } & 0 & \frac{36 \sqrt{5}}{\pi
} & 0 & -\frac{18 \sqrt{35}}{\pi } \\
 0 & \frac{60}{\pi } & 0 & 0 & 0 & 0 & -\frac{45 \sqrt{21}}{2 \pi } & 0 & \frac{45 \sqrt{3}}{2 \pi } & 0 & 0 & 0 & 0 & 0 & 0 \\
 0 & 0 & -\frac{30}{\pi } & 0 & 0 & 0 & 0 & -\frac{45 \sqrt{\frac{21}{2}}}{\pi } & 0 & \frac{45 \sqrt{\frac{3}{2}}}{\pi } & 0 & 0 & 0 & 0 & 0 \\
 -\frac{4 \sqrt{5}}{\pi } & 0 & 0 & -\frac{20}{\pi } & 0 & -\frac{10 \sqrt{3}}{\pi } & 0 & 0 & 0 & 0 & \frac{81 \sqrt{5}}{2 \pi } & 0 & -\frac{90}{\pi
} & 0 & -\frac{45 \sqrt{7}}{2 \pi } \\
 0 & 0 & 0 & 0 & \frac{60}{\pi } & 0 & 0 & 0 & 0 & 0 & 0 & \frac{135 \sqrt{\frac{3}{2}}}{2 \pi } & 0 & -\frac{45 \sqrt{\frac{21}{2}}}{2 \pi } & 0
\\
 \frac{4 \sqrt{15}}{\pi } & 0 & 0 & -\frac{10 \sqrt{3}}{\pi } & 0 & 0 & 0 & 0 & 0 & 0 & \frac{9 \sqrt{15}}{2 \pi } & 0 & 0 & 0 & -\frac{45 \sqrt{21}}{2
\pi } \\
 0 & -\frac{45 \sqrt{21}}{2 \pi } & 0 & 0 & 0 & 0 & 0 & 0 & -\frac{405 \sqrt{7}}{2 \pi } & 0 & 0 & 0 & 0 & 0 & 0 \\
 0 & 0 & -\frac{45 \sqrt{\frac{21}{2}}}{\pi } & 0 & 0 & 0 & 0 & -\frac{2835}{4 \pi } & 0 & \frac{405 \sqrt{7}}{4 \pi } & 0 & 0 & 0 & 0 & 0 \\
\end{array}
\right)\]

\[\left(
\begin{array}{ccccccccccccccc}
 0 & \frac{20 \sqrt{15}}{\pi } & 0 & 0 & 0 & 0 & -\frac{18 \sqrt{35}}{\pi } & 0 & \frac{54 \sqrt{5}}{\pi } & 0 & 0 & 0 & 0 & 0 & 0 \\
 \frac{20 \sqrt{15}}{\pi } & 0 & 0 & -\frac{20 \sqrt{3}}{\pi } & 0 & -\frac{30}{\pi } & 0 & 0 & 0 & 0 & -\frac{45 \sqrt{15}}{2 \pi } & 0 & \frac{225
\sqrt{3}}{\pi } & 0 & \frac{45 \sqrt{21}}{2 \pi } \\
 0 & 0 & 0 & 0 & \frac{75}{\pi } & 0 & 0 & 0 & 0 & 0 & 0 & \frac{180 \sqrt{6}}{\pi } & 0 & \frac{45 \sqrt{42}}{\pi } & 0 \\
 0 & -\frac{20 \sqrt{3}}{\pi } & 0 & 0 & 0 & 0 & -\frac{45 \sqrt{7}}{2 \pi } & 0 & -\frac{675}{2 \pi } & 0 & 0 & 0 & 0 & 0 & 0 \\
 0 & 0 & \frac{75}{\pi } & 0 & 0 & 0 & 0 & -\frac{135 \sqrt{\frac{21}{2}}}{2 \pi } & 0 & -\frac{225 \sqrt{\frac{3}{2}}}{2 \pi } & 0 & 0 & 0 & 0 &
0 \\
 0 & -\frac{30}{\pi } & 0 & 0 & 0 & 0 & -\frac{45 \sqrt{21}}{2 \pi } & 0 & -\frac{495 \sqrt{3}}{2 \pi } & 0 & 0 & 0 & 0 & 0 & 0 \\
 -\frac{18 \sqrt{35}}{\pi } & 0 & 0 & -\frac{45 \sqrt{7}}{2 \pi } & 0 & -\frac{45 \sqrt{21}}{2 \pi } & 0 & 0 & 0 & 0 & \frac{81 \sqrt{35}}{\pi }
& 0 & -\frac{405 \sqrt{7}}{2 \pi } & 0 & 0 \\
 0 & 0 & 0 & 0 & -\frac{135 \sqrt{\frac{21}{2}}}{2 \pi } & 0 & 0 & 0 & 0 & 0 & 0 & -\frac{405 \sqrt{7}}{2 \pi } & 0 & 0 & 0 \\
\end{array}
\right)\]

\[\left(
\begin{array}{ccccccccccccccc}
 0 & -\frac{20 \sqrt{15}}{\pi } & 0 & 0 & 0 & 0 & -\frac{18 \sqrt{35}}{\pi } & 0 & -\frac{54 \sqrt{5}}{\pi } & 0 & 0 & 0 & 0 & 0 & 0 \\
 -\frac{20 \sqrt{15}}{\pi } & 0 & 0 & \frac{20 \sqrt{3}}{\pi } & 0 & -\frac{30}{\pi } & 0 & 0 & 0 & 0 & \frac{45 \sqrt{15}}{2 \pi } & 0 & \frac{225
\sqrt{3}}{\pi } & 0 & -\frac{45 \sqrt{21}}{2 \pi } \\
 0 & 0 & 0 & 0 & -\frac{75}{\pi } & 0 & 0 & 0 & 0 & 0 & 0 & \frac{225 \sqrt{\frac{3}{2}}}{2 \pi } & 0 & -\frac{135 \sqrt{\frac{21}{2}}}{2 \pi } &
0 \\
 0 & \frac{20 \sqrt{3}}{\pi } & 0 & 0 & 0 & 0 & -\frac{45 \sqrt{7}}{2 \pi } & 0 & \frac{675}{2 \pi } & 0 & 0 & 0 & 0 & 0 & 0 \\
 0 & 0 & -\frac{75}{\pi } & 0 & 0 & 0 & 0 & \frac{45 \sqrt{42}}{\pi } & 0 & -\frac{180 \sqrt{6}}{\pi } & 0 & 0 & 0 & 0 & 0 \\
 0 & -\frac{30}{\pi } & 0 & 0 & 0 & 0 & \frac{45 \sqrt{21}}{2 \pi } & 0 & -\frac{495 \sqrt{3}}{2 \pi } & 0 & 0 & 0 & 0 & 0 & 0 \\
 -\frac{18 \sqrt{35}}{\pi } & 0 & 0 & -\frac{45 \sqrt{7}}{2 \pi } & 0 & \frac{45 \sqrt{21}}{2 \pi } & 0 & 0 & 0 & 0 & \frac{81 \sqrt{35}}{\pi } &
0 & \frac{405 \sqrt{7}}{2 \pi } & 0 & 0 \\
 0 & 0 & 0 & 0 & \frac{45 \sqrt{42}}{\pi } & 0 & 0 & 0 & 0 & 0 & 0 & -\frac{405 \sqrt{7}}{2 \pi } & 0 & 0 & 0 \\
\end{array}
\right)\]

\[\left(
\begin{array}{ccccccccccccccc}
 0 & 0 & -\frac{4 \sqrt{15}}{\pi } & 0 & 0 & 0 & 0 & -\frac{9 \sqrt{70}}{\pi } & 0 & \frac{9 \sqrt{10}}{\pi } & 0 & 0 & 0 & 0 & 0 \\
 0 & 0 & 0 & 0 & \frac{45}{\pi } & 0 & 0 & 0 & 0 & 0 & 0 & -\frac{135 \sqrt{\frac{3}{2}}}{2 \pi } & 0 & \frac{225 \sqrt{\frac{21}{2}}}{2 \pi } &
0 \\
 -\frac{4 \sqrt{15}}{\pi } & 0 & 0 & \frac{25 \sqrt{3}}{\pi } & 0 & -\frac{75}{\pi } & 0 & 0 & 0 & 0 & -\frac{27 \sqrt{15}}{\pi } & 0 & \frac{90
\sqrt{3}}{\pi } & 0 & -\frac{45 \sqrt{21}}{\pi } \\
 0 & 0 & \frac{25 \sqrt{3}}{\pi } & 0 & 0 & 0 & 0 & \frac{225 \sqrt{\frac{7}{2}}}{2 \pi } & 0 & -\frac{225}{2 \sqrt{2} \pi } & 0 & 0 & 0 & 0 & 0
\\
 0 & \frac{45}{\pi } & 0 & 0 & 0 & 0 & \frac{45 \sqrt{21}}{\pi } & 0 & -\frac{180 \sqrt{3}}{\pi } & 0 & 0 & 0 & 0 & 0 & 0 \\
 0 & 0 & -\frac{75}{\pi } & 0 & 0 & 0 & 0 & -\frac{225 \sqrt{\frac{21}{2}}}{2 \pi } & 0 & \frac{225 \sqrt{\frac{3}{2}}}{2 \pi } & 0 & 0 & 0 & 0 &
0 \\
 0 & 0 & 0 & 0 & \frac{45 \sqrt{21}}{\pi } & 0 & 0 & 0 & 0 & 0 & 0 & -\frac{405 \sqrt{\frac{7}{2}}}{2 \pi } & 0 & \frac{2835}{2 \sqrt{2} \pi } &
0 \\
 -\frac{9 \sqrt{70}}{\pi } & 0 & 0 & \frac{225 \sqrt{\frac{7}{2}}}{2 \pi } & 0 & -\frac{225 \sqrt{\frac{21}{2}}}{2 \pi } & 0 & 0 & 0 & 0 & -\frac{243
\sqrt{\frac{35}{2}}}{2 \pi } & 0 & \frac{405 \sqrt{\frac{7}{2}}}{\pi } & 0 & -\frac{2835}{2 \sqrt{2} \pi } \\
\end{array}
\right)\]

\[\left(
\begin{array}{ccccccccccccccc}
 0 & 0 & 0 & \frac{12 \sqrt{5}}{\pi } & 0 & \frac{4 \sqrt{15}}{\pi } & 0 & 0 & 0 & 0 & -\frac{90}{\pi } & 0 & \frac{36 \sqrt{5}}{\pi } & 0 & \frac{18
\sqrt{35}}{\pi } \\
 0 & -\frac{90}{\pi } & 0 & 0 & 0 & 0 & -\frac{45 \sqrt{21}}{2 \pi } & 0 & -\frac{225 \sqrt{3}}{2 \pi } & 0 & 0 & 0 & 0 & 0 & 0 \\
 0 & 0 & 0 & 0 & 0 & 0 & 0 & -\frac{45 \sqrt{\frac{21}{2}}}{2 \pi } & 0 & -\frac{315 \sqrt{\frac{3}{2}}}{2 \pi } & 0 & 0 & 0 & 0 & 0 \\
 \frac{12 \sqrt{5}}{\pi } & 0 & 0 & \frac{30}{\pi } & 0 & -\frac{10 \sqrt{3}}{\pi } & 0 & 0 & 0 & 0 & -\frac{153 \sqrt{5}}{2 \pi } & 0 & -\frac{90}{\pi
} & 0 & \frac{45 \sqrt{7}}{2 \pi } \\
 0 & 0 & 0 & 0 & \frac{90}{\pi } & 0 & 0 & 0 & 0 & 0 & 0 & -\frac{135 \sqrt{\frac{3}{2}}}{\pi } & 0 & -\frac{45 \sqrt{\frac{21}{2}}}{\pi } & 0 \\
 \frac{4 \sqrt{15}}{\pi } & 0 & 0 & -\frac{10 \sqrt{3}}{\pi } & 0 & -\frac{30}{\pi } & 0 & 0 & 0 & 0 & \frac{9 \sqrt{15}}{2 \pi } & 0 & -\frac{90
\sqrt{3}}{\pi } & 0 & -\frac{45 \sqrt{21}}{2 \pi } \\
 0 & -\frac{45 \sqrt{21}}{2 \pi } & 0 & 0 & 0 & 0 & 0 & 0 & -\frac{405 \sqrt{7}}{2 \pi } & 0 & 0 & 0 & 0 & 0 & 0 \\
 0 & 0 & -\frac{45 \sqrt{\frac{21}{2}}}{2 \pi } & 0 & 0 & 0 & 0 & -\frac{2835}{4 \pi } & 0 & -\frac{1215 \sqrt{7}}{4 \pi } & 0 & 0 & 0 & 0 & 0 \\
\end{array}
\right)\]

\[\left(
\begin{array}{ccccccccccccccc}
 0 & 0 & 0 & 0 & -\frac{8 \sqrt{15}}{\pi } & 0 & 0 & 0 & 0 & 0 & 0 & \frac{18 \sqrt{10}}{\pi } & 0 & \frac{18 \sqrt{70}}{\pi } & 0 \\
 0 & 0 & -\frac{90}{\pi } & 0 & 0 & 0 & 0 & \frac{135 \sqrt{\frac{21}{2}}}{\pi } & 0 & \frac{135 \sqrt{\frac{3}{2}}}{\pi } & 0 & 0 & 0 & 0 & 0 \\
 0 & -\frac{90}{\pi } & 0 & 0 & 0 & 0 & -\frac{90 \sqrt{21}}{\pi } & 0 & \frac{45 \sqrt{3}}{\pi } & 0 & 0 & 0 & 0 & 0 & 0 \\
 0 & 0 & 0 & 0 & -\frac{10 \sqrt{3}}{\pi } & 0 & 0 & 0 & 0 & 0 & 0 & \frac{45}{\sqrt{2} \pi } & 0 & -\frac{225 \sqrt{\frac{7}{2}}}{\pi } & 0 \\
 -\frac{8 \sqrt{15}}{\pi } & 0 & 0 & -\frac{10 \sqrt{3}}{\pi } & 0 & -\frac{30}{\pi } & 0 & 0 & 0 & 0 & \frac{36 \sqrt{15}}{\pi } & 0 & \frac{225
\sqrt{3}}{\pi } & 0 & -\frac{90 \sqrt{21}}{\pi } \\
 0 & 0 & 0 & 0 & -\frac{30}{\pi } & 0 & 0 & 0 & 0 & 0 & 0 & \frac{45 \sqrt{\frac{3}{2}}}{\pi } & 0 & \frac{135 \sqrt{\frac{21}{2}}}{\pi } & 0 \\
 0 & 0 & -\frac{90 \sqrt{21}}{\pi } & 0 & 0 & 0 & 0 & 0 & 0 & \frac{405 \sqrt{\frac{7}{2}}}{\pi } & 0 & 0 & 0 & 0 & 0 \\
 0 & \frac{135 \sqrt{\frac{21}{2}}}{\pi } & 0 & 0 & 0 & 0 & 0 & 0 & -\frac{405 \sqrt{\frac{7}{2}}}{2 \pi } & 0 & 0 & 0 & 0 & 0 & 0 \\
\end{array}
\right)\]

\[\left(
\begin{array}{ccccccccccccccc}
 0 & 0 & 0 & 0 & \frac{12 \sqrt{15}}{\pi } & 0 & 0 & 0 & 0 & 0 & 0 & -\frac{27 \sqrt{10}}{\pi } & 0 & \frac{9 \sqrt{70}}{\pi } & 0 \\
 0 & 0 & -\frac{15}{\pi } & 0 & 0 & 0 & 0 & -\frac{45 \sqrt{\frac{21}{2}}}{2 \pi } & 0 & \frac{45 \sqrt{\frac{3}{2}}}{2 \pi } & 0 & 0 & 0 & 0 & 0
\\
 0 & -\frac{15}{\pi } & 0 & 0 & 0 & 0 & -\frac{45 \sqrt{21}}{\pi } & 0 & -\frac{45 \sqrt{3}}{\pi } & 0 & 0 & 0 & 0 & 0 & 0 \\
 0 & 0 & 0 & 0 & -\frac{15 \sqrt{3}}{\pi } & 0 & 0 & 0 & 0 & 0 & 0 & \frac{135}{2 \sqrt{2} \pi } & 0 & -\frac{225 \sqrt{\frac{7}{2}}}{2 \pi } & 0
\\
 \frac{12 \sqrt{15}}{\pi } & 0 & 0 & -\frac{15 \sqrt{3}}{\pi } & 0 & \frac{15}{\pi } & 0 & 0 & 0 & 0 & -\frac{9 \sqrt{15}}{\pi } & 0 & \frac{45 \sqrt{3}}{\pi
} & 0 & -\frac{45 \sqrt{21}}{\pi } \\
 0 & 0 & 0 & 0 & \frac{15}{\pi } & 0 & 0 & 0 & 0 & 0 & 0 & -\frac{45 \sqrt{\frac{3}{2}}}{2 \pi } & 0 & -\frac{45 \sqrt{\frac{21}{2}}}{2 \pi } & 0
\\
 0 & 0 & -\frac{45 \sqrt{21}}{\pi } & 0 & 0 & 0 & 0 & -\frac{2835}{2 \sqrt{2} \pi } & 0 & \frac{405 \sqrt{\frac{7}{2}}}{2 \pi } & 0 & 0 & 0 & 0 &
0 \\
 0 & -\frac{45 \sqrt{\frac{21}{2}}}{2 \pi } & 0 & 0 & 0 & 0 & -\frac{2835}{2 \sqrt{2} \pi } & 0 & -\frac{405 \sqrt{\frac{7}{2}}}{2 \pi } & 0 & 0
& 0 & 0 & 0 & 0 \\
\end{array}
\right)\]

\[\left(
\begin{array}{ccccccccccccccc}
 0 & 0 & \frac{8 \sqrt{15}}{\pi } & 0 & 0 & 0 & 0 & \frac{18 \sqrt{70}}{\pi } & 0 & -\frac{18 \sqrt{10}}{\pi } & 0 & 0 & 0 & 0 & 0 \\
 0 & 0 & 0 & 0 & \frac{90}{\pi } & 0 & 0 & 0 & 0 & 0 & 0 & -\frac{135 \sqrt{\frac{3}{2}}}{\pi } & 0 & \frac{135 \sqrt{\frac{21}{2}}}{\pi } & 0 \\
 \frac{8 \sqrt{15}}{\pi } & 0 & 0 & \frac{10 \sqrt{3}}{\pi } & 0 & -\frac{30}{\pi } & 0 & 0 & 0 & 0 & -\frac{36 \sqrt{15}}{\pi } & 0 & \frac{225
\sqrt{3}}{\pi } & 0 & \frac{90 \sqrt{21}}{\pi } \\
 0 & 0 & \frac{10 \sqrt{3}}{\pi } & 0 & 0 & 0 & 0 & -\frac{225 \sqrt{\frac{7}{2}}}{\pi } & 0 & -\frac{45}{\sqrt{2} \pi } & 0 & 0 & 0 & 0 & 0 \\
 0 & \frac{90}{\pi } & 0 & 0 & 0 & 0 & -\frac{90 \sqrt{21}}{\pi } & 0 & -\frac{45 \sqrt{3}}{\pi } & 0 & 0 & 0 & 0 & 0 & 0 \\
 0 & 0 & -\frac{30}{\pi } & 0 & 0 & 0 & 0 & -\frac{135 \sqrt{\frac{21}{2}}}{\pi } & 0 & \frac{45 \sqrt{\frac{3}{2}}}{\pi } & 0 & 0 & 0 & 0 & 0 \\
 0 & 0 & 0 & 0 & -\frac{90 \sqrt{21}}{\pi } & 0 & 0 & 0 & 0 & 0 & 0 & \frac{405 \sqrt{\frac{7}{2}}}{\pi } & 0 & 0 & 0 \\
 \frac{18 \sqrt{70}}{\pi } & 0 & 0 & -\frac{225 \sqrt{\frac{7}{2}}}{\pi } & 0 & -\frac{135 \sqrt{\frac{21}{2}}}{\pi } & 0 & 0 & 0 & 0 & \frac{243
\sqrt{\frac{35}{2}}}{\pi } & 0 & \frac{405 \sqrt{\frac{7}{2}}}{2 \pi } & 0 & 0 \\
\end{array}
\right)\]

\[\left(
\begin{array}{ccccccccccccccc}
 0 & -\frac{8 \sqrt{15}}{\pi } & 0 & 0 & 0 & 0 & 0 & 0 & -\frac{72 \sqrt{5}}{\pi } & 0 & 0 & 0 & 0 & 0 & 0 \\
 -\frac{8 \sqrt{15}}{\pi } & 0 & 0 & \frac{20 \sqrt{3}}{\pi } & 0 & 0 & 0 & 0 & 0 & 0 & -\frac{9 \sqrt{15}}{\pi } & 0 & -\frac{315 \sqrt{3}}{\pi
} & 0 & -\frac{45 \sqrt{21}}{\pi } \\
 0 & 0 & 0 & 0 & -\frac{90}{\pi } & 0 & 0 & 0 & 0 & 0 & 0 & -\frac{675 \sqrt{\frac{3}{2}}}{2 \pi } & 0 & -\frac{135 \sqrt{\frac{21}{2}}}{2 \pi }
& 0 \\
 0 & \frac{20 \sqrt{3}}{\pi } & 0 & 0 & 0 & 0 & 0 & 0 & \frac{180}{\pi } & 0 & 0 & 0 & 0 & 0 & 0 \\
 0 & 0 & -\frac{90}{\pi } & 0 & 0 & 0 & 0 & \frac{135 \sqrt{\frac{21}{2}}}{2 \pi } & 0 & \frac{585 \sqrt{\frac{3}{2}}}{2 \pi } & 0 & 0 & 0 & 0 &
0 \\
 0 & 0 & 0 & 0 & 0 & 0 & \frac{45 \sqrt{21}}{\pi } & 0 & \frac{315 \sqrt{3}}{\pi } & 0 & 0 & 0 & 0 & 0 & 0 \\
 0 & 0 & 0 & 0 & 0 & \frac{45 \sqrt{21}}{\pi } & 0 & 0 & 0 & 0 & 0 & 0 & \frac{405 \sqrt{7}}{\pi } & 0 & 0 \\
 0 & 0 & 0 & 0 & \frac{135 \sqrt{\frac{21}{2}}}{2 \pi } & 0 & 0 & 0 & 0 & 0 & 0 & \frac{405 \sqrt{7}}{2 \pi } & 0 & 0 & 0 \\
 -\frac{72 \sqrt{5}}{\pi } & 0 & 0 & \frac{180}{\pi } & 0 & \frac{315 \sqrt{3}}{\pi } & 0 & 0 & 0 & 0 & -\frac{81 \sqrt{5}}{\pi } & 0 & 0 & 0 & -\frac{405
\sqrt{7}}{\pi } \\
 0 & 0 & 0 & 0 & \frac{585 \sqrt{\frac{3}{2}}}{2 \pi } & 0 & 0 & 0 & 0 & 0 & 0 & \frac{405}{\pi } & 0 & -\frac{405 \sqrt{7}}{2 \pi } & 0 \\
 0 & -\frac{9 \sqrt{15}}{\pi } & 0 & 0 & 0 & 0 & 0 & 0 & -\frac{81 \sqrt{5}}{\pi } & 0 & 0 & 0 & 0 & 0 & 0 \\
 0 & 0 & -\frac{675 \sqrt{\frac{3}{2}}}{2 \pi } & 0 & 0 & 0 & 0 & \frac{405 \sqrt{7}}{2 \pi } & 0 & \frac{405}{\pi } & 0 & 0 & 0 & 0 & 0 \\
 0 & -\frac{315 \sqrt{3}}{\pi } & 0 & 0 & 0 & 0 & \frac{405 \sqrt{7}}{\pi } & 0 & 0 & 0 & 0 & 0 & 0 & 0 & 0 \\
 0 & 0 & -\frac{135 \sqrt{\frac{21}{2}}}{2 \pi } & 0 & 0 & 0 & 0 & 0 & 0 & -\frac{405 \sqrt{7}}{2 \pi } & 0 & 0 & 0 & 0 & 0 \\
 0 & -\frac{45 \sqrt{21}}{\pi } & 0 & 0 & 0 & 0 & 0 & 0 & -\frac{405 \sqrt{7}}{\pi } & 0 & 0 & 0 & 0 & 0 & 0 \\
\end{array}
\right)\]

\[\left(
\begin{array}{ccccccccccccccc}
 \frac{64}{\pi } & 0 & 0 & -\frac{20 \sqrt{5}}{\pi } & 0 & -\frac{28 \sqrt{15}}{\pi } & 0 & 0 & 0 & 0 & -\frac{18}{\pi } & 0 & 0 & 0 & \frac{18 \sqrt{35}}{\pi
} \\
 0 & \frac{90}{\pi } & 0 & 0 & 0 & 0 & -\frac{45 \sqrt{21}}{2 \pi } & 0 & \frac{225 \sqrt{3}}{2 \pi } & 0 & 0 & 0 & 0 & 0 & 0 \\
 0 & 0 & 0 & 0 & 0 & 0 & 0 & \frac{135 \sqrt{\frac{21}{2}}}{2 \pi } & 0 & \frac{945 \sqrt{\frac{3}{2}}}{2 \pi } & 0 & 0 & 0 & 0 & 0 \\
 -\frac{20 \sqrt{5}}{\pi } & 0 & 0 & -\frac{70}{\pi } & 0 & \frac{10 \sqrt{3}}{\pi } & 0 & 0 & 0 & 0 & \frac{315 \sqrt{5}}{2 \pi } & 0 & \frac{405}{\pi
} & 0 & \frac{45 \sqrt{7}}{2 \pi } \\
 0 & 0 & 0 & 0 & -\frac{30}{\pi } & 0 & 0 & 0 & 0 & 0 & 0 & \frac{45 \sqrt{\frac{3}{2}}}{\pi } & 0 & \frac{45 \sqrt{\frac{21}{2}}}{\pi } & 0 \\
 -\frac{28 \sqrt{15}}{\pi } & 0 & 0 & \frac{10 \sqrt{3}}{\pi } & 0 & \frac{150}{\pi } & 0 & 0 & 0 & 0 & \frac{117 \sqrt{15}}{2 \pi } & 0 & \frac{135
\sqrt{3}}{\pi } & 0 & -\frac{45 \sqrt{21}}{2 \pi } \\
 0 & -\frac{45 \sqrt{21}}{2 \pi } & 0 & 0 & 0 & 0 & 0 & 0 & -\frac{405 \sqrt{7}}{2 \pi } & 0 & 0 & 0 & 0 & 0 & 0 \\
 0 & 0 & \frac{135 \sqrt{\frac{21}{2}}}{2 \pi } & 0 & 0 & 0 & 0 & -\frac{2835}{4 \pi } & 0 & -\frac{2025 \sqrt{7}}{4 \pi } & 0 & 0 & 0 & 0 & 0 \\
\end{array}
\right)\]

\[\left(
\begin{array}{ccccccccccccccc}
 0 & 0 & -\frac{20 \sqrt{15}}{\pi } & 0 & 0 & 0 & 0 & -\frac{9 \sqrt{70}}{\pi } & 0 & \frac{45 \sqrt{10}}{\pi } & 0 & 0 & 0 & 0 & 0 \\
 0 & 0 & 0 & 0 & -\frac{75}{\pi } & 0 & 0 & 0 & 0 & 0 & 0 & \frac{225 \sqrt{\frac{3}{2}}}{2 \pi } & 0 & -\frac{315 \sqrt{\frac{21}{2}}}{2 \pi } &
0 \\
 -\frac{20 \sqrt{15}}{\pi } & 0 & 0 & \frac{5 \sqrt{3}}{\pi } & 0 & \frac{45}{\pi } & 0 & 0 & 0 & 0 & \frac{45 \sqrt{15}}{\pi } & 0 & -\frac{180
\sqrt{3}}{\pi } & 0 & -\frac{45 \sqrt{21}}{\pi } \\
 0 & 0 & \frac{5 \sqrt{3}}{\pi } & 0 & 0 & 0 & 0 & \frac{225 \sqrt{\frac{7}{2}}}{2 \pi } & 0 & -\frac{45}{2 \sqrt{2} \pi } & 0 & 0 & 0 & 0 & 0 \\
 0 & -\frac{75}{\pi } & 0 & 0 & 0 & 0 & \frac{45 \sqrt{21}}{\pi } & 0 & \frac{90 \sqrt{3}}{\pi } & 0 & 0 & 0 & 0 & 0 & 0 \\
 0 & 0 & \frac{45}{\pi } & 0 & 0 & 0 & 0 & \frac{315 \sqrt{\frac{21}{2}}}{2 \pi } & 0 & -\frac{135 \sqrt{\frac{3}{2}}}{2 \pi } & 0 & 0 & 0 & 0 &
0 \\
 0 & 0 & 0 & 0 & \frac{45 \sqrt{21}}{\pi } & 0 & 0 & 0 & 0 & 0 & 0 & -\frac{405 \sqrt{\frac{7}{2}}}{2 \pi } & 0 & \frac{2835}{2 \sqrt{2} \pi } &
0 \\
 -\frac{9 \sqrt{70}}{\pi } & 0 & 0 & \frac{225 \sqrt{\frac{7}{2}}}{2 \pi } & 0 & \frac{315 \sqrt{\frac{21}{2}}}{2 \pi } & 0 & 0 & 0 & 0 & -\frac{243
\sqrt{\frac{35}{2}}}{2 \pi } & 0 & 0 & 0 & -\frac{2835}{2 \sqrt{2} \pi } \\
\end{array}
\right)\]

\[\left(
\begin{array}{ccccccccccccccc}
 0 & 0 & 0 & 0 & \frac{4 \sqrt{15}}{\pi } & 0 & 0 & 0 & 0 & 0 & 0 & -\frac{9 \sqrt{10}}{\pi } & 0 & -\frac{9 \sqrt{70}}{\pi } & 0 \\
 0 & 0 & -\frac{45}{\pi } & 0 & 0 & 0 & 0 & \frac{225 \sqrt{\frac{21}{2}}}{2 \pi } & 0 & \frac{135 \sqrt{\frac{3}{2}}}{2 \pi } & 0 & 0 & 0 & 0 &
0 \\
 0 & -\frac{45}{\pi } & 0 & 0 & 0 & 0 & \frac{45 \sqrt{21}}{\pi } & 0 & \frac{180 \sqrt{3}}{\pi } & 0 & 0 & 0 & 0 & 0 & 0 \\
 0 & 0 & 0 & 0 & -\frac{25 \sqrt{3}}{\pi } & 0 & 0 & 0 & 0 & 0 & 0 & \frac{225}{2 \sqrt{2} \pi } & 0 & \frac{225 \sqrt{\frac{7}{2}}}{2 \pi } & 0
\\
 \frac{4 \sqrt{15}}{\pi } & 0 & 0 & -\frac{25 \sqrt{3}}{\pi } & 0 & -\frac{75}{\pi } & 0 & 0 & 0 & 0 & \frac{27 \sqrt{15}}{\pi } & 0 & \frac{90 \sqrt{3}}{\pi
} & 0 & \frac{45 \sqrt{21}}{\pi } \\
 0 & 0 & 0 & 0 & -\frac{75}{\pi } & 0 & 0 & 0 & 0 & 0 & 0 & \frac{225 \sqrt{\frac{3}{2}}}{2 \pi } & 0 & \frac{225 \sqrt{\frac{21}{2}}}{2 \pi } &
0 \\
 0 & 0 & \frac{45 \sqrt{21}}{\pi } & 0 & 0 & 0 & 0 & -\frac{2835}{2 \sqrt{2} \pi } & 0 & -\frac{405 \sqrt{\frac{7}{2}}}{2 \pi } & 0 & 0 & 0 & 0 &
0 \\
 0 & \frac{225 \sqrt{\frac{21}{2}}}{2 \pi } & 0 & 0 & 0 & 0 & -\frac{2835}{2 \sqrt{2} \pi } & 0 & -\frac{405 \sqrt{\frac{7}{2}}}{\pi } & 0 & 0 &
0 & 0 & 0 & 0 \\
\end{array}
\right)\]

\[\left(
\begin{array}{ccccccccccccccc}
 -\frac{64}{\pi } & 0 & 0 & \frac{32 \sqrt{5}}{\pi } & 0 & -\frac{24 \sqrt{15}}{\pi } & 0 & 0 & 0 & 0 & -\frac{72}{\pi } & 0 & \frac{36 \sqrt{5}}{\pi
} & 0 & 0 \\
 0 & 0 & 0 & 0 & 0 & 0 & \frac{135 \sqrt{21}}{\pi } & 0 & 0 & 0 & 0 & 0 & 0 & 0 & 0 \\
 0 & 0 & \frac{30}{\pi } & 0 & 0 & 0 & 0 & \frac{45 \sqrt{\frac{21}{2}}}{\pi } & 0 & -\frac{45 \sqrt{\frac{3}{2}}}{\pi } & 0 & 0 & 0 & 0 & 0 \\
 \frac{32 \sqrt{5}}{\pi } & 0 & 0 & -\frac{80}{\pi } & 0 & \frac{60 \sqrt{3}}{\pi } & 0 & 0 & 0 & 0 & \frac{36 \sqrt{5}}{\pi } & 0 & -\frac{90}{\pi
} & 0 & 0 \\
 0 & 0 & 0 & 0 & -\frac{90}{\pi } & 0 & 0 & 0 & 0 & 0 & 0 & \frac{135 \sqrt{\frac{3}{2}}}{\pi } & 0 & \frac{45 \sqrt{\frac{21}{2}}}{\pi } & 0 \\
 -\frac{24 \sqrt{15}}{\pi } & 0 & 0 & \frac{60 \sqrt{3}}{\pi } & 0 & 0 & 0 & 0 & 0 & 0 & -\frac{27 \sqrt{15}}{\pi } & 0 & 0 & 0 & \frac{135 \sqrt{21}}{\pi
} \\
 0 & \frac{135 \sqrt{21}}{\pi } & 0 & 0 & 0 & 0 & \frac{2835}{\pi } & 0 & -\frac{405 \sqrt{7}}{2 \pi } & 0 & 0 & 0 & 0 & 0 & 0 \\
 0 & 0 & \frac{45 \sqrt{\frac{21}{2}}}{\pi } & 0 & 0 & 0 & 0 & \frac{2835}{4 \pi } & 0 & -\frac{405 \sqrt{7}}{4 \pi } & 0 & 0 & 0 & 0 & 0 \\
\end{array}
\right)\]

\[\left(
\begin{array}{ccccccccccccccc}
 0 & -\frac{48 \sqrt{15}}{\pi } & 0 & 0 & 0 & 0 & 0 & 0 & \frac{72 \sqrt{5}}{\pi } & 0 & 0 & 0 & 0 & 0 & 0 \\
 -\frac{48 \sqrt{15}}{\pi } & 0 & 0 & \frac{120 \sqrt{3}}{\pi } & 0 & 0 & 0 & 0 & 0 & 0 & -\frac{54 \sqrt{15}}{\pi } & 0 & 0 & 0 & -\frac{270 \sqrt{21}}{\pi
} \\
 0 & 0 & 0 & 0 & -\frac{120}{\pi } & 0 & 0 & 0 & 0 & 0 & 0 & \frac{90 \sqrt{6}}{\pi } & 0 & -\frac{45 \sqrt{42}}{\pi } & 0 \\
 0 & \frac{120 \sqrt{3}}{\pi } & 0 & 0 & 0 & 0 & 0 & 0 & -\frac{180}{\pi } & 0 & 0 & 0 & 0 & 0 & 0 \\
 0 & 0 & -\frac{120}{\pi } & 0 & 0 & 0 & 0 & \frac{45 \sqrt{42}}{\pi } & 0 & \frac{90 \sqrt{6}}{\pi } & 0 & 0 & 0 & 0 & 0 \\
 0 & 0 & 0 & 0 & 0 & 0 & \frac{270 \sqrt{21}}{\pi } & 0 & 0 & 0 & 0 & 0 & 0 & 0 & 0 \\
 0 & 0 & 0 & 0 & 0 & \frac{270 \sqrt{21}}{\pi } & 0 & 0 & 0 & 0 & 0 & 0 & -\frac{405 \sqrt{7}}{\pi } & 0 & 0 \\
 0 & 0 & 0 & 0 & \frac{45 \sqrt{42}}{\pi } & 0 & 0 & 0 & 0 & 0 & 0 & -\frac{405 \sqrt{7}}{2 \pi } & 0 & 0 & 0 \\
 \frac{72 \sqrt{5}}{\pi } & 0 & 0 & -\frac{180}{\pi } & 0 & 0 & 0 & 0 & 0 & 0 & \frac{81 \sqrt{5}}{\pi } & 0 & 0 & 0 & \frac{405 \sqrt{7}}{\pi }
\\
 0 & 0 & 0 & 0 & \frac{90 \sqrt{6}}{\pi } & 0 & 0 & 0 & 0 & 0 & 0 & -\frac{405}{\pi } & 0 & \frac{405 \sqrt{7}}{2 \pi } & 0 \\
 0 & -\frac{54 \sqrt{15}}{\pi } & 0 & 0 & 0 & 0 & 0 & 0 & \frac{81 \sqrt{5}}{\pi } & 0 & 0 & 0 & 0 & 0 & 0 \\
 0 & 0 & \frac{90 \sqrt{6}}{\pi } & 0 & 0 & 0 & 0 & -\frac{405 \sqrt{7}}{2 \pi } & 0 & -\frac{405}{\pi } & 0 & 0 & 0 & 0 & 0 \\
 0 & 0 & 0 & 0 & 0 & 0 & -\frac{405 \sqrt{7}}{\pi } & 0 & 0 & 0 & 0 & 0 & 0 & 0 & 0 \\
 0 & 0 & -\frac{45 \sqrt{42}}{\pi } & 0 & 0 & 0 & 0 & 0 & 0 & \frac{405 \sqrt{7}}{2 \pi } & 0 & 0 & 0 & 0 & 0 \\
 0 & -\frac{270 \sqrt{21}}{\pi } & 0 & 0 & 0 & 0 & 0 & 0 & \frac{405 \sqrt{7}}{\pi } & 0 & 0 & 0 & 0 & 0 & 0 \\
\end{array}
\right)\]

\[\left(
\begin{array}{ccccccccccccccc}
 -\frac{64}{\pi } & 0 & 0 & \frac{32 \sqrt{5}}{\pi } & 0 & \frac{24 \sqrt{15}}{\pi } & 0 & 0 & 0 & 0 & -\frac{72}{\pi } & 0 & -\frac{36 \sqrt{5}}{\pi
} & 0 & 0 \\
 0 & 0 & 0 & 0 & 0 & 0 & -\frac{135 \sqrt{21}}{\pi } & 0 & 0 & 0 & 0 & 0 & 0 & 0 & 0 \\
 0 & 0 & -\frac{90}{\pi } & 0 & 0 & 0 & 0 & -\frac{45 \sqrt{\frac{21}{2}}}{\pi } & 0 & \frac{135 \sqrt{\frac{3}{2}}}{\pi } & 0 & 0 & 0 & 0 & 0 \\
 \frac{32 \sqrt{5}}{\pi } & 0 & 0 & -\frac{80}{\pi } & 0 & -\frac{60 \sqrt{3}}{\pi } & 0 & 0 & 0 & 0 & \frac{36 \sqrt{5}}{\pi } & 0 & \frac{90}{\pi
} & 0 & 0 \\
 0 & 0 & 0 & 0 & \frac{30}{\pi } & 0 & 0 & 0 & 0 & 0 & 0 & -\frac{45 \sqrt{\frac{3}{2}}}{\pi } & 0 & -\frac{45 \sqrt{\frac{21}{2}}}{\pi } & 0 \\
 \frac{24 \sqrt{15}}{\pi } & 0 & 0 & -\frac{60 \sqrt{3}}{\pi } & 0 & 0 & 0 & 0 & 0 & 0 & \frac{27 \sqrt{15}}{\pi } & 0 & 0 & 0 & -\frac{135 \sqrt{21}}{\pi
} \\
 0 & -\frac{135 \sqrt{21}}{\pi } & 0 & 0 & 0 & 0 & \frac{2835}{\pi } & 0 & \frac{405 \sqrt{7}}{2 \pi } & 0 & 0 & 0 & 0 & 0 & 0 \\
 0 & 0 & -\frac{45 \sqrt{\frac{21}{2}}}{\pi } & 0 & 0 & 0 & 0 & \frac{2835}{4 \pi } & 0 & \frac{405 \sqrt{7}}{4 \pi } & 0 & 0 & 0 & 0 & 0 \\
\end{array}
\right)\]

}

\section{Defining Equations of the Octahedral Variety}
Below, we reproduce explicitly the $15$ symmetric $10 \times 10$ matrices $P_i$ in the equations
(9) cutting out the octahedral variety.

\[\left(
\begin{array}{cccccccccc}
 \frac{756}{\pi } & 0 & 0 & 0 & 0 & \frac{54 \sqrt{21}}{\pi } & 0 & \frac{27 \sqrt{105}}{\pi } & 0 & 0 \\
 0 & 0 & 0 & 0 & 0 & 0 & 0 & 0 & 0 & 0 \\
 0 & 0 & 0 & 0 & \frac{405 \sqrt{7}}{2 \pi } & 0 & 0 & 0 & 0 & 0 \\
 0 & 0 & 0 & -\frac{810}{\pi } & 0 & 0 & 0 & 0 & 0 & 0 \\
 0 & 0 & \frac{405 \sqrt{7}}{2 \pi } & 0 & -\frac{405}{\pi } & 0 & 0 & 0 & 0 & 0 \\
 \frac{54 \sqrt{21}}{\pi } & 0 & 0 & 0 & 0 & -\frac{1944}{\pi } & 0 & \frac{648 \sqrt{5}}{\pi } & 0 & 0 \\
 0 & 0 & 0 & 0 & 0 & 0 & -\frac{2835}{\pi } & 0 & \frac{405 \sqrt{7}}{2 \pi } & 0 \\
 \frac{27 \sqrt{105}}{\pi } & 0 & 0 & 0 & 0 & \frac{648 \sqrt{5}}{\pi } & 0 & -\frac{810}{\pi } & 0 & 0 \\
 0 & 0 & 0 & 0 & 0 & 0 & \frac{405 \sqrt{7}}{2 \pi } & 0 & 0 & 0 \\
 0 & 0 & 0 & 0 & 0 & 0 & 0 & 0 & 0 & 0 \\
\end{array}
\right)\]

\[\left(
\begin{array}{cccccccccc}
 0 & 0 & 0 & 0 & 0 & 0 & \frac{27 \sqrt{\frac{105}{2}}}{\pi } & 0 & \frac{189 \sqrt{\frac{15}{2}}}{\pi } & 0 \\
 0 & 0 & 0 & 0 & \frac{405 \sqrt{\frac{7}{2}}}{\pi } & 0 & 0 & 0 & 0 & 0 \\
 0 & 0 & 0 & -\frac{405 \sqrt{\frac{7}{2}}}{2 \pi } & 0 & 0 & 0 & 0 & 0 & 0 \\
 0 & 0 & -\frac{405 \sqrt{\frac{7}{2}}}{2 \pi } & 0 & -\frac{405}{2 \sqrt{2} \pi } & 0 & 0 & 0 & 0 & 0 \\
 0 & \frac{405 \sqrt{\frac{7}{2}}}{\pi } & 0 & -\frac{405}{2 \sqrt{2} \pi } & 0 & 0 & 0 & 0 & 0 & 0 \\
 0 & 0 & 0 & 0 & 0 & 0 & -\frac{81 \sqrt{10}}{\pi } & 0 & \frac{243 \sqrt{\frac{35}{2}}}{\pi } & 0 \\
 \frac{27 \sqrt{\frac{105}{2}}}{\pi } & 0 & 0 & 0 & 0 & -\frac{81 \sqrt{10}}{\pi } & 0 & -\frac{2025}{2 \sqrt{2} \pi } & 0 & \frac{405 \sqrt{\frac{7}{2}}}{\pi
} \\
 0 & 0 & 0 & 0 & 0 & 0 & -\frac{2025}{2 \sqrt{2} \pi } & 0 & -\frac{405 \sqrt{\frac{7}{2}}}{2 \pi } & 0 \\
 \frac{189 \sqrt{\frac{15}{2}}}{\pi } & 0 & 0 & 0 & 0 & \frac{243 \sqrt{\frac{35}{2}}}{\pi } & 0 & -\frac{405 \sqrt{\frac{7}{2}}}{2 \pi } & 0 & 0
\\
 0 & 0 & 0 & 0 & 0 & 0 & \frac{405 \sqrt{\frac{7}{2}}}{\pi } & 0 & 0 & 0 \\
\end{array}
\right)\]

\[\left(
\begin{array}{cccccccccc}
 0 & 0 & 0 & \frac{27 \sqrt{105}}{\pi } & 0 & 0 & 0 & 0 & 0 & 0 \\
 0 & 0 & 0 & 0 & 0 & 0 & 0 & 0 & 0 & 0 \\
 0 & 0 & 0 & 0 & 0 & 0 & \frac{405 \sqrt{7}}{2 \pi } & 0 & 0 & 0 \\
 \frac{27 \sqrt{105}}{\pi } & 0 & 0 & 0 & 0 & \frac{648 \sqrt{5}}{\pi } & 0 & 0 & 0 & 0 \\
 0 & 0 & 0 & 0 & 0 & 0 & -\frac{1215}{\pi } & 0 & -\frac{405 \sqrt{7}}{2 \pi } & 0 \\
 0 & 0 & 0 & \frac{648 \sqrt{5}}{\pi } & 0 & 0 & 0 & 0 & 0 & 0 \\
 0 & 0 & \frac{405 \sqrt{7}}{2 \pi } & 0 & -\frac{1215}{\pi } & 0 & 0 & 0 & 0 & 0 \\
 0 & 0 & 0 & 0 & 0 & 0 & 0 & 0 & 0 & 0 \\
 0 & 0 & 0 & 0 & -\frac{405 \sqrt{7}}{2 \pi } & 0 & 0 & 0 & 0 & 0 \\
 0 & 0 & 0 & 0 & 0 & 0 & 0 & 0 & 0 & 0 \\
\end{array}
\right)\]

\[\left(
\begin{array}{cccccccccc}
 0 & 0 & \frac{189 \sqrt{\frac{15}{2}}}{\pi } & 0 & \frac{27 \sqrt{\frac{105}{2}}}{\pi } & 0 & 0 & 0 & 0 & 0 \\
 0 & 0 & 0 & 0 & 0 & 0 & \frac{405 \sqrt{\frac{7}{2}}}{\pi } & 0 & 0 & 0 \\
 \frac{189 \sqrt{\frac{15}{2}}}{\pi } & 0 & 0 & 0 & 0 & \frac{243 \sqrt{\frac{35}{2}}}{\pi } & 0 & -\frac{405 \sqrt{\frac{7}{2}}}{2 \pi } & 0 & 0
\\
 0 & 0 & 0 & 0 & 0 & 0 & -\frac{2025}{2 \sqrt{2} \pi } & 0 & \frac{405 \sqrt{\frac{7}{2}}}{2 \pi } & 0 \\
 \frac{27 \sqrt{\frac{105}{2}}}{\pi } & 0 & 0 & 0 & 0 & -\frac{81 \sqrt{10}}{\pi } & 0 & \frac{405}{2 \sqrt{2} \pi } & 0 & -\frac{405 \sqrt{\frac{7}{2}}}{\pi
} \\
 0 & 0 & \frac{243 \sqrt{\frac{35}{2}}}{\pi } & 0 & -\frac{81 \sqrt{10}}{\pi } & 0 & 0 & 0 & 0 & 0 \\
 0 & \frac{405 \sqrt{\frac{7}{2}}}{\pi } & 0 & -\frac{2025}{2 \sqrt{2} \pi } & 0 & 0 & 0 & 0 & 0 & 0 \\
 0 & 0 & -\frac{405 \sqrt{\frac{7}{2}}}{2 \pi } & 0 & \frac{405}{2 \sqrt{2} \pi } & 0 & 0 & 0 & 0 & 0 \\
 0 & 0 & 0 & \frac{405 \sqrt{\frac{7}{2}}}{2 \pi } & 0 & 0 & 0 & 0 & 0 & 0 \\
 0 & 0 & 0 & 0 & -\frac{405 \sqrt{\frac{7}{2}}}{\pi } & 0 & 0 & 0 & 0 & 0 \\
\end{array}
\right)\]

\[\left(
\begin{array}{cccccccccc}
 0 & 0 & \frac{189 \sqrt{\frac{15}{2}}}{\pi } & 0 & -\frac{27 \sqrt{\frac{105}{2}}}{\pi } & 0 & 0 & 0 & 0 & 0 \\
 0 & 0 & 0 & 0 & 0 & 0 & \frac{405 \sqrt{\frac{7}{2}}}{\pi } & 0 & 0 & 0 \\
 \frac{189 \sqrt{\frac{15}{2}}}{\pi } & 0 & 0 & 0 & 0 & \frac{243 \sqrt{\frac{35}{2}}}{\pi } & 0 & \frac{405 \sqrt{\frac{7}{2}}}{2 \pi } & 0 & 0
\\
 0 & 0 & 0 & 0 & 0 & 0 & \frac{405}{2 \sqrt{2} \pi } & 0 & -\frac{405 \sqrt{\frac{7}{2}}}{2 \pi } & 0 \\
 -\frac{27 \sqrt{\frac{105}{2}}}{\pi } & 0 & 0 & 0 & 0 & \frac{81 \sqrt{10}}{\pi } & 0 & -\frac{2025}{2 \sqrt{2} \pi } & 0 & -\frac{405 \sqrt{\frac{7}{2}}}{\pi
} \\
 0 & 0 & \frac{243 \sqrt{\frac{35}{2}}}{\pi } & 0 & \frac{81 \sqrt{10}}{\pi } & 0 & 0 & 0 & 0 & 0 \\
 0 & \frac{405 \sqrt{\frac{7}{2}}}{\pi } & 0 & \frac{405}{2 \sqrt{2} \pi } & 0 & 0 & 0 & 0 & 0 & 0 \\
 0 & 0 & \frac{405 \sqrt{\frac{7}{2}}}{2 \pi } & 0 & -\frac{2025}{2 \sqrt{2} \pi } & 0 & 0 & 0 & 0 & 0 \\
 0 & 0 & 0 & -\frac{405 \sqrt{\frac{7}{2}}}{2 \pi } & 0 & 0 & 0 & 0 & 0 & 0 \\
 0 & 0 & 0 & 0 & -\frac{405 \sqrt{\frac{7}{2}}}{\pi } & 0 & 0 & 0 & 0 & 0 \\
\end{array}
\right)\]

\[\left(
\begin{array}{cccccccccc}
 \frac{756}{\pi } & 0 & 0 & 0 & 0 & \frac{54 \sqrt{21}}{\pi } & 0 & -\frac{27 \sqrt{105}}{\pi } & 0 & 0 \\
 0 & 0 & 0 & 0 & 0 & 0 & 0 & 0 & 0 & 0 \\
 0 & 0 & 0 & 0 & -\frac{405 \sqrt{7}}{2 \pi } & 0 & 0 & 0 & 0 & 0 \\
 0 & 0 & 0 & -\frac{810}{\pi } & 0 & 0 & 0 & 0 & 0 & 0 \\
 0 & 0 & -\frac{405 \sqrt{7}}{2 \pi } & 0 & -\frac{2835}{\pi } & 0 & 0 & 0 & 0 & 0 \\
 \frac{54 \sqrt{21}}{\pi } & 0 & 0 & 0 & 0 & -\frac{1944}{\pi } & 0 & -\frac{648 \sqrt{5}}{\pi } & 0 & 0 \\
 0 & 0 & 0 & 0 & 0 & 0 & -\frac{405}{\pi } & 0 & -\frac{405 \sqrt{7}}{2 \pi } & 0 \\
 -\frac{27 \sqrt{105}}{\pi } & 0 & 0 & 0 & 0 & -\frac{648 \sqrt{5}}{\pi } & 0 & -\frac{810}{\pi } & 0 & 0 \\
 0 & 0 & 0 & 0 & 0 & 0 & -\frac{405 \sqrt{7}}{2 \pi } & 0 & 0 & 0 \\
 0 & 0 & 0 & 0 & 0 & 0 & 0 & 0 & 0 & 0 \\
\end{array}
\right)\]

\[\left(
\begin{array}{cccccccccc}
 0 & 0 & 0 & 0 & 0 & 0 & \frac{27 \sqrt{\frac{105}{2}}}{\pi } & 0 & -\frac{189 \sqrt{\frac{15}{2}}}{\pi } & 0 \\
 0 & 0 & 0 & 0 & -\frac{405 \sqrt{\frac{7}{2}}}{\pi } & 0 & 0 & 0 & 0 & 0 \\
 0 & 0 & 0 & -\frac{405 \sqrt{\frac{7}{2}}}{2 \pi } & 0 & 0 & 0 & 0 & 0 & 0 \\
 0 & 0 & -\frac{405 \sqrt{\frac{7}{2}}}{2 \pi } & 0 & -\frac{2025}{2 \sqrt{2} \pi } & 0 & 0 & 0 & 0 & 0 \\
 0 & -\frac{405 \sqrt{\frac{7}{2}}}{\pi } & 0 & -\frac{2025}{2 \sqrt{2} \pi } & 0 & 0 & 0 & 0 & 0 & 0 \\
 0 & 0 & 0 & 0 & 0 & 0 & -\frac{81 \sqrt{10}}{\pi } & 0 & -\frac{243 \sqrt{\frac{35}{2}}}{\pi } & 0 \\
 \frac{27 \sqrt{\frac{105}{2}}}{\pi } & 0 & 0 & 0 & 0 & -\frac{81 \sqrt{10}}{\pi } & 0 & -\frac{405}{2 \sqrt{2} \pi } & 0 & -\frac{405 \sqrt{\frac{7}{2}}}{\pi
} \\
 0 & 0 & 0 & 0 & 0 & 0 & -\frac{405}{2 \sqrt{2} \pi } & 0 & -\frac{405 \sqrt{\frac{7}{2}}}{2 \pi } & 0 \\
 -\frac{189 \sqrt{\frac{15}{2}}}{\pi } & 0 & 0 & 0 & 0 & -\frac{243 \sqrt{\frac{35}{2}}}{\pi } & 0 & -\frac{405 \sqrt{\frac{7}{2}}}{2 \pi } & 0 &
0 \\
 0 & 0 & 0 & 0 & 0 & 0 & -\frac{405 \sqrt{\frac{7}{2}}}{\pi } & 0 & 0 & 0 \\
\end{array}
\right)\]

\[\left(
\begin{array}{cccccccccc}
 \frac{756}{\pi } & 0 & 0 & 0 & 0 & -\frac{27 \sqrt{21}}{2 \pi } & 0 & 0 & 0 & \frac{189 \sqrt{15}}{2 \pi } \\
 0 & 0 & 0 & \frac{405 \sqrt{7}}{2 \pi } & 0 & 0 & 0 & 0 & 0 & 0 \\
 0 & 0 & -\frac{2835}{4 \pi } & 0 & \frac{405 \sqrt{7}}{4 \pi } & 0 & 0 & 0 & 0 & 0 \\
 0 & \frac{405 \sqrt{7}}{2 \pi } & 0 & -\frac{405}{\pi } & 0 & 0 & 0 & 0 & 0 & 0 \\
 0 & 0 & \frac{405 \sqrt{7}}{4 \pi } & 0 & -\frac{405}{4 \pi } & 0 & 0 & 0 & 0 & 0 \\
 -\frac{27 \sqrt{21}}{2 \pi } & 0 & 0 & 0 & 0 & -\frac{1134}{\pi } & 0 & \frac{1215 \sqrt{5}}{2 \pi } & 0 & -\frac{81 \sqrt{35}}{\pi } \\
 0 & 0 & 0 & 0 & 0 & 0 & -\frac{8505}{4 \pi } & 0 & \frac{2025 \sqrt{7}}{4 \pi } & 0 \\
 0 & 0 & 0 & 0 & 0 & \frac{1215 \sqrt{5}}{2 \pi } & 0 & -\frac{1620}{\pi } & 0 & \frac{405 \sqrt{7}}{2 \pi } \\
 0 & 0 & 0 & 0 & 0 & 0 & \frac{2025 \sqrt{7}}{4 \pi } & 0 & -\frac{2835}{4 \pi } & 0 \\
 \frac{189 \sqrt{15}}{2 \pi } & 0 & 0 & 0 & 0 & -\frac{81 \sqrt{35}}{\pi } & 0 & \frac{405 \sqrt{7}}{2 \pi } & 0 & 0 \\
\end{array}
\right)\]

\[\left(
\begin{array}{cccccccccc}
 0 & \frac{189 \sqrt{15}}{2 \pi } & 0 & -\frac{27 \sqrt{105}}{2 \pi } & 0 & 0 & 0 & 0 & 0 & 0 \\
 \frac{189 \sqrt{15}}{2 \pi } & 0 & 0 & 0 & 0 & -\frac{81 \sqrt{35}}{\pi } & 0 & \frac{405 \sqrt{7}}{2 \pi } & 0 & 0 \\
 0 & 0 & 0 & 0 & 0 & 0 & \frac{405 \sqrt{7}}{\pi } & 0 & 0 & 0 \\
 -\frac{27 \sqrt{105}}{2 \pi } & 0 & 0 & 0 & 0 & \frac{567 \sqrt{5}}{2 \pi } & 0 & -\frac{1215}{2 \pi } & 0 & -\frac{405 \sqrt{7}}{2 \pi } \\
 0 & 0 & 0 & 0 & 0 & 0 & -\frac{405}{\pi } & 0 & 0 & 0 \\
 0 & -\frac{81 \sqrt{35}}{\pi } & 0 & \frac{567 \sqrt{5}}{2 \pi } & 0 & 0 & 0 & 0 & 0 & 0 \\
 0 & 0 & \frac{405 \sqrt{7}}{\pi } & 0 & -\frac{405}{\pi } & 0 & 0 & 0 & 0 & 0 \\
 0 & \frac{405 \sqrt{7}}{2 \pi } & 0 & -\frac{1215}{2 \pi } & 0 & 0 & 0 & 0 & 0 & 0 \\
 0 & 0 & 0 & 0 & 0 & 0 & 0 & 0 & 0 & 0 \\
 0 & 0 & 0 & -\frac{405 \sqrt{7}}{2 \pi } & 0 & 0 & 0 & 0 & 0 & 0 \\
\end{array}
\right)\]

\[\left(
\begin{array}{cccccccccc}
 \frac{378}{\pi } & 0 & 0 & 0 & 0 & -\frac{81 \sqrt{21}}{2 \pi } & 0 & \frac{27 \sqrt{105}}{\pi } & 0 & -\frac{189 \sqrt{15}}{2 \pi } \\
 0 & 0 & 0 & -\frac{405 \sqrt{7}}{2 \pi } & 0 & 0 & 0 & 0 & 0 & 0 \\
 0 & 0 & -\frac{2835}{4 \pi } & 0 & \frac{405 \sqrt{7}}{4 \pi } & 0 & 0 & 0 & 0 & 0 \\
 0 & -\frac{405 \sqrt{7}}{2 \pi } & 0 & -\frac{1215}{\pi } & 0 & 0 & 0 & 0 & 0 & 0 \\
 0 & 0 & \frac{405 \sqrt{7}}{4 \pi } & 0 & -\frac{405}{4 \pi } & 0 & 0 & 0 & 0 & 0 \\
 -\frac{81 \sqrt{21}}{2 \pi } & 0 & 0 & 0 & 0 & -\frac{162}{\pi } & 0 & \frac{81 \sqrt{5}}{2 \pi } & 0 & \frac{81 \sqrt{35}}{\pi } \\
 0 & 0 & 0 & 0 & 0 & 0 & -\frac{2025}{4 \pi } & 0 & -\frac{1215 \sqrt{7}}{4 \pi } & 0 \\
 \frac{27 \sqrt{105}}{\pi } & 0 & 0 & 0 & 0 & \frac{81 \sqrt{5}}{2 \pi } & 0 & 0 & 0 & -\frac{405 \sqrt{7}}{2 \pi } \\
 0 & 0 & 0 & 0 & 0 & 0 & -\frac{1215 \sqrt{7}}{4 \pi } & 0 & -\frac{2835}{4 \pi } & 0 \\
 -\frac{189 \sqrt{15}}{2 \pi } & 0 & 0 & 0 & 0 & \frac{81 \sqrt{35}}{\pi } & 0 & -\frac{405 \sqrt{7}}{2 \pi } & 0 & 0 \\
\end{array}
\right)\]

\[\left(
\begin{array}{cccccccccc}
 0 & -\frac{189 \sqrt{15}}{2 \pi } & 0 & -\frac{81 \sqrt{105}}{2 \pi } & 0 & 0 & 0 & 0 & 0 & 0 \\
 -\frac{189 \sqrt{15}}{2 \pi } & 0 & 0 & 0 & 0 & \frac{81 \sqrt{35}}{\pi } & 0 & \frac{405 \sqrt{7}}{2 \pi } & 0 & 0 \\
 0 & 0 & 0 & 0 & 0 & 0 & -\frac{405 \sqrt{7}}{2 \pi } & 0 & 0 & 0 \\
 -\frac{81 \sqrt{105}}{2 \pi } & 0 & 0 & 0 & 0 & -\frac{729 \sqrt{5}}{2 \pi } & 0 & \frac{1215}{2 \pi } & 0 & -\frac{405 \sqrt{7}}{2 \pi } \\
 0 & 0 & 0 & 0 & 0 & 0 & \frac{810}{\pi } & 0 & -\frac{405 \sqrt{7}}{2 \pi } & 0 \\
 0 & \frac{81 \sqrt{35}}{\pi } & 0 & -\frac{729 \sqrt{5}}{2 \pi } & 0 & 0 & 0 & 0 & 0 & 0 \\
 0 & 0 & -\frac{405 \sqrt{7}}{2 \pi } & 0 & \frac{810}{\pi } & 0 & 0 & 0 & 0 & 0 \\
 0 & \frac{405 \sqrt{7}}{2 \pi } & 0 & \frac{1215}{2 \pi } & 0 & 0 & 0 & 0 & 0 & 0 \\
 0 & 0 & 0 & 0 & -\frac{405 \sqrt{7}}{2 \pi } & 0 & 0 & 0 & 0 & 0 \\
 0 & 0 & 0 & -\frac{405 \sqrt{7}}{2 \pi } & 0 & 0 & 0 & 0 & 0 & 0 \\
\end{array}%
\right)\]

\bibliographystyle{ACM-Reference-Format}
\bibliography{bibliography}